\documentclass[%
 reprint,
 amsmath,amssymb,
 aps,
]{revtex4-1}
\usepackage{float}
\usepackage{graphicx}
\usepackage{dcolumn}
\usepackage{bm}
\usepackage[utf8]{inputenc}
\usepackage{graphicx}
\usepackage{colortbl,array} 
\usepackage{multirow,bigdelim}
\usepackage{booktabs}
\usepackage[flushleft]{threeparttable}
\usepackage{float}
\usepackage{subcaption}
\usepackage{bigdelim}
\usepackage{tabularx}
\usepackage{amsmath}
\usepackage{multirow}
\usepackage{amssymb,bm,mathrsfs,bbm,amscd}
\usepackage{ragged2e}
\usepackage{hhline}
\usepackage{tabularx}
\usepackage{caption}
\usepackage{multirow}
\usepackage{tablefootnote}
\usepackage{placeins}
\usepackage{array}
\usepackage{booktabs}
\makeatletter
\newcommand{\thickhline}{%
    \noalign {\ifnum 0=`}\fi \hrule height 1pt
    \futurelet \reserved@a \@xhline
}
\newcolumntype{"}{@{\hskip\tabcolsep\vrule width 1pt\hskip\tabcolsep}}
\makeatother

\def\be{\begin{equation}}
\def\ee{\end{equation}}
\begin{document}

\preprint{APS/123-QED}

\title{Systematic study of the rotational bands in the $A\sim 250$ mass region}

\author{Anshul Dadwal}
\email{dadwal.anshul@gmail.com}
\author{Xiao-Tao He}%
\affiliation{%
Department of Nuclear Science and Technology,\\
College of Materials Science and Technology,\\
Nanjing University of Aeronautics and Astronautics, Nanjing 210016, China.
}%


%

\date{\today}

\begin{abstract}

For the first time, we conducted a comprehensive analysis of the rotational bands in the \( A\sim250 \) mass region. Utilizing a variety of rotational energy models and formulas, we have extracted free parameters for 36 rotational bands within this mass region, encompassing neutron numbers from N = 148 to 152. A significant enhancement has been made to the vibrational distortion model, leading to an exceptional match with the experimental data on dynamic moment of inertia across most of the rotational bands studied. Through the application of this revised model, we have categorized the dynamic moment of inertia into three distinct groups. The implications of the modified vibrational distortion term and how it evolves with changes in rotational frequency is also investigated. Our systematic investigation brings to light the peculiar behavior of Pu isotopes, especially in ground state bands of \( ^{242,244}\text{Pu}\), within this mass region. We provide a thorough discussion on various aspects such as the role of shape-fluctuation energy, the softness parameter, band-head and average moment of inertia, the prominence of the anti-pairing effect, and the transition from pairing to anti-pairing effects in Pu isotopes. Moreover, these findings are compared with other isotones, offering a comprehensive understanding of their unique characteristics.

\begin{description}
\item[PACS numbers]
21.10.Hw, 21.10.Re, 21.60.Ev.
\end{description}
\end{abstract}

\pacs{Valid PACS appear here}
\maketitle

\section{Introduction}
Since the moment Mendeleev first organized the elements into the periodic table, a key question in the field of natural sciences has been centered on determining the heaviest chemical element that can be encountered in nature or fabricated by human intervention.
The stability of atomic nuclei hinges on the specific quantity of neutrons and protons they contain. Notably, the heaviest known nuclei possessing a ``magic" number of both protons and neutrons is $^{208}$Pb. Coulomb repulsion between the protons grows significantly beyond proton number Z=82. The examination of the next proton and neutron shell closures and, consequently, an island of greater stability occupies a central position in the field of nuclear structural studies. It is proposed that to generate a chemical element, the composite nuclear system must survive for $10^{-14}$ seconds. This much time is necessary for the creation of the complex nucleus.
A superheavy element (SHE) is defined as one in which a macroscopic fission barrier computed would lead to a lifetime even lower than this time limit. The existence of SHE is purely predicated on the shell effects, which provide it with increased stability and finite lifetimes. The quest for superheavy elements (SHE) receives a boost through the identification of elements with Z=110-116 
\cite{Og, Og-2}. The stability of the SHE is determined by the shell-correction energy, which lowers the ground state and consequently creates a barrier against fission. The clustering of single-particle orbitals and the regions of low-level density are responsible for the appliance of shell-correction energy.\par
The synthesis of super-heavy elements (SHE) faces significant hindrance due to the limited cross-sections, resulting in insufficient experimental data to support theoretical predictions. However, because of the increasing sensitivity of setups at the ANL (Argonne), GSI (Darmstadt), JYFL (Yyv\"{a}skyl\"{a}), GANIL (Caen), and FLNR (Dubna) laboratories, the measure of $\alpha-\gamma$ or $\alpha$ conversion-electron coincidence is now attainable \cite{Reiter1,Reiter2,Reiter3,Hump,Butler,Herzberg,Bastin}. Various theoretical approaches have been employed in the past to obtain the next spherical shell gaps (or magic numbers). Beyond doubly magic nucleus $^{208}$Pb with proton number Z=82 and neutron number N=126. According to the Woods-Saxon potential, Z = 114 is the next proton spherical magic number \cite{CWIOK}, however, relativistic mean-field theory \cite{Rutz} and Hartree-Fock-Bogolyubov \cite{Bender} computations predicted magic numbers Z = 120 and 126, respectively. Microscopic-macroscopic \cite{Sobiczewski} studies predicted Z = 100 and N = 152 as shell closures, but the self-consistent approaches \cite{BENDER2,Afanasjev} predicted Z = 96, 98, 104, and N = 150. The discrepancies between these approaches could be because of the location of the high-j orbital near sphericity. The observation of high-quality spectroscopic data is crucial for theoretical predictions. The “in-beam” studies have allowed experiments
to be performed on the SHE, giving new experimental data on the rotational bands, moment of inertia (MoI), and alignment properties. For almost all the rotational bands, a fit to the bands based on the experimental kinematic MoI is used to assign spin values to the observed states \cite{Green_PhysRevLett.109.012501}. It is quite established that the MoI is sensitive to nuclear properties such as pairing strength and specific orbitals that are active near to the Fermi surface. A comprehensive investigation of the numerous parameters, such as spins, kinematic and dynamic MoI, pairing gap parameters, alignment, etc. can therefore provide invaluable information on such features in these heavy nuclei. The particle-number conserving cranked shell model (PNC-CSM) has been highly successful in characterizing the $A\sim250$ mass region \cite{He1,He200945}. Nowadays, multiple rotational bands have been observed in the heavy nuclei \cite{nndc}. The accuracy and predictive strength of the theoretical models must be evaluated for the rotational bands of the nuclei, ranging from light to the heaviest mass regions.\par

Our investigation enhances the current understanding of rotational bands within the mass spectrum around $A\sim250$, specifically focusing on isotones featuring neutron numbers from N=148 to N=152. This study extends the understanding of the evolution of the moment of inertia, the role of vibrations, shape variations and pairing, and the anti-pairing effects present.
\section{Rotational energy formulae}

\subsection{The vibration distortion model}
In the domain of low-lying nuclear energy spectra, the coupling between vibration and rotation is critically significant. Utilizing an analogous effect observed in the vibration and rotation of nuclei and molecules, the vibrational distortion model has been introduced \cite{roy}. In scenarios involving molecules with rotational and vibrational degrees of freedom, the excitation energy can be formulated as follows:
\be
F_{\nu}=(B_{\nu}-D_{\nu}I(I+1))I(I+1),
\ee
where \(\nu\) represents the vibrational frequency. Drawing inspiration from this expression and experimental evidence of octupole vibrational modes, the moment of inertia can be expressed as a function that integrates both rotational and vibrational distortions. This approach is formulated as follows:
\be
\Im^{(2)}=\Im^{(2)}_{c}\pm \Im^{(2)}_{vib}[\omega_{max}-\omega/\omega_{max}]^{2},
\label{eq1}
\ee
where \(\Im^{(2)}_{c}\) and \(\Im^{(2)}_{vib}\) represent the constant and the vibrational components of the dynamic MoI, respectively. The term accounting for frequency-dependent vibrational distortion can be ascribed to a cluster-like or binary structure within the nucleus at elevated spin levels. Direct observation of this phenomenon is feasible through the measurement of the dynamic MoI, providing insights into these complex nuclear behaviors.

\begin{table}
\caption{The parameters obtained from the least-squares fitting for $36$ rotational bands of the N=148 to 152
isotones using vibrational distortion model. The $E_{\gamma}$ represents the first intraband $\gamma-$transition
used (observed) in the calculation taken from Ref.\cite{nndc}}\label{tb1}
\begin{tabular}{c@{\hspace{1.5em}}c@{\hspace{1.5em}}c@{\hspace{1.5em}}c@{\hspace{1.5em}}c}
\thickhline
Isotope & Band  & {$E_{\gamma}$}& $\Im_{c}^{(2)}$               & $\Im_{vib}^{(2)}$             \\ [0.5ex]
        &       & (keV)         &  $\mathrm{\hbar^{2}MeV^{-1}}$ & $\mathrm{\hbar^{2}MeV^{-1}}$   \\ [1.5ex]
\hline\noalign{\smallskip}
N=148 	    &		&		    &		    &	      \\[0.1cm]\hline
$^{240}$U	&	1	&	45.0	&	50.24	&	12.26 \\[0.1cm]
$^{242}$Pu	&	1	&	44.6	&	59.06	&	9.19  \\[0.1cm]
	        &	2	&	52.3	&	95.58	&	2.01  \\[0.1cm]
	        &	3	&	201.4	&	84.72	&	3.16  \\[0.1cm]
	        &	4	&	219.2	&	85.41	&	2.20  \\[0.1cm]
$^{244}$Cm	&	1	&	43.0	&	63.82	&	5.07  \\[0.1cm]
$^{248}$Fm	&	1	&	46.0	&	54.49	&	9.37  \\[0.1cm]
\hline
N=149	    &		&		    &		    &	       \\[0.1cm]
\hline
$^{243}$Pu	&	1	&	125.0	&	90.26	&	0.82  \\[0.1cm]
	        &	2	&	149.0	&	85.55	&	1.71  \\[0.1cm]
\hline
N=150	    &		&		    &		    &	      \\[0.1cm]
\hline
$^{244}$Pu	&	1	&	44.2	&	64.97	&	4.58  \\[0.1cm]
	        &	2	&	189.0	&	84.48	&	1.90  \\[0.1cm]
	        &	3	&	231.0	&	91.88	&	0.35  \\[0.1cm]
	        &	4	&	254.0	&	88.27	&	0.97  \\[0.1cm]
$^{246}$Cm	&	1	&	42.9	&	52.33	&	14.20 \\[0.1cm]
$^{248}$Cf	&	1	&	41.5	&	62.26	&	8.57  \\[0.1cm]
$^{250}$Fm	&	1	&	44.0	&	58.89	&	8.65  \\[0.1cm]
$^{252}$No	&	1	&	46.4	&	58.30	&	5.85  \\[0.1cm]
	        &	3	&	224.0	&	57.82	&	13.52 \\[0.1cm]
	        &	4	&	247.0	&	84.19	&	1.80  \\[0.1cm]
\hline
N=151	    &		&		    &		    &	       \\[0.1cm]
\hline
$^{245}$Pu	&	1	&	138.0	&	81.78	&	1.24\\[0.1cm]
	        &	2	&	162.0	&	82.79	&	0.99\\[0.1cm]
	        &	3	&	154.0	&	82.56	&	0.60\\[0.1cm]
	        &	4	&	129.0	&	87.42	&	0.24\\[0.1cm]
$^{247}$Cm	&	1	&	134.7	&	77.73	&	5.01\\[0.1cm]
	        &	2	&	157.0	&	72.13	&	7.57\\[0.1cm]
$^{249}$Cf	&	1	&	136.0	&	82.97	&	2.23\\[0.1cm]
	        &	2	&	158.0	&	81.03	&	2.85\\[0.1cm]
$^{253}$No	&	1	&	132.0	&	44.07	&	26.47\\[0.1cm]
	        &	2	&	156.0	&	50.99	&	18.73\\[0.1cm]
\hline
N=152	    &		&		    &		    &	     \\[0.1cm]
\hline
$^{248}$Cm	&	1	&	43.4	&	52.95	&	12.07\\[0.1cm]
	        &	2	&	260.0	&	71.12	&	11.66\\[0.1cm]
	        &	3	&	167.9	&	89.46	&	0.75\\[0.1cm]
	        &	4	&	186.0	&	79.84	&	5.50\\[0.1cm]
$^{250}$Cf	&	1	&	42.7	&	53.53	&	13.48\\[0.1cm]
$^{254}$No	&	1	&	44.2	&	59.36	&	8.52\\[0.1cm]
$^{256}$Rf	&	1	&	44.0	&	50.44	&	13.90\\[0.1cm]

\hhline{=====}
\end{tabular}
\end{table}
\begin{figure}
   \centering
   \includegraphics[width=9cm]{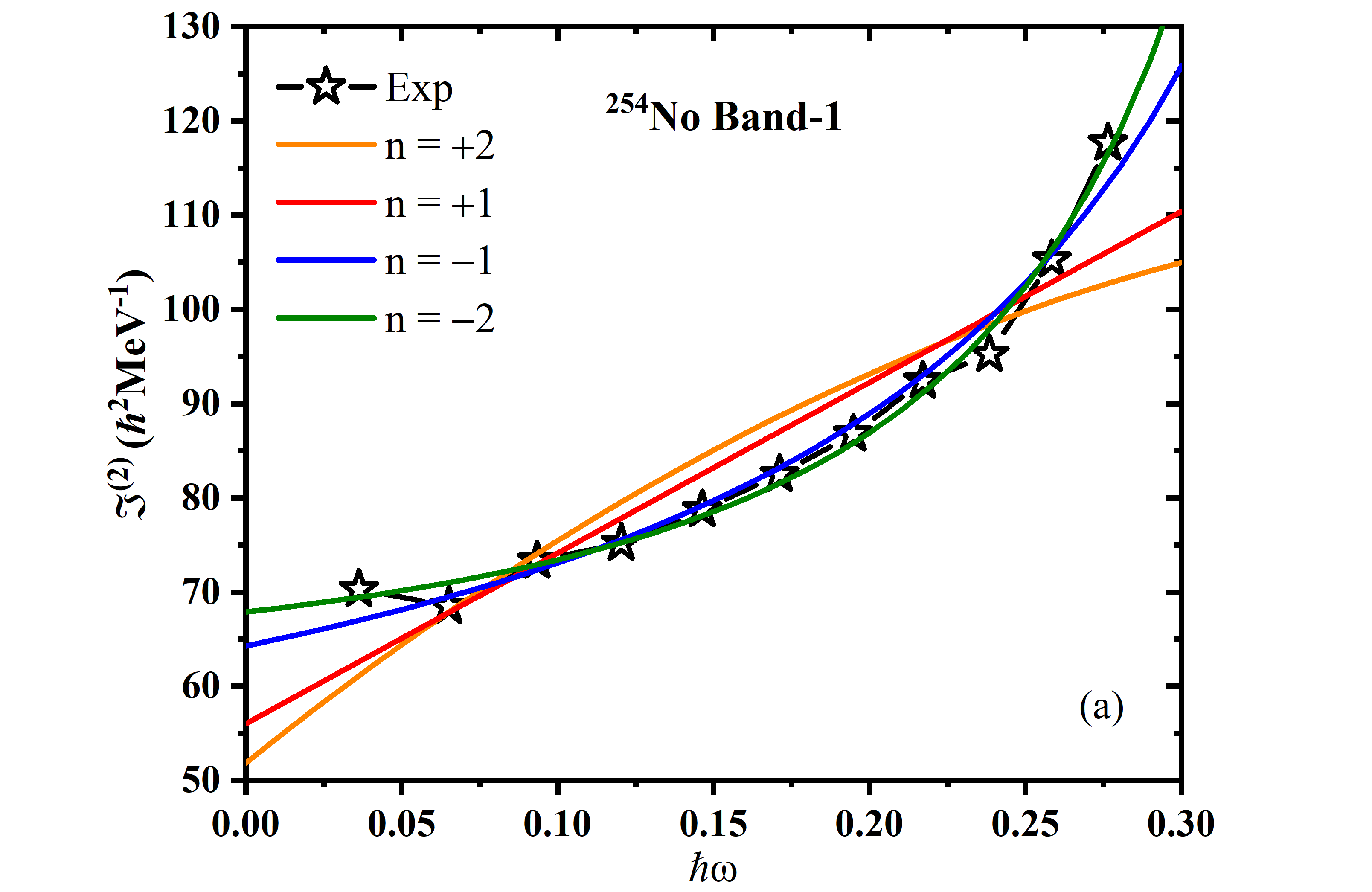}
   \includegraphics[width=9cm]{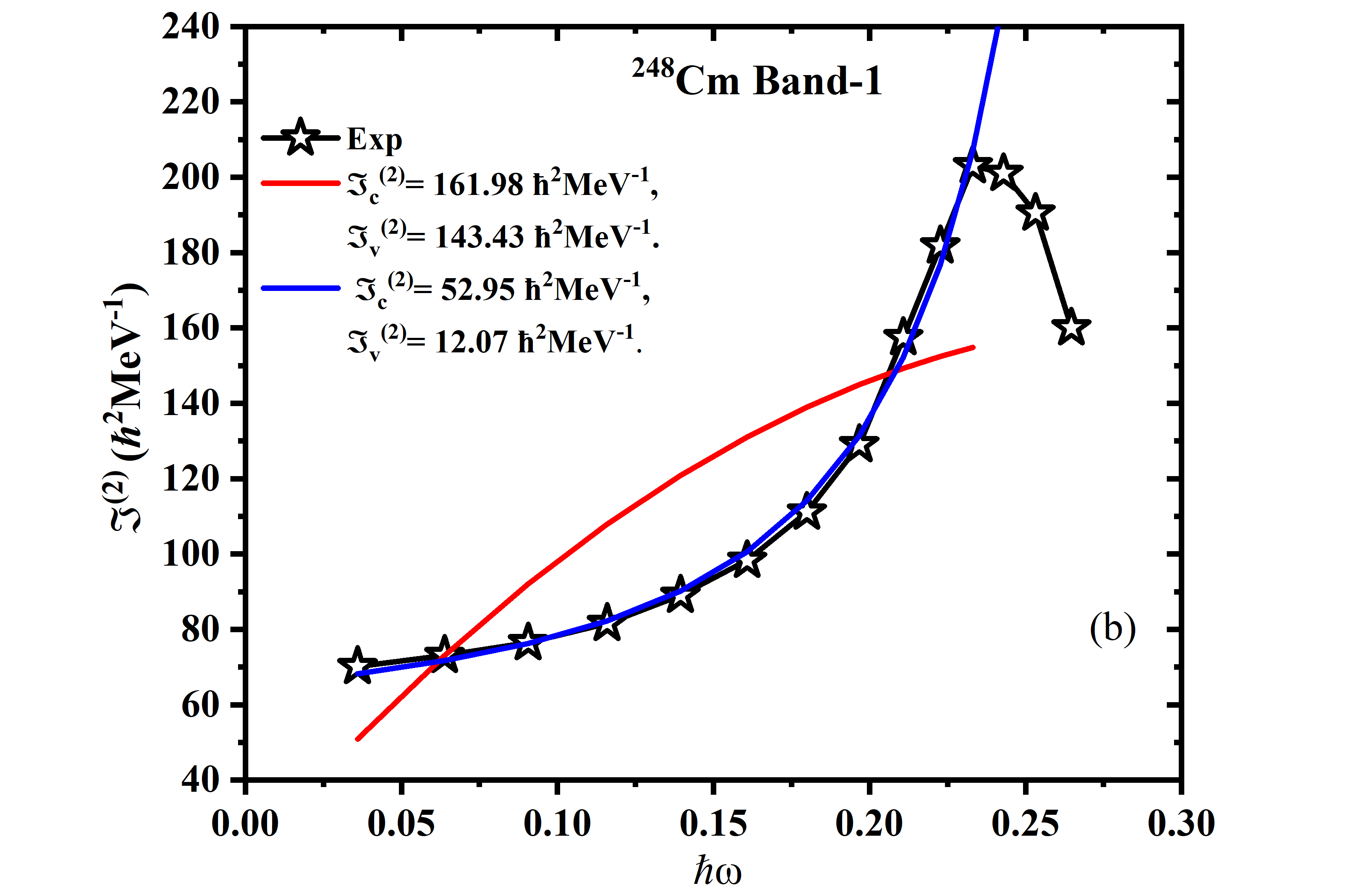}
   \captionsetup{justification=raggedright, singlelinecheck=false} 
\caption{(a) The calculated dynamic moment of inertia and comparison with experimental data of
$^{254}$No band-1 at different values of $n$. (b) The calculated dynamic moment of inertia and
comparison with experimental data of $^{248}$Cm band-1 at $n=-2$ (blue curve) and $n=+2$ (red curve).}
\label{fig1}
\end{figure}

\subsection{Shape fluctuation model}

The Shape Fluctuation Model (SF model) was introduced by incorporating the spin dependence of the intrinsic wave function. The SF model, as proposed by Sathpathy \cite{sathpathy}, offers a robust framework for quantifying the variation in the intrinsic shape within the ground-state band. The energy expression of the SF model is given as
\begin{eqnarray}
\nonumber E(I) &=&  E_{0}+ E^{'}\phi^{'}I+ (B_{0}+\phi^{'}B^{'})I(I+1), \\
   &=& B_{0} I(I+1)+ E^{'}\phi^{'}I+ \phi^{'}B^{'}I^{2}(I+1),
   \label{eq2}
\end{eqnarray}
In this model, \(E\) represents the Hartree-Fock energy, while \(B\) is the inverse of twice the moment of inertia (MoI). The term \(E_{ROT}\) denotes the rotational energy, which arises from the rotation of the stable core and is included in the first term of the equation. The additional rotational and intrinsic energies, caused by fluctuations of the core, are represented in the second and third terms, respectively. The rotational spectrum, captured by the third term and defined by a spin-dependent factor akin to the inverse of MoI, parallels the phonon spectrum indicated in the second term. Consequently, the shape fluctuation energy (\(E_{SF}\)) quantifies the combined contributions of these rotational and phonon spectra.



\begin{figure*}
\includegraphics[width=1\textwidth]{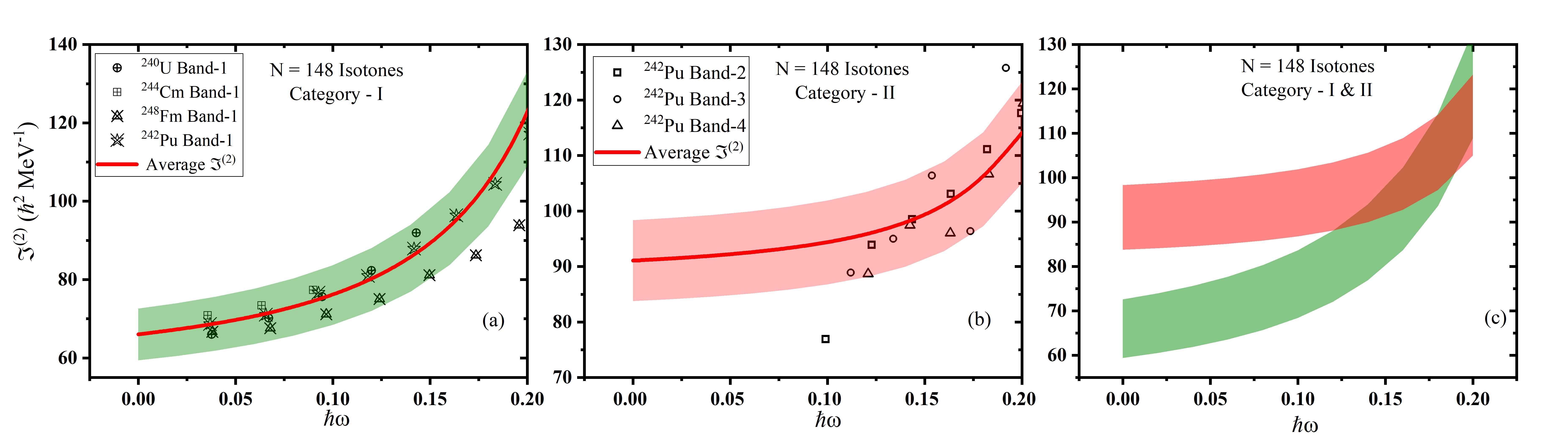}
\captionsetup{justification=raggedright, singlelinecheck=false} 
  \caption{(a) The variation between $\Im^{(2)}$ and $\hbar\omega$ for category-I bands in N=148 isotones. This graph features a red line representing the average trend, characterized by parameters $\Im_{c}^{(2)}= 57.0\hbar^{2}MeV^{-1}$ and $\Im_{vib}^{(2)}= 9.0\hbar^{2}MeV^{-1}$. The green shaded area encompasses a range of $\pm 10\%$ variation in these parameter values. (b) Illustration of the $\Im^{(2)}$ variation with $\hbar\omega$ for category-II bands in N=148 isotones. Here, the red line denotes the average trend, with parameters set at $\Im_{c}^{(2)}= 88.6\hbar^{2}MeV^{-1}$ and $\Im_{vib}^{(2)}= 2.46\hbar^{2}MeV^{-1}$. The red shaded region indicates a fluctuation of $\pm 8\%$ around the values of these two parameters. (c) Graph depicting the variation of $\Im^{(2)}$ with $\hbar\omega$ for both category-I and II bands in N=148 isotones, providing a comparative analysis of these categories.}
\vspace*{0cm}       
\label{fig2}       
\end{figure*}

\begin{figure*}
\centering
\includegraphics[width=1\textwidth]{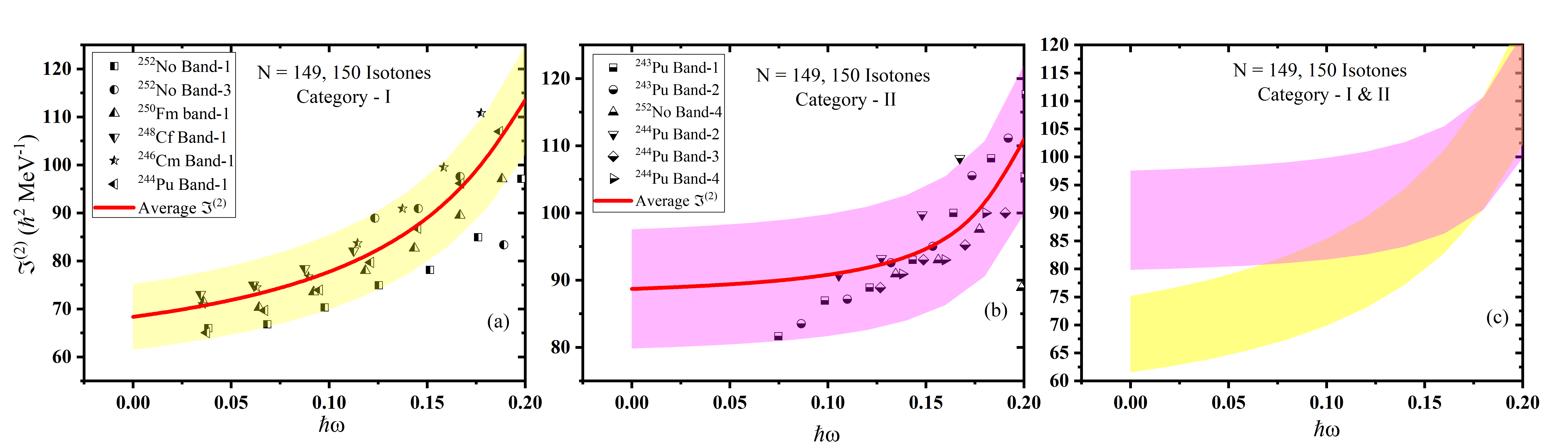}
\captionsetup{justification=raggedright, singlelinecheck=false} 
\caption{(a) The variation between $\Im^{(2)}$ and $\hbar\omega$ for category-I bands in  N=149,150 isotones. This graph features a red line representing the average trend, characterized by parameters $\Im_{c}^{(2)}= 59.1\hbar^{2}MeV^{-1}$ and $\Im_{vib}^{(2)}= 9.23\hbar^{2}MeV^{-1}$. The yellow shaded area encompasses a range of $\pm 10\%$ variation in these parameter values. (b) Illustration of the $\Im^{(2)}$ variation with $\hbar\omega$ for category-II bands in N=149, 150 isotones. Here, the red line denotes the average trend, with parameters set at $\Im_{c}^{(2)}= 87.44\hbar^{2}MeV^{-1}$ and $\Im_{vib}^{(2)}= 1.25\hbar^{2}MeV^{-1}$. The pink shaded region indicates a fluctuation of $\pm 10\%$ around the values of these two parameters. (c) Graph depicting the variation of $\Im^{(2)}$ with $\hbar\omega$ for both category-I and II bands in N=149, 150 isotones, providing a comparative analysis of these categories.}
\vspace*{0cm}       
\label{fig3}       
\end{figure*}

\begin{figure*}
\centering
\includegraphics[width=1\textwidth]{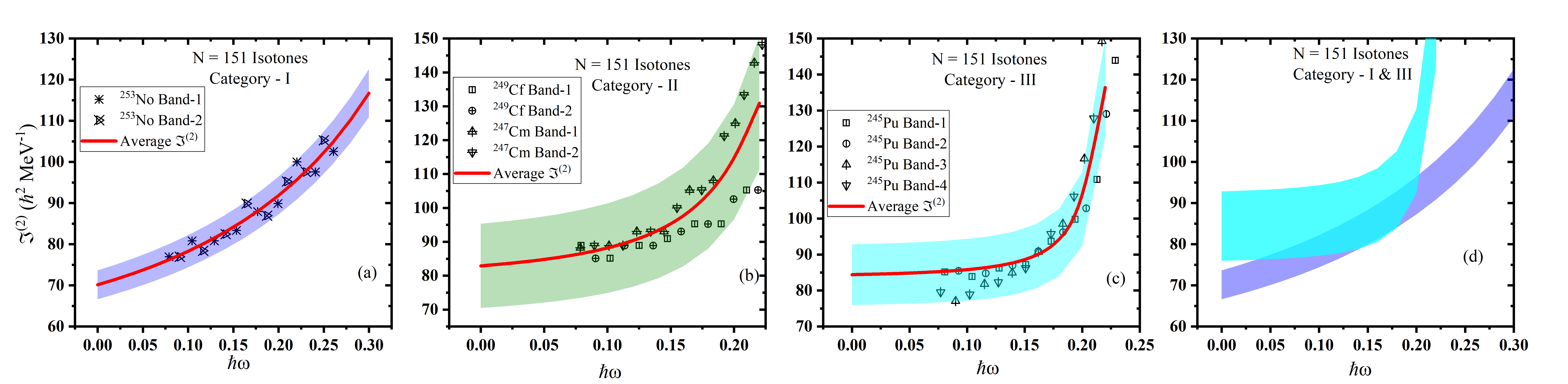}
\captionsetup{justification=raggedright, singlelinecheck=false} 

\caption{(a) The variation between $\Im^{(2)}$ and $\hbar\omega$ for category-I bands in  N=151 isotones. This graph features a red line representing the average trend, characterized by parameters $\Im_{c}^{(2)}= 47.53\hbar^{2}MeV^{-1}$ and $\Im_{vib}^{(2)}= 22.6 \hbar^{2}MeV^{-1}$. The blue shaded area encompasses a range of $\pm 5\%$ variation in these parameter values. (b) Illustration of the $\Im^{(2)}$ variation with $\hbar\omega$ for category-II bands in N=151 isotones. Here, the red line denotes the average trend, with parameters set at $\Im_{c}^{(2)}= 78.47\hbar^{2}MeV^{-1}$ and $\Im_{vib}^{(2)}= 4.42\hbar^{2}MeV^{-1}$. The green shaded region indicates a fluctuation of $\pm 15\%$ around the values of these two parameters. (c) The variation of $\Im^{(2)}$ with $\hbar\omega$ for category-III bands in N=151 isotones. The red line shows the average curve with parameters $\Im_{c}^{(2)}= 83.64\hbar^{2}MeV^{-1}$ and $\Im_{vib}^{(2)}= 0.76\hbar^{2}MeV^{-1}$. The cyan region spans the $\pm 10\%$ of the value of the two parameters. (d) Graph depicting the variation of $\Im^{(2)}$ with $\hbar\omega$ for both category-I and III bands in N=151 isotones, providing a comparative analysis of these categories.}

\vspace*{0cm}       
\label{fig4}       
\end{figure*}

\begin{figure*}
\includegraphics[width=1\textwidth]{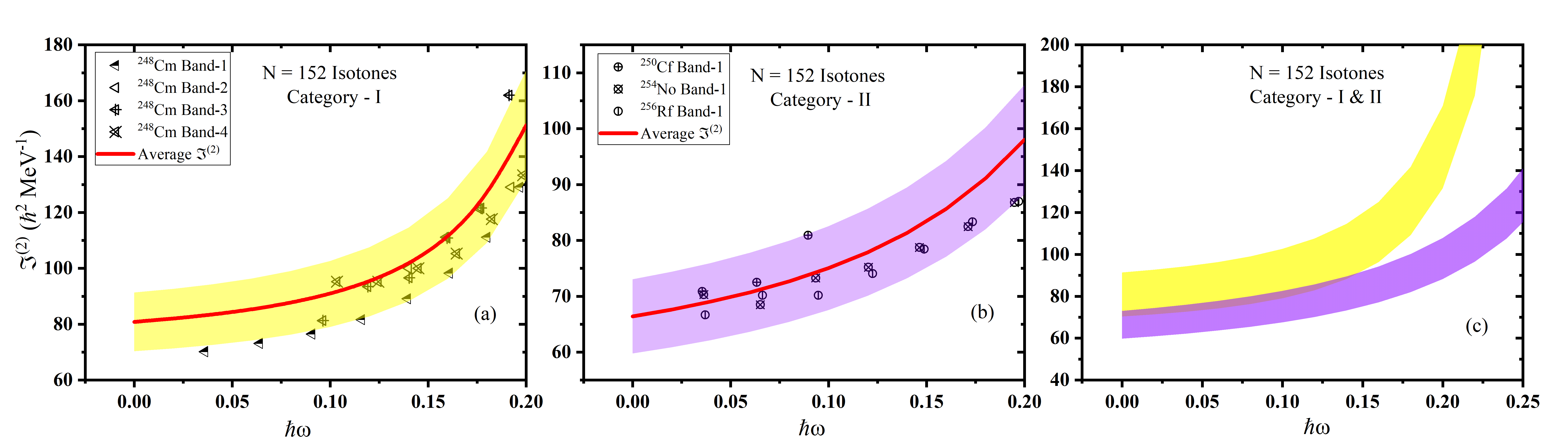}
  \captionsetup{justification=raggedright, singlelinecheck=false} 
  \caption{(a) The variation between $\Im^{(2)}$ and $\hbar\omega$ for category-I bands in  N=152 isotones. This graph features a red line representing the average trend, characterized by parameters $\Im_{c}^{(2)}= 73.34\hbar^{2}MeV^{-1}$
  and $\Im_{vib}^{(2)}= 7.49 \hbar^{2}MeV^{-1}$. The yellow shaded area encompasses a range of $\pm 13\%$ variation in these parameter values. (b) Illustration of the $\Im^{(2)}$ variation with $\hbar\omega$ for category-II bands in N=152 isotones. Here, the red line denotes the average trend, with parameters set at  $\Im_{c}^{(2)}= 54.44\hbar^{2}MeV^{-1}$ and $\Im_{vib}^{(2)}= 11.97\hbar^{2}MeV^{-1}$. The violet shaded region indicates a fluctuation of $\pm 10\%$ around the values of these two parameters. (c) Graph depicting the variation of $\Im^{(2)}$ with $\hbar\omega$ for both category-I and III bands in N=152 isotones, providing a comparative analysis of these categories.}

\vspace*{0cm}       
\label{fig5}       
\end{figure*}

\subsection{Nuclear Softness formula}

Gupta \cite{gupta} introduced a Nuclear Softness (NS) formula, particularly for the ground state bands of even-even nuclei, emphasizing the fluctuation of the moment of inertia in relation to spin (\(I\)). In a similar way, for well-deformed and transitional nuclei, Brentano \textit{et al.} \cite{brentano} proposed a parallel expression. They further developed the ``Soft-Rotor Formula" (SRF/ NS), suggesting that the moment of inertia (\(\Im\)) also exhibits a linear dependency on the excitation energy, which can be represented as \(\Im=\Im_{0}(1+\sigma I+\beta E)\). The rigid rotor energy formula is given by
\be
E=\frac{\hbar^{2}}{2\Im}I(I+1).
\label{eq3}
\ee
The observed experimental energy values consistently exceeded those predicted by the rigid rotor formula. To account for this discrepancy, the variation of the MoI with angular momentum was introduced \cite{gupta}. Consequently, Equation (\ref{eq3}) was modified to incorporate this adjustment, resulting in the following revised expression:
\be
E= \frac{\hbar^{2}}{2\Im_{I}}I(I+1).
\label{eq4}
\ee
After the Taylor series expansion of $\Im_{I}$ about its ground state value (band-head MoI) $\Im_{0}$ for $I=0$, we get
\begin{align}
E_{I}  & = \frac{\hbar^{2}}{2} \left(\left[\frac{1}{\Im_{0}}-\left(\frac{1}{\Im^{2}_{I}}\frac{\partial \Im_{I}}{\partial I}\right)\right]_{I=0}I+\right.\nonumber\\[1mm]
 & \qquad \left.\left[\frac{2}{\Im^{3}_{I}}\left(\frac{\partial \Im_{I}}{\partial I}\right)^{2}-\frac{1}{\Im^{2}_{I}}\frac{\partial^{2} \Im_{I}}{\partial I^{2}}\right]_{I=0}\frac{I^{2}}{2!}+...\right)I(I+1).
\label{eq5}
\end{align}

Morinaga \cite{morinaga} stated that, as the spin (I) of the nuclei increases, the moment of inertia approaches the ``rigid'' value and this percentage increase in the moment of inertia per unit change in angular momentum ($(1/\Im_{0})\times \Delta\Im_{0}/\Delta I$)
is defined as the ``softness parameter''. The nuclear softness parameter thus serves as an indicator of the nucleus's rigidity, which intensifies with increasing deformation. Consequently, Equation (\ref{eq5}) is reformulated to reflect this understanding
\begin{align}
E_{I} & = \frac{\hbar^{2}I(I+1)}{2\Im_{0}}\frac{1}{(1+\sigma_{1} I)}\times \nonumber\\[1mm]
 & \qquad \left(1-\frac{\sigma_{2}I^{^2}}{(1+\sigma_{1}I+\sigma_{2}I^{2})}-
\frac{\sigma_{3}I^{^3}}{(1+\sigma_{1}I+\sigma_{3}I^{3})}+...\right),
\label{eq6}
\end{align}
where,
\be
\sigma_{1}=\frac{1}{\Im_{0}}\frac{\Delta\Im_{0}}{\Delta I},
\sigma_{2}=\frac{1}{2! \Im_{0}}\frac{\partial^{2}\Im_{0}}{\partial I^{2}},
\sigma_{3}=\frac{1}{3! \Im_{0}}\frac{\partial^{3}\Im_{0}}{\partial I^{3}}....
\label{eq7}
\ee
are the constants of first, second, third etc., orders of the ``nuclear softness''. Keeping the nuclear softness to only first order i.e. putting $\sigma_{2}, \sigma_{3}.... = 0$, we get a two parameter formula. In this case, Eq. (\ref{eq6}) can be written as ($\sigma_1=\sigma$),
\be
E = \frac{\hbar^2}{2\Im_0}\times\frac{I(I+1)}{(1+\sigma I)},
\label{eq8}
\ee
\justify
where $\Im_{0}$ and $\sigma$ are the fitting parameters.
\subsection{The semiclassical particle rotor model}
In the semiclassical-PRM, the axially symmetric Hamiltonian for the rotor plus one valence particle is \cite{jain}
\begin{equation*}
H=Qj_{3}^{2}+A(I-j)^{2}.
\end{equation*}
Here total angular momentum $I$ is the sum of the core rotational angular momentum $R$ and the angular momentum of the valence particle $j$. The rotor energy formula obtained is \cite{dudeja}
\be
E_{rot}(I)=\frac{\hbar^{2}}{2\Im_{(avg)}}\left[I(I+1)-2iI+i(i+1)\right],
\label{eq9}
\ee
where $\Im_{(avg)}$ (the average value of moment of inertia over the whole band) and $i_{avg}$ (the average alignment over a limited range of angular momentum) are the fitting parameters.

\subsection{The variable moment of inertia inspired interacting boson model}

The Hamiltonian of the variable moment of inertia (VMI) inspired interacting boson model (IBM) is \cite{Yuxin_JPG}
\begin{equation}
H=E_{0}+\kappa \hat{Q}^{(2)}.\hat{Q}^{(2)}+\frac{C_{0}}{1+f\hat{L}.\hat{L}}\hat{L}.\hat{L}
\label{eq10}
\end{equation}
where $\hat{Q}^{(2)}$ and $ \hat{L}$ are the quadrupole and angular momentum operator, respectively. The term $f$ is identified as the Arima coefficient. It has been underscored that the expansion of the Arima coefficient plays a critical role in characterizing the evolving nature of the dynamic MoI. Consequently, the energy expression in the framework of the VMI model can be written as \cite{Yuxin_JPG}
\begin{align}\label{eq11}
  E= & E_{0}(N_{B},N_{F}) \nonumber\\[1mm]
   & +\frac{C_{0}}{1+f_{1}I(I+1)+f_{2}I^{2}(I+1)^{2}}I(I+1).
\end{align}
This expression is the one given by a core with the $SU(3)$ symmetry plus a pseudospin $S$. To describe the $\Delta I=4$ bifurcation, the SU(3) symmetry must be broken and the interaction $SU_{sdg}(5)$ symmetry as a perturbation was taken into account \cite{yuxin_prc_1}. Hence, the energy of the state can be written as
\begin{align}
    E &= E_{0}(N_{B}, N_{F}) \nonumber \\[2mm]
    &\quad + A \biggl[ n_{1}(n_{1}+4) + n_{2}(n_{1}+2) \nonumber \\[2mm]
    &\quad + n_{3}^{2} + n_{4}(n_{4}-2) - \frac{1}{5}(n_{1}+n_{2}+n_{3}+n_{4})^{2} \biggr] \nonumber \\[2mm]
    &\quad + B[\tau_{1}(\tau_{1}+3) + \tau_{2}(\tau_{2}+1)] \nonumber \\[2mm]
    &\quad + \frac{C_{0}}{1+f_{1}I(I+1)+f_{2}I^{2}(I+1)^{2}} I(I+1).
    \label{eq12}
\end{align}
The perturbed SU(3) limit of the $sdg$ IBM can describe the rotational bands. Moreover, the $SU_{sdg}(5)$ limit of $sdg$ IBM is relevant for deformed nuclei as well as the $SU_{sdg}$(3) limit. The calculation of the hexadecupole deformation parameter $\beta_{4}$,
the two-nucleon transfer cross section, and the energy spectra illustrated that the $SU_{sdg}(5)$ limit has almost the same property in describing deformed rotational nuclear spectra as the $SU_{sdg}(3)$ does. Such findings suggest that the $SU_{sdg}(5)$ symmetry within the $sdg$ IBM framework is indicative of a dual presence of shape coexistence and shape phase transition, influenced primarily by hexadecapole deformation and angular momentum dynamics. Since the irreducible representation (irrep) ($\lambda,\mu$), the irrep $[n_{1},n_{2},n_{3},n_{4}]$ of $SU_{sdg}(5)$ contributes nothing to the excitation energy of the states in the band. Hence, only the contribution of the perturbation to the energy of the SD bands is with the $SO_{sdg}(5)$ symmetry. Now the Eq.(\ref{eq12}) can be written as
\begin{align}
E & =E_{0}(N_{B},N_{F})\nonumber\\[1mm]
& \qquad +B[\tau_{1}(\tau_{1}+3)+\tau_{2}(\tau_{2}+1)]\nonumber\\[1mm]
 & \qquad +\frac{C_{0}}{1+f_{1}I(I+1)+f_{2}I^{2}(I+1)^{2}}I(I+1)
\label{eq13}
\end{align}
where $I=I-i,(\tau_{1},\tau_{2})$ is the irrep of the SO(5) group.
In more realistic calculation, the irrep $(\tau_{1},\tau_{2})$ are given as
\begin{equation}
(\tau_{1}, \tau_{2}) =
\begin{cases}
    \begin{aligned}
        &\biggl(\dfrac{L}{2}, 0\biggr), \\
        &\text{if } L = 4k, 4k + 1 \quad (k = 0, 1, \ldots)
    \end{aligned} \\
    \begin{aligned}
        &\biggl(\dfrac{L}{2} - 1, 2\biggr), \\
        &\text{if } L = 4k + 2, 4k + 3 \quad (k = 0, 1, \ldots)
    \end{aligned}
\end{cases}
\end{equation}
Here $[L/2]$ denoted the integer part of the $L$ and $B$, $C_{0}$, $f_{1}$ and $f_{2}$ are the free parameters \cite{Yuxin_JPG,yuxin_prc_1,yuxin_prc_2,yuxin_prc_3,yuxin_prc_4,yuxin_prc_5,yuxin_prc_6,ZhangDaLi,ZhangDaLi2}.\par
\begin{figure}
\centering

   \includegraphics[width=8.5cm]{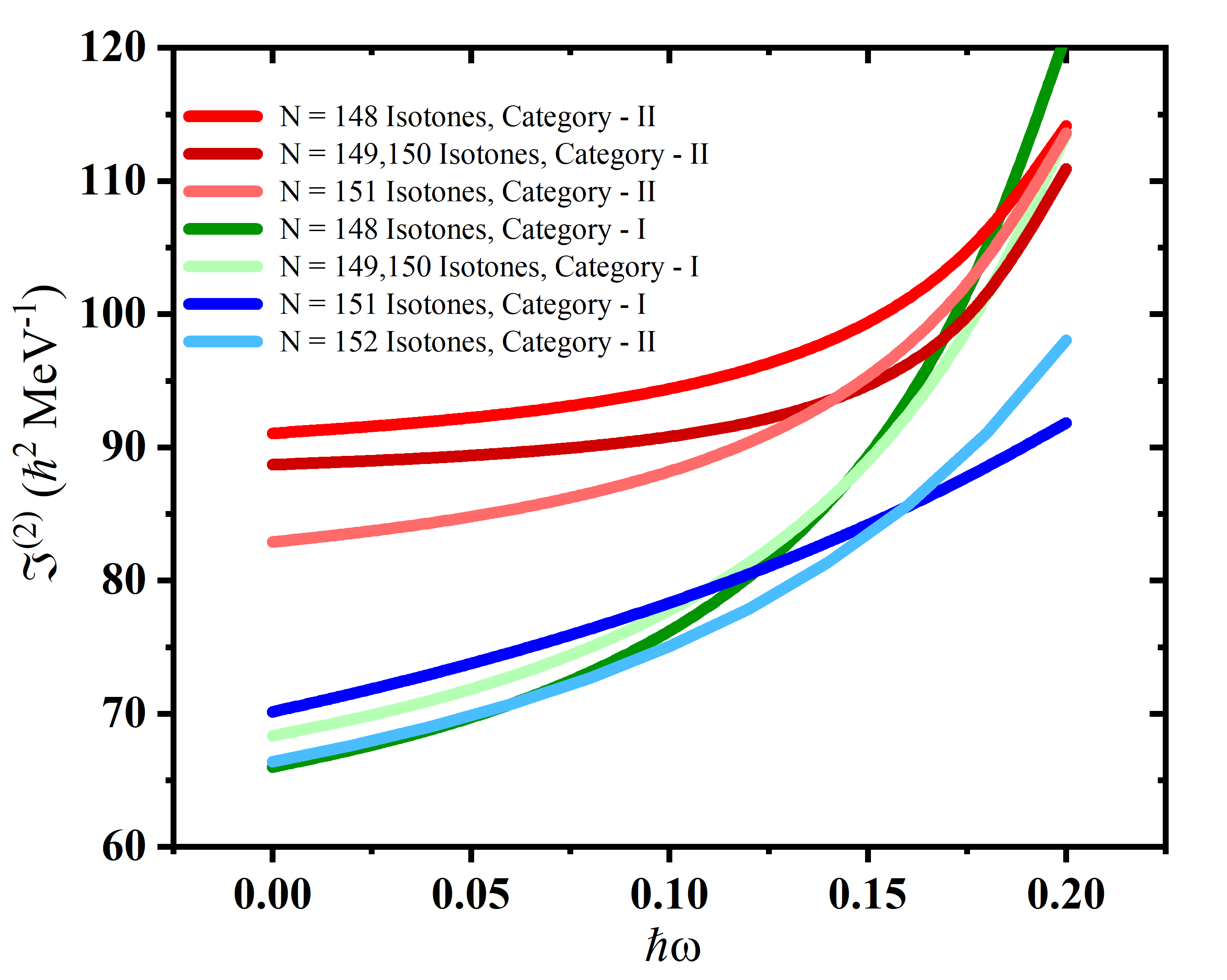}
\captionsetup{justification=raggedright, singlelinecheck=false}
\caption{The variation of average $\Im^{(2)}$ with $\hbar\omega$ for all categories combined of rotational bands in N= 148 to 152 isotones.}
\label{fig6}
\end{figure}
\begin{figure}
\centering

   \includegraphics[width=6.5cm]{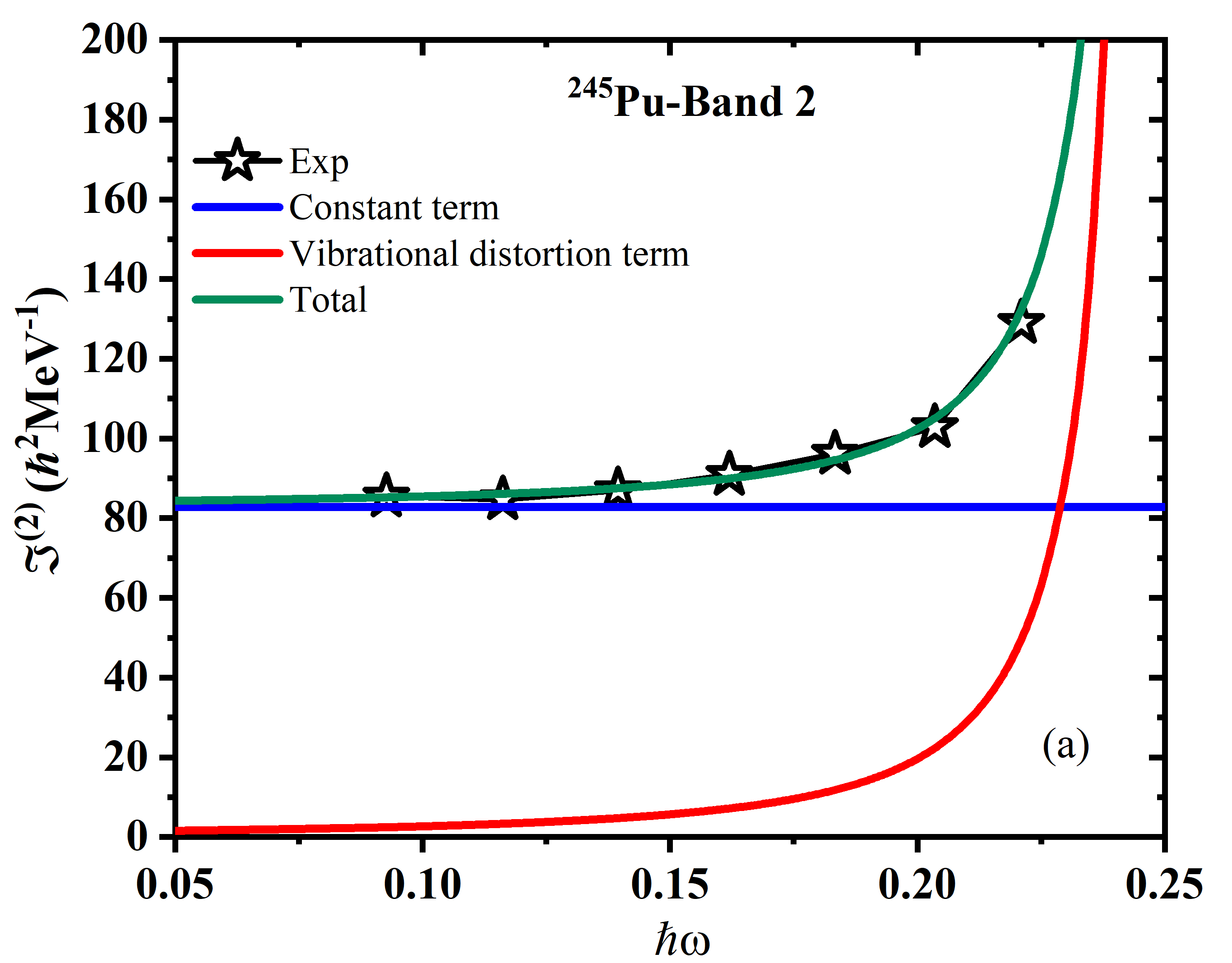}

   \includegraphics[width=6.5cm]{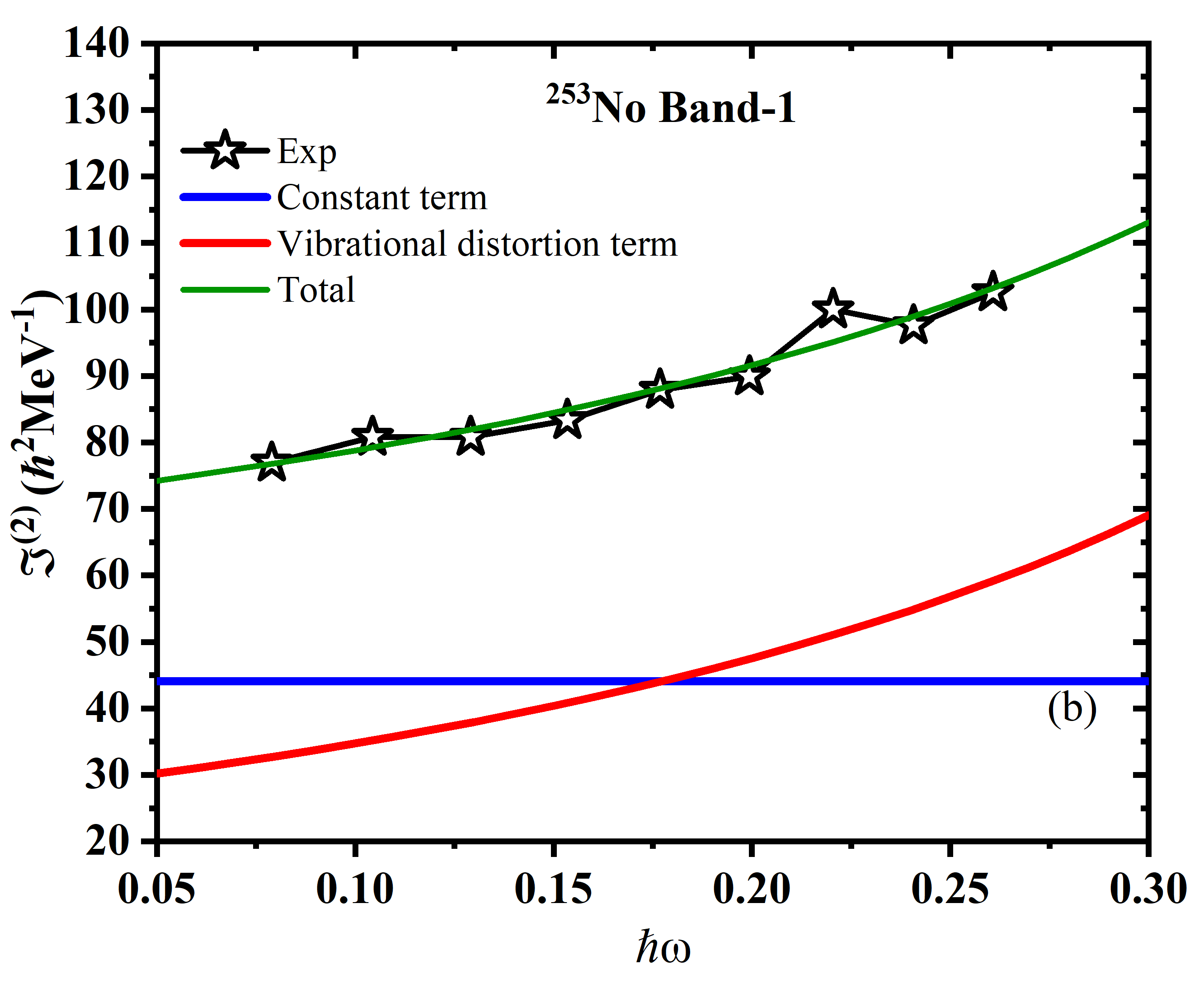}

   \includegraphics[width=6.5cm]{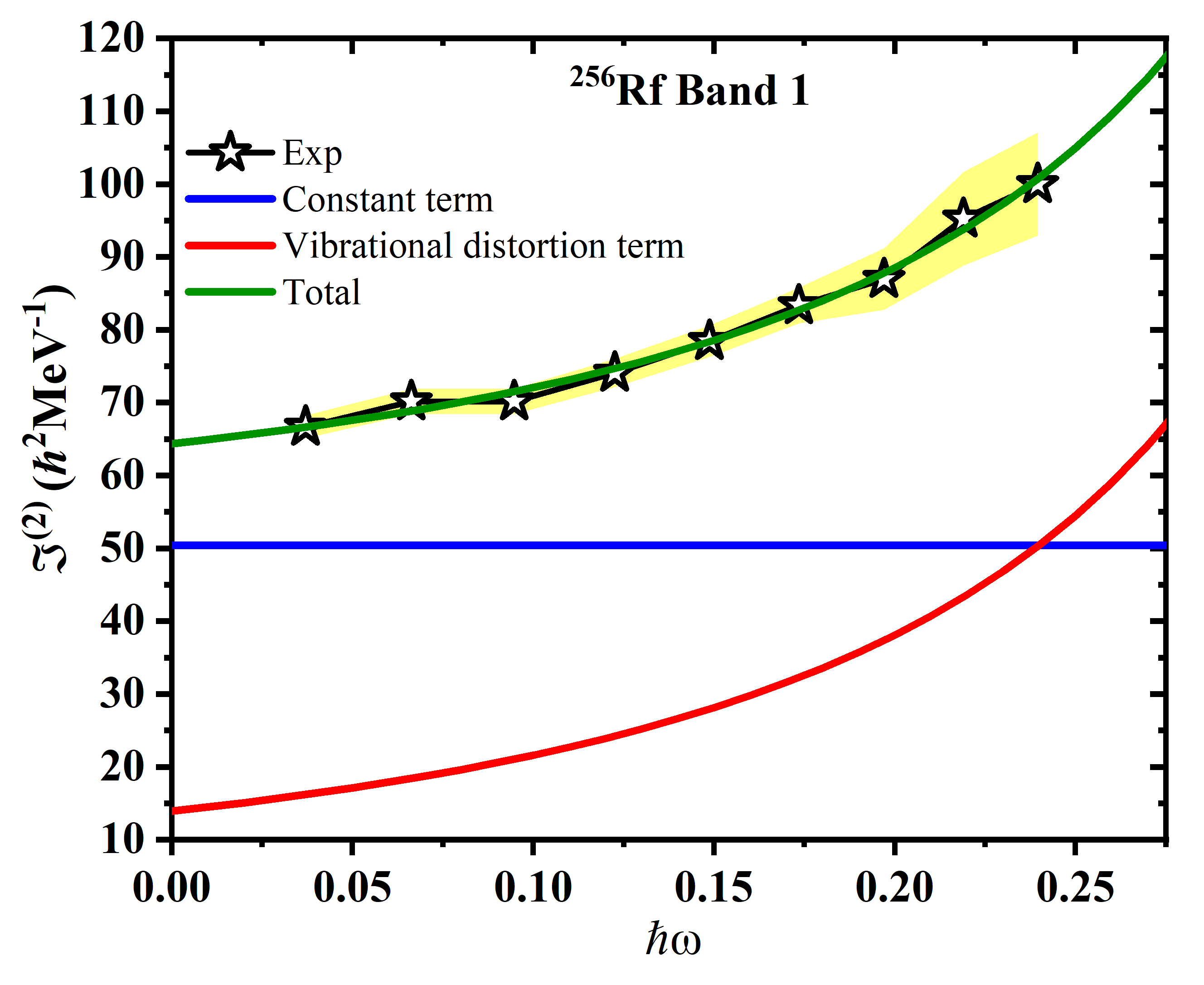}
\captionsetup{justification=raggedright, singlelinecheck=false}
\caption{(a)The variation of $\Im^{(2)}$ with $\hbar\omega$ for $^{245}$Pu band-2.
The contribution from the constant term ($\Im^{(2)}_{c}$),
vibrational distortion term ($\Im^{(2)}_{vib}$) and total ($\Im^{(2)}_{c}$+$\Im^{(2)}_{vib}$)
is shown by blue, red and green curves, respectively. (b)  Same as (a) but for $^{253}$No band-1.
(c) same as (a) but for $^{256}$Rf band-1.}
\label{fig7}
\end{figure}
\begin{figure}
\centering

   \includegraphics[width=8cm]{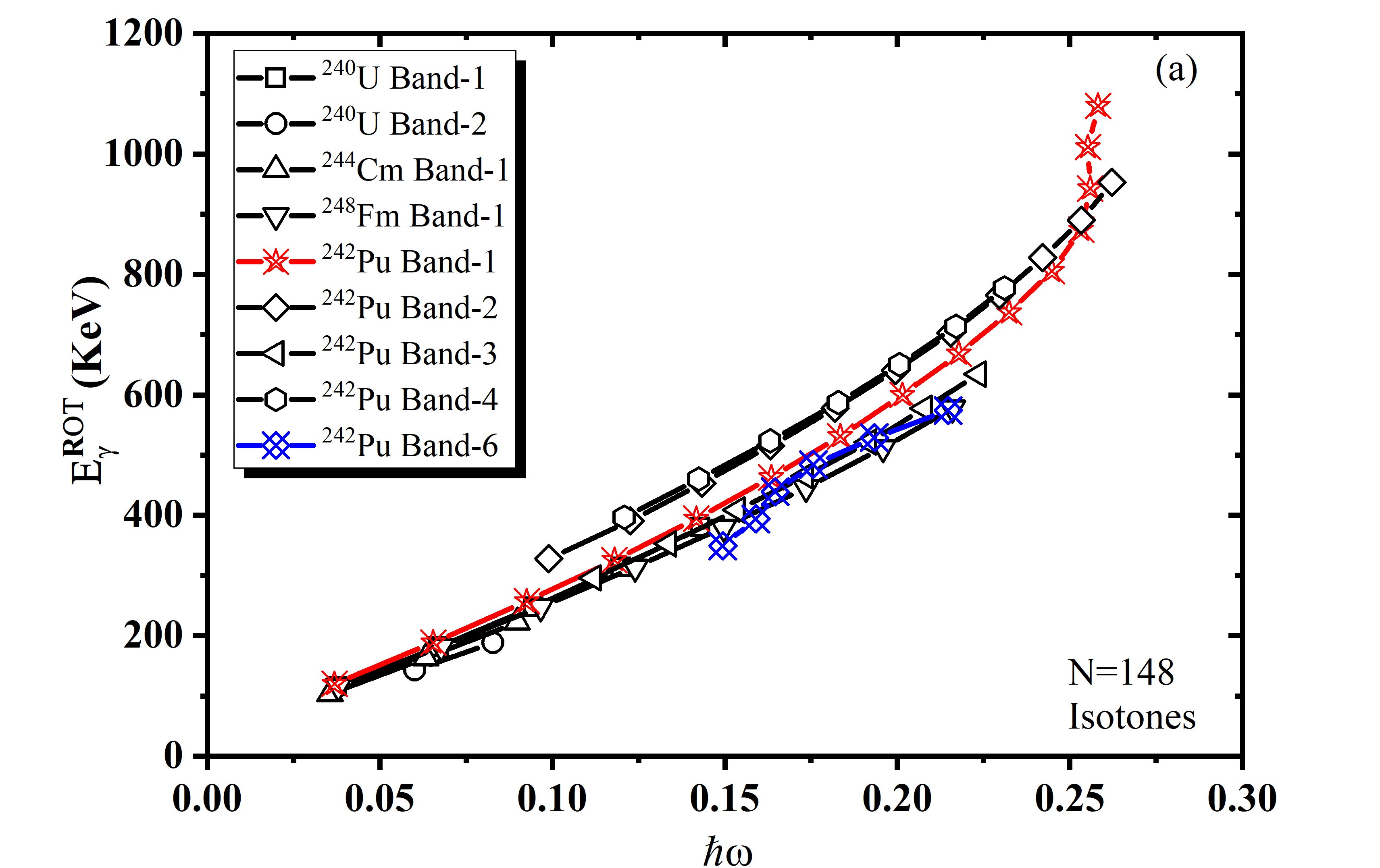}

   \includegraphics[width=8cm]{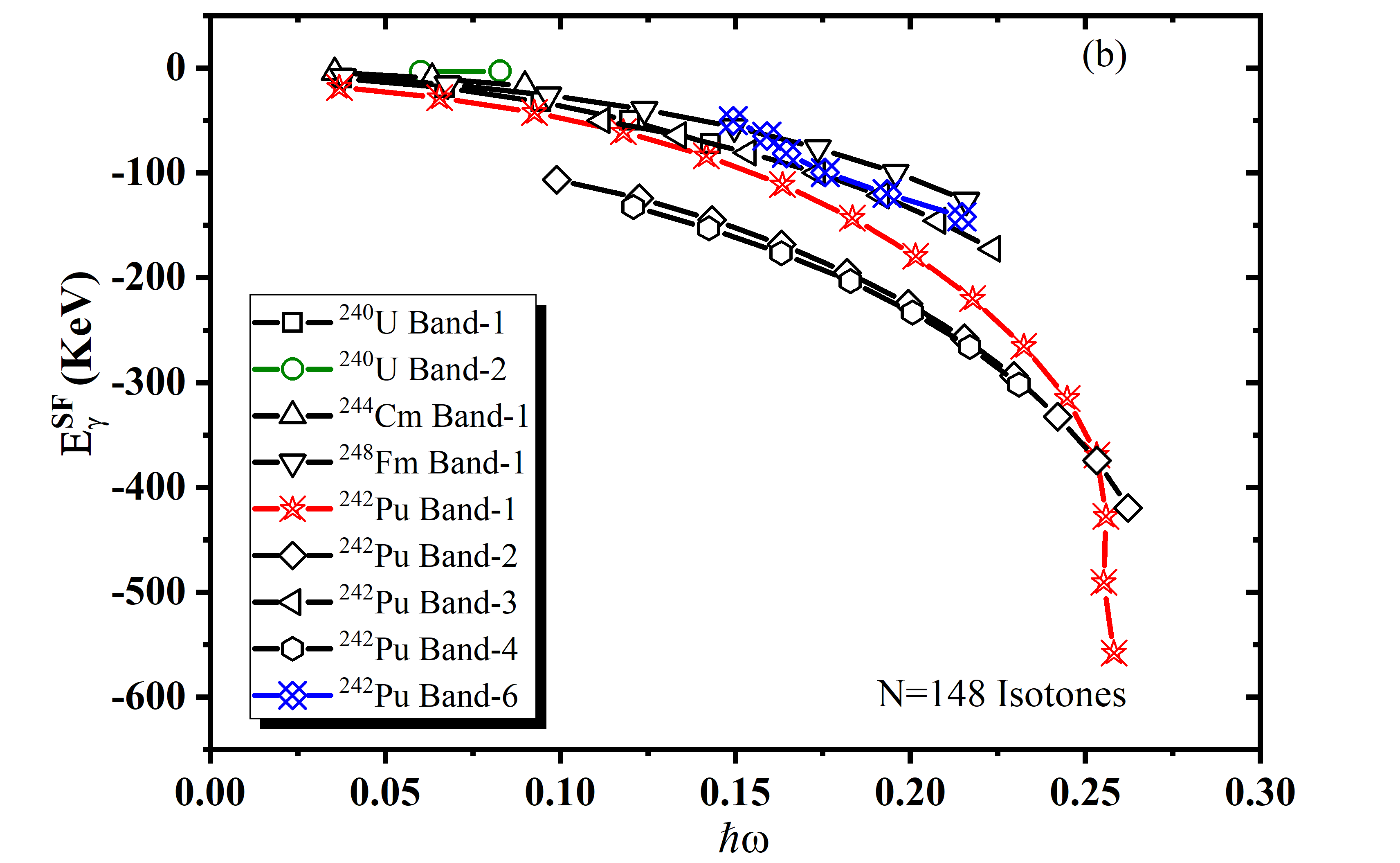}

\captionsetup{justification=raggedright, singlelinecheck=false}
\caption{(a) The variation of rotational energy (ROTE/$E_{\gamma}^{ROT}$) with rotational frequency for N=148 isotones.
(b) The variation of shape fluctuation energy (SFE/$E_{\gamma}^{SF}$) with rotational frequency for N=148 isotones. }
\label{fig8}
\end{figure}

%
\begin{figure}
\centering

   \includegraphics[width=8cm]{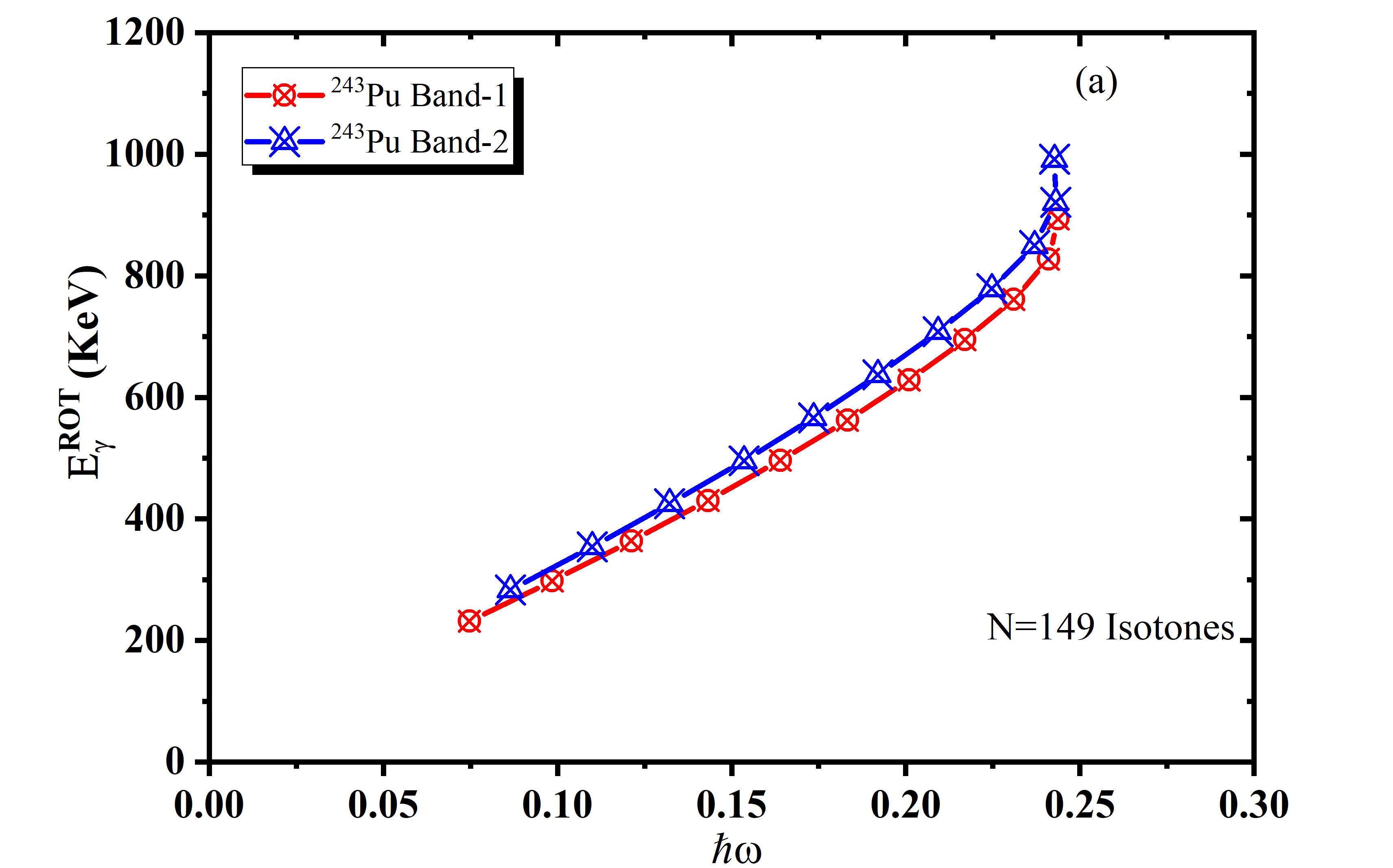}

   \includegraphics[width=8cm]{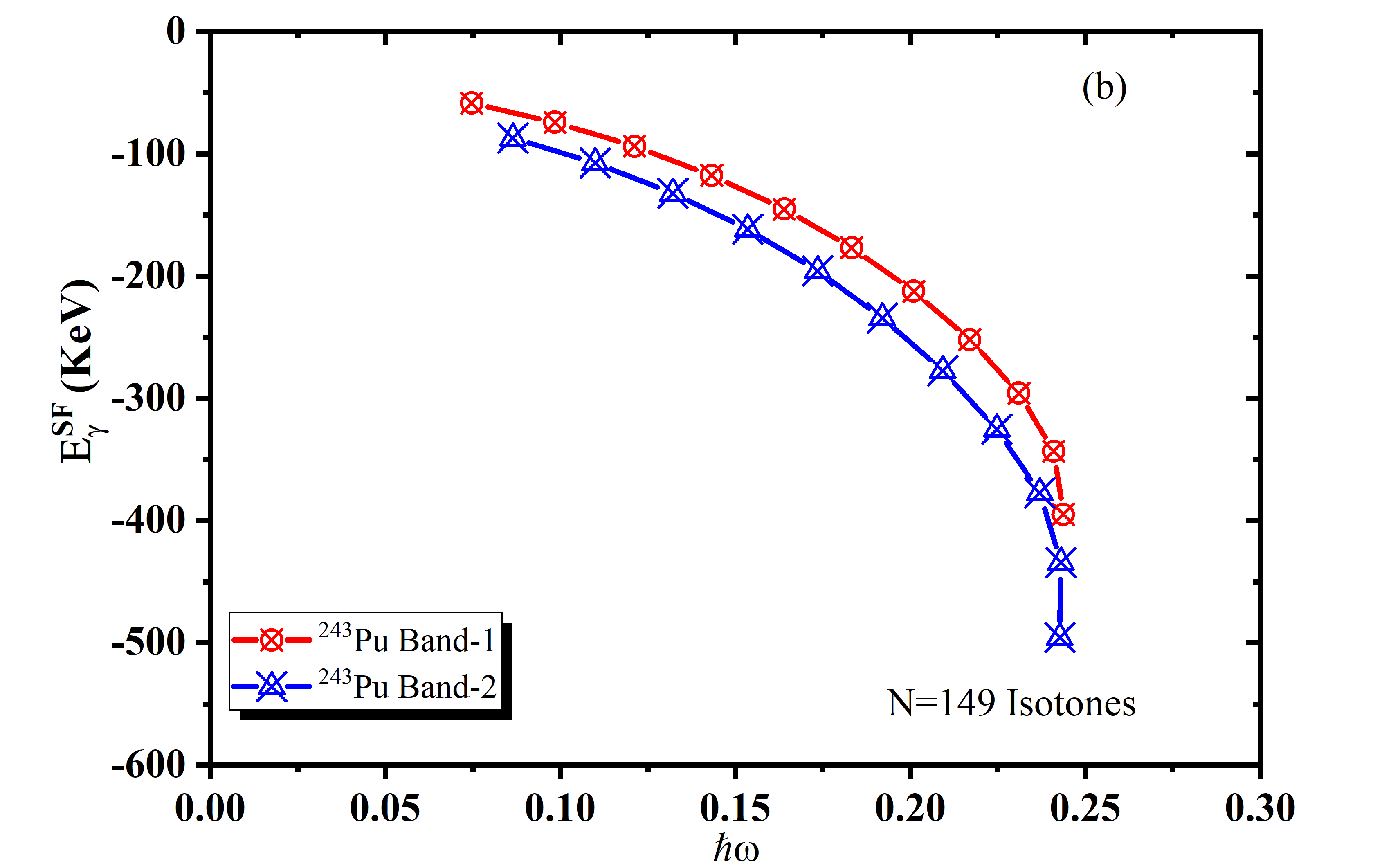}

\captionsetup{justification=raggedright, singlelinecheck=false}
\caption{(a) The variation of rotational energy (ROTE/$E_{\gamma}^{ROT}$) with rotational frequency for N=149 isotones.
(b) The variation of shape fluctuation energy (SFE/$E_{\gamma}^{SF}$) with rotational frequency for N=149 isotones. }
\label{fig9}
\end{figure}

\begin{figure}
\centering

   \includegraphics[width=8cm]{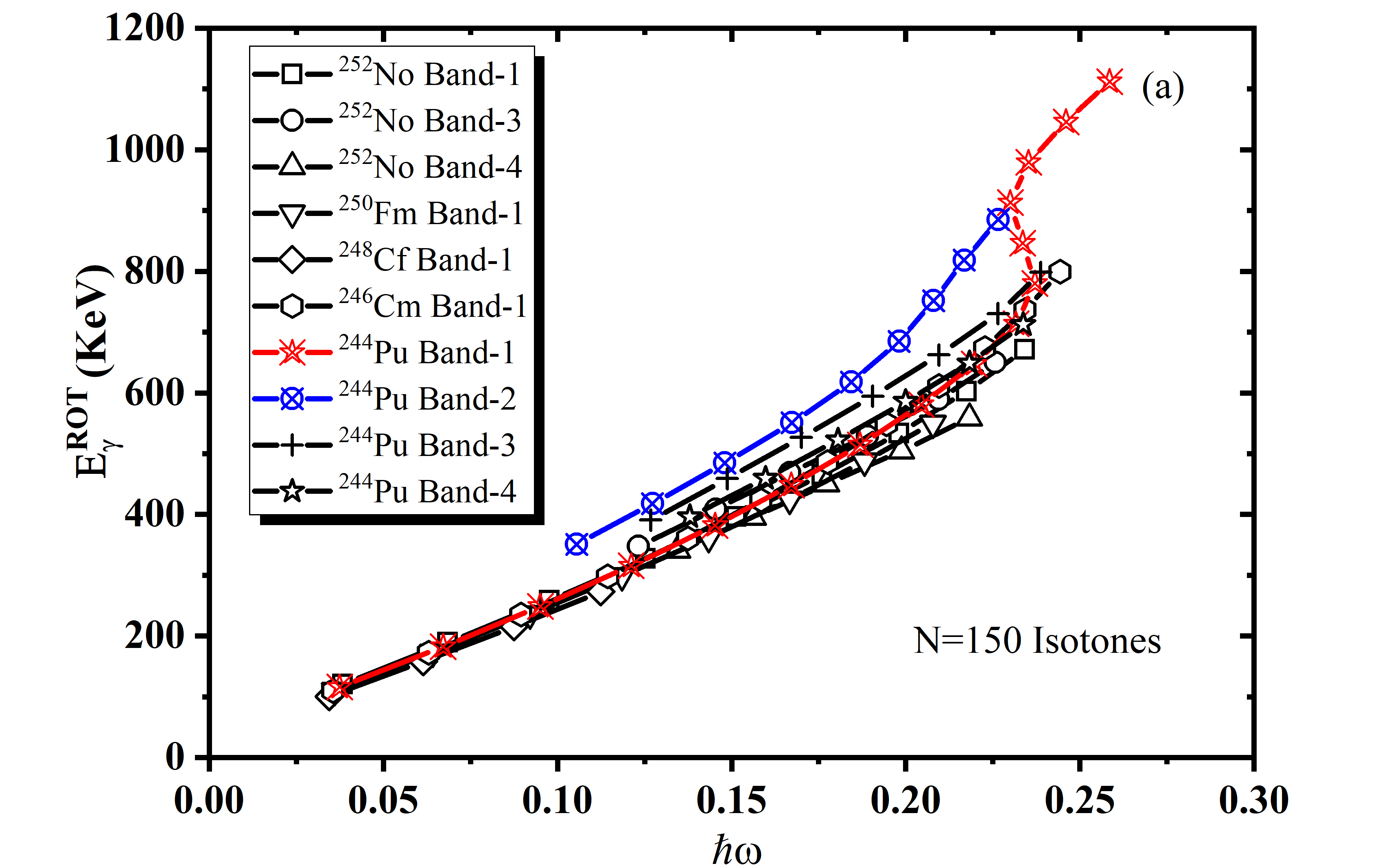}

   \includegraphics[width=8cm]{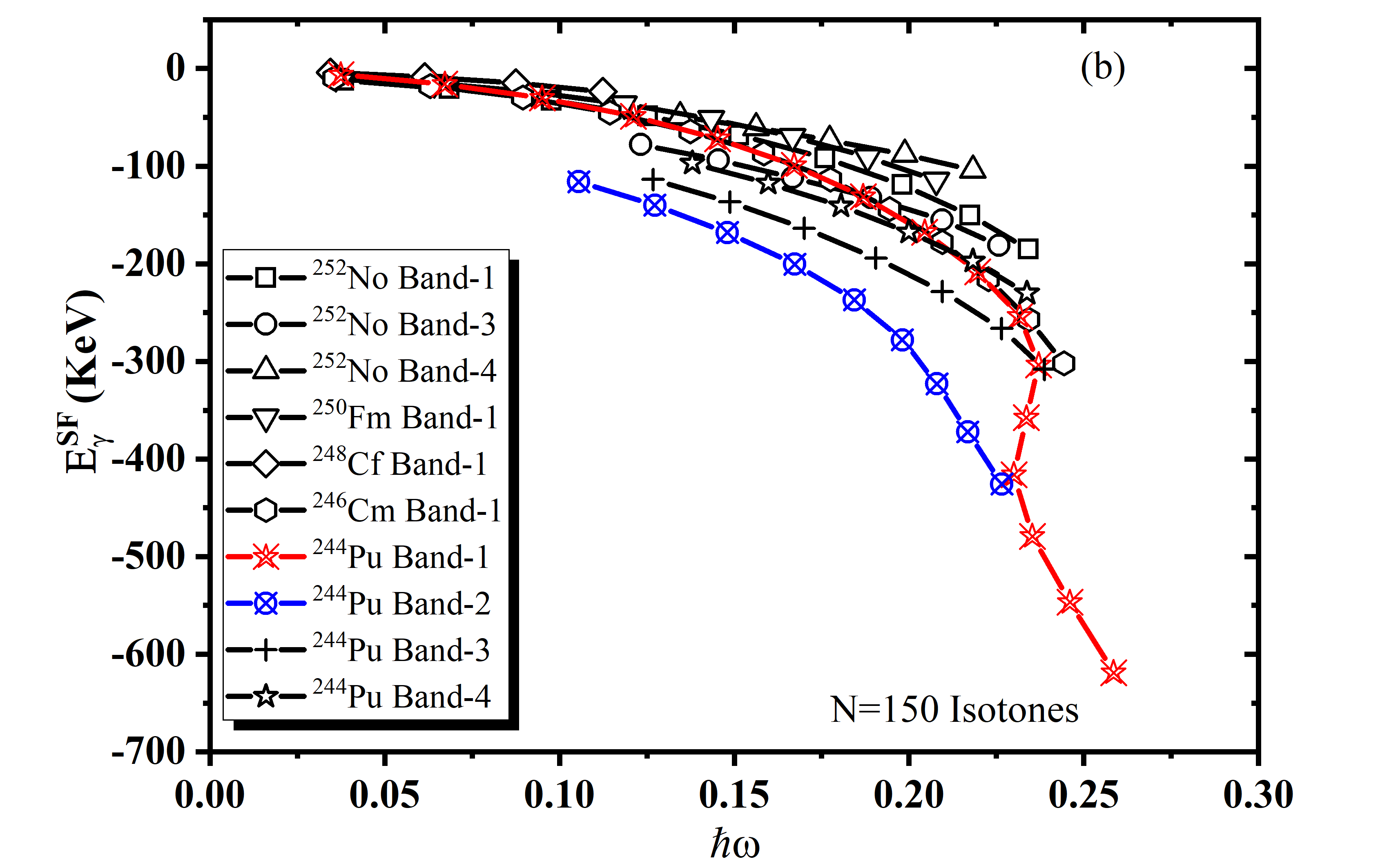}

\captionsetup{justification=raggedright, singlelinecheck=false}
\caption{(a) The variation of rotational energy (ROTE/$E_{\gamma}^{ROT}$) with
rotational frequency for N=150 isotones.
(b) The variation of shape fluctuation energy
(SFE/$E_{\gamma}^{SF}$) with rotational frequency for N=150 isotones. }
\label{fig10}
\end{figure}
\begin{figure}
\centering

   \includegraphics[width=7.5cm]{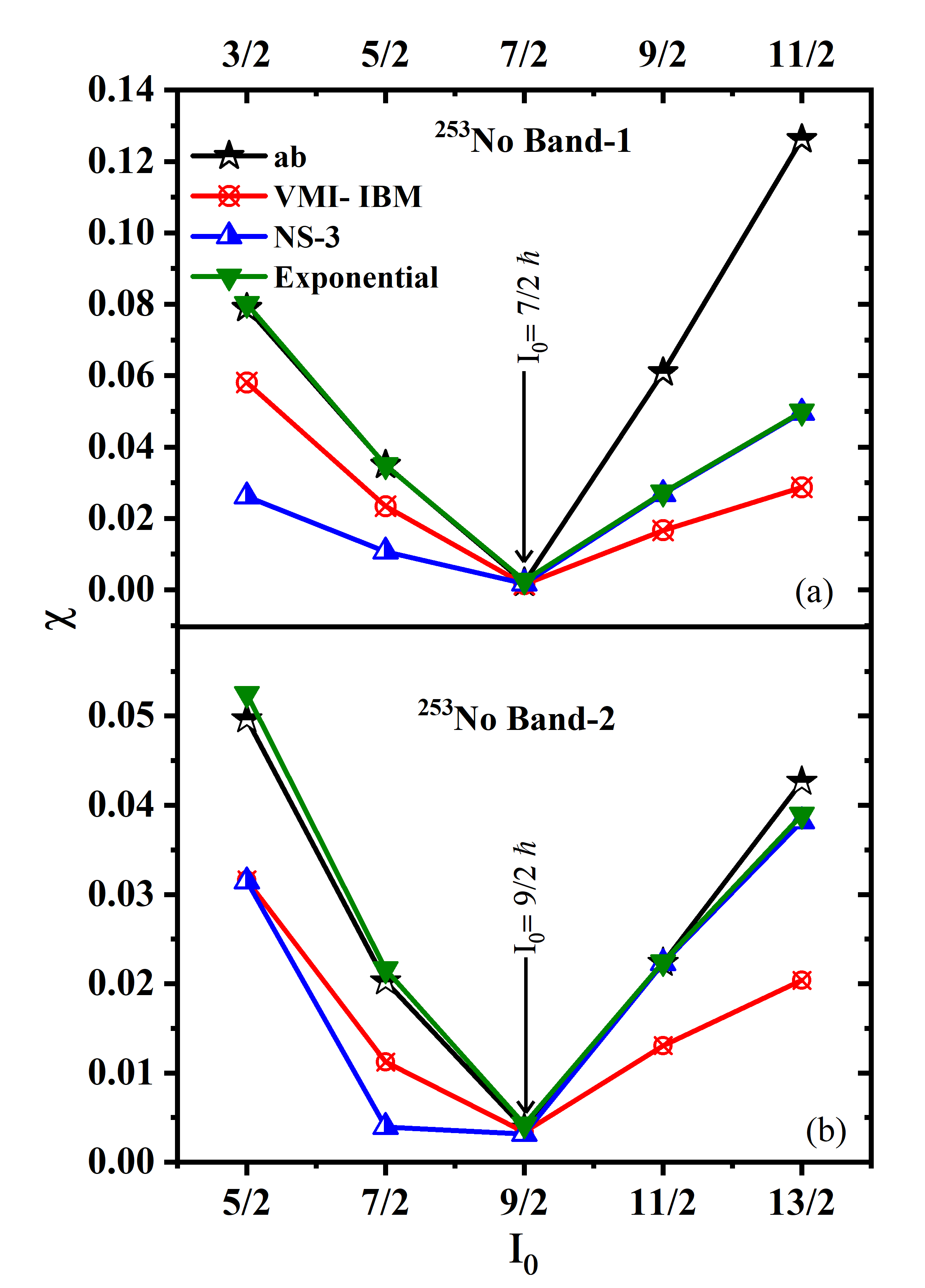}
    \captionsetup{justification=raggedright, singlelinecheck=false}
\caption{(a) The RMS deviation plot at different spins assignments for $^{253}$No band-1.
The $I_{0}$ corresponds to the spin value of the lowest level observed level.
The star ($\star$), circle ($\otimes$), up-triangle ($\bigtriangleup$) and
down-triangle ($\bigtriangledown$) corresponds to values obtained from
$ab$ formula \cite{WuPhysRevC.45.261}, VMI-IBM \cite{Yuxin_JPG}, nuclear-softness \cite{dadwal3} and exponential model with
pairing attenuation \cite{ZhouPhysRevC.55.2324}, respectively.(b) The same as (a) but for $^{253}$No band-2. }
\label{fig11}
\end{figure}

\begin{figure}
\centering

   \includegraphics[width=7cm]{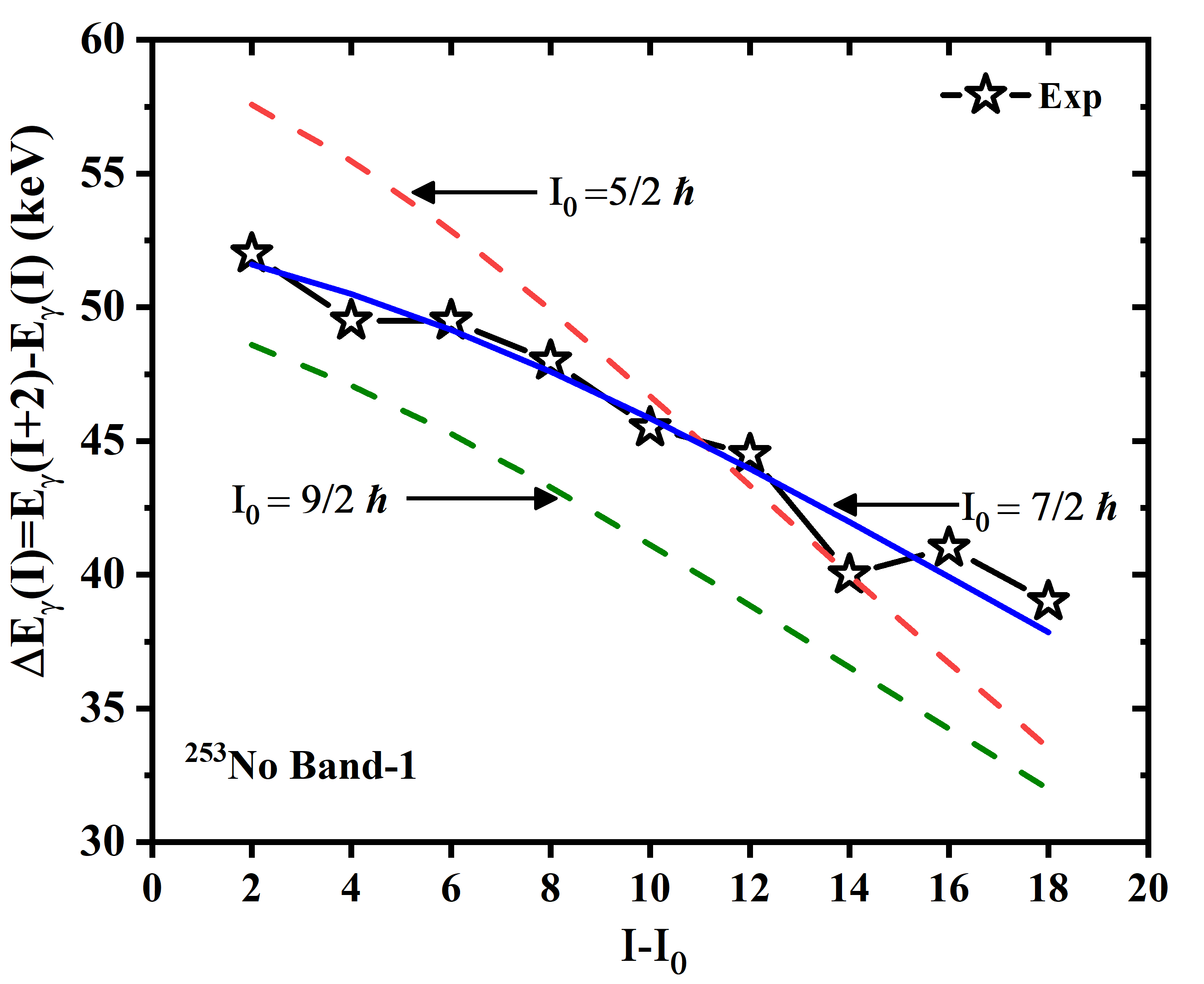}
   \captionsetup{justification=raggedright, singlelinecheck=false}
\caption{The variation of $\Delta E_{\gamma}(I)\equiv E_{\gamma}(I+2)-E_{\gamma}(I)$ vs $I-I_{0}$ for $^{253}$No band-1.}
\label{fig12}
\end{figure}

\begin{figure}
\centering

   \includegraphics[width=7.5cm]{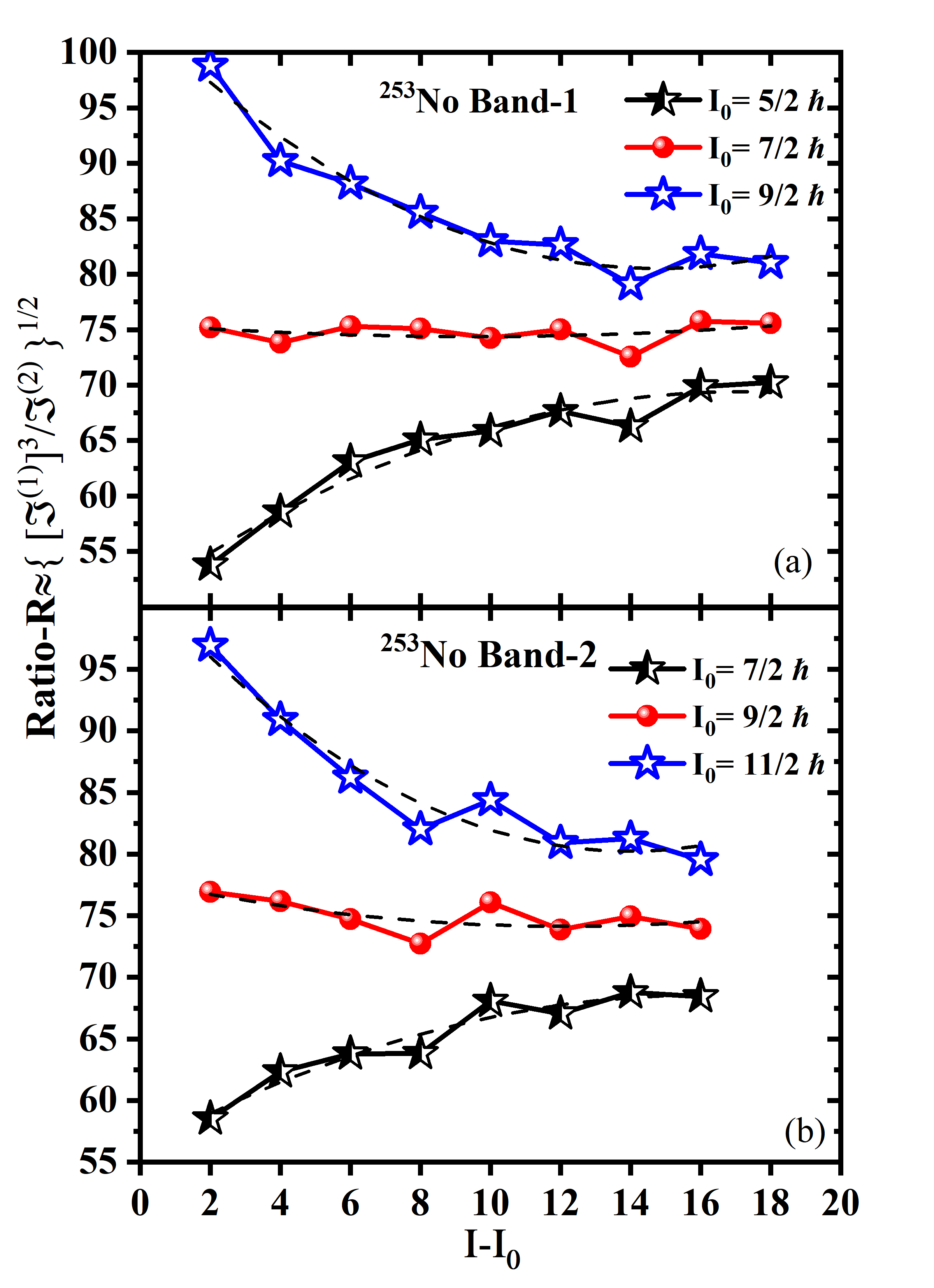}
    \captionsetup{justification=raggedright, singlelinecheck=false}
\caption{(a) The variation of ratio-R$\equiv\sqrt{[\Im^{(1)}]^{3}/\Im^{(2)}}$ at
different spin values for $^{253}$No band-1. (b) The same as (a) but for $^{253}$No band-2.}
\label{fig13}
\end{figure}
\begin{figure}
\centering

   \includegraphics[width=8cm]{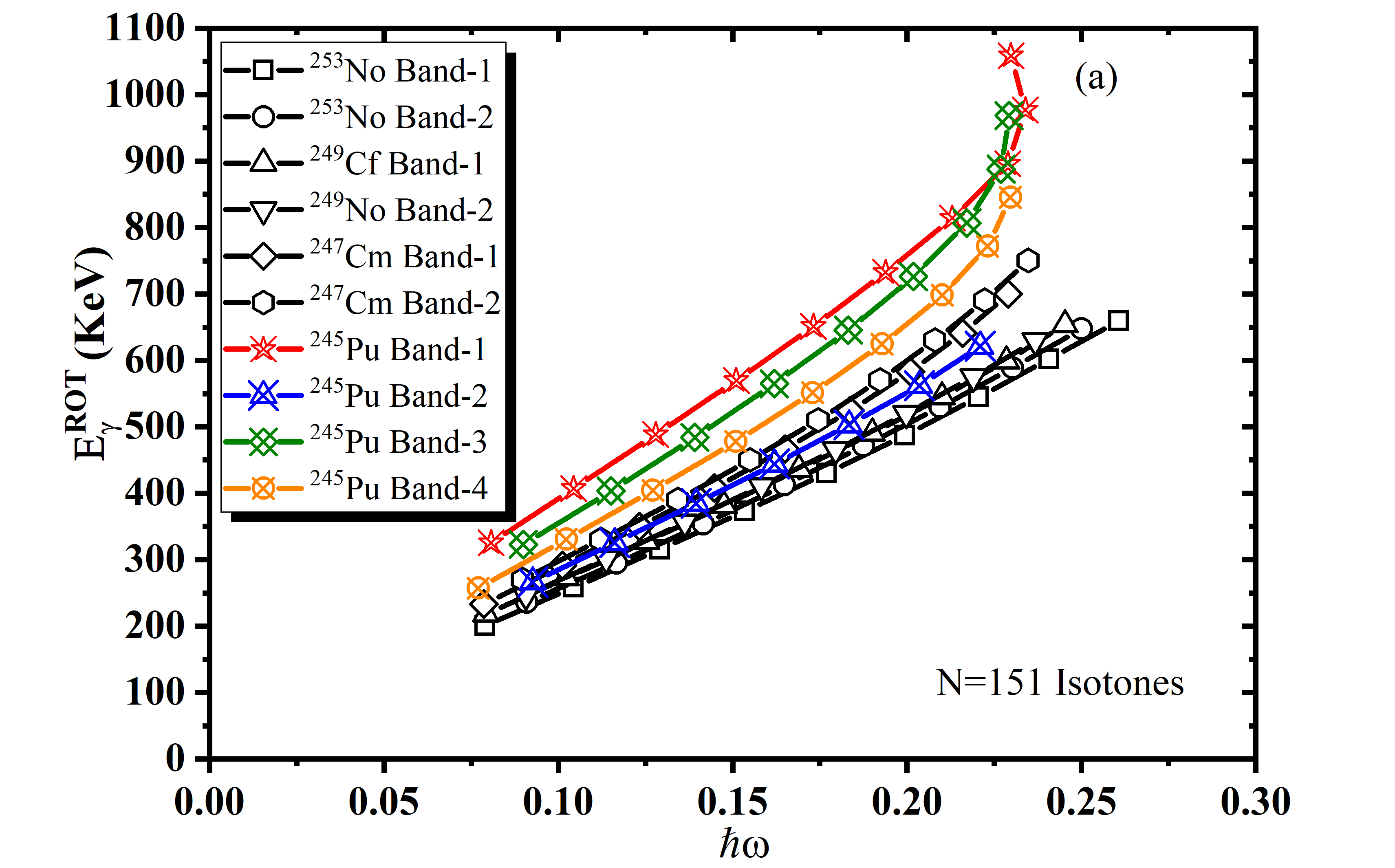}

   \includegraphics[width=8cm]{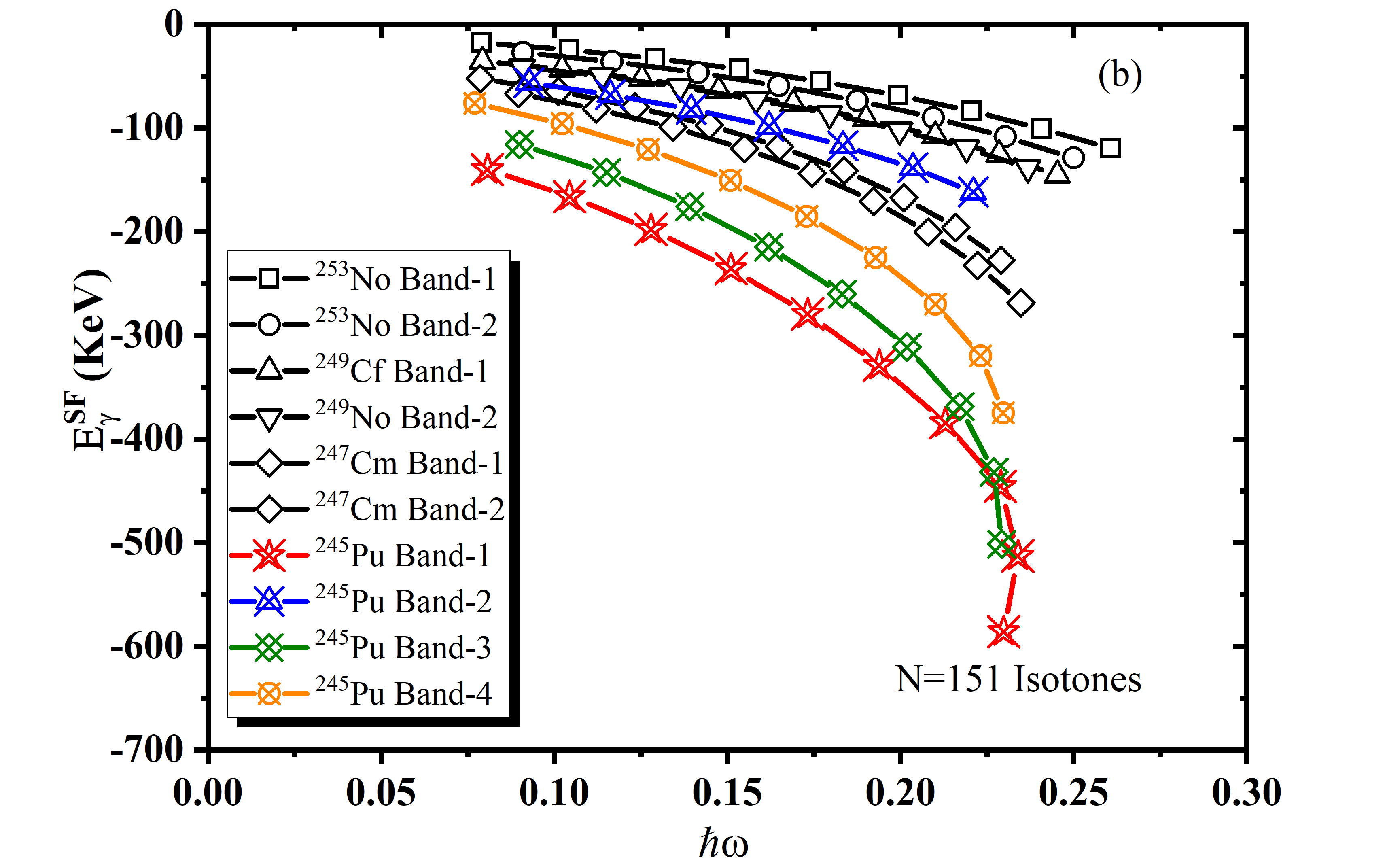}
\captionsetup{justification=raggedright, singlelinecheck=false}
\caption{(a) The variation of rotational energy (ROTE/$E_{\gamma}^{ROT}$)
with rotational frequency for N=151 isotones. (b) The variation of
shape fluctuation energy (SFE/$E_{\gamma}^{SF}$) with rotational frequency for N=151 isotones. }
\label{fig14}
\end{figure}
\begin{figure}
\centering

   \includegraphics[width=8cm]{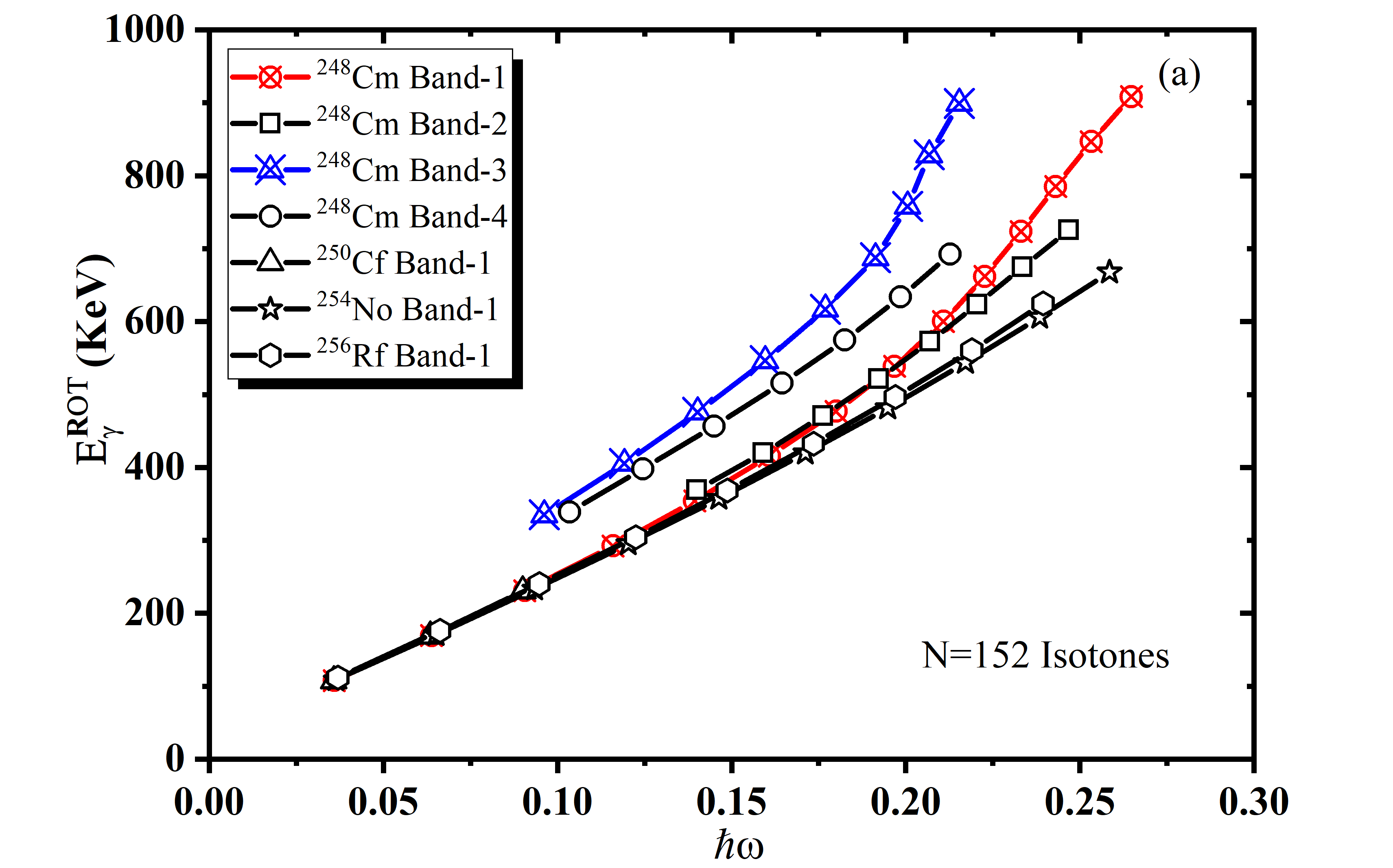}

   \includegraphics[width=8cm]{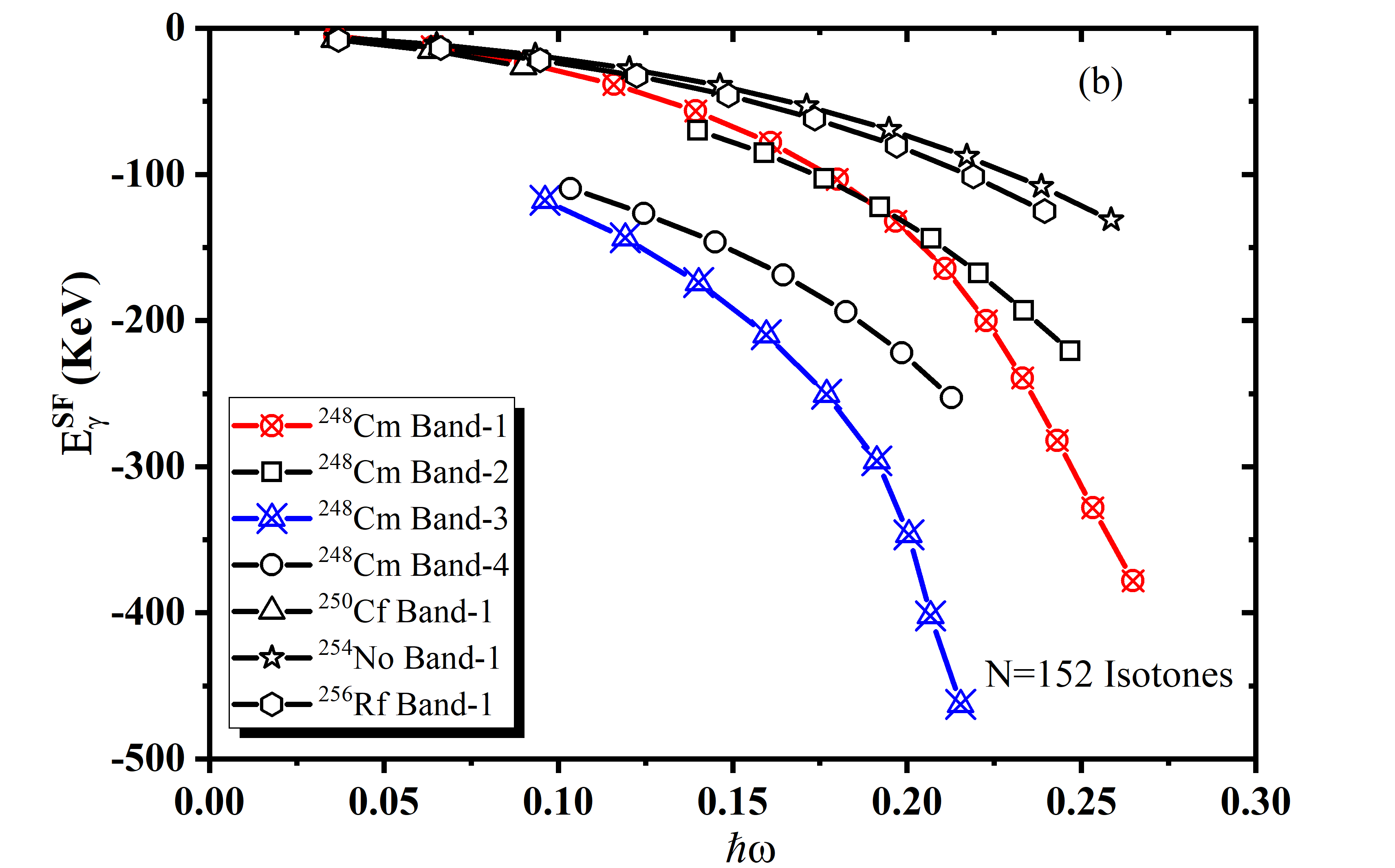}
\captionsetup{justification=raggedright, singlelinecheck=false}
\caption{(a) The variation of rotational energy (ROTE/$E_{\gamma}^{ROT}$) with rotational frequency for N=152 isotones.
 (b) The variation of shape fluctuation energy (SFE/$E_{\gamma}^{SF}$) with rotational frequency for N=152 isotones. }
\label{fig15}
\end{figure}
\begin{figure}
\centering

   \includegraphics[width=8cm]{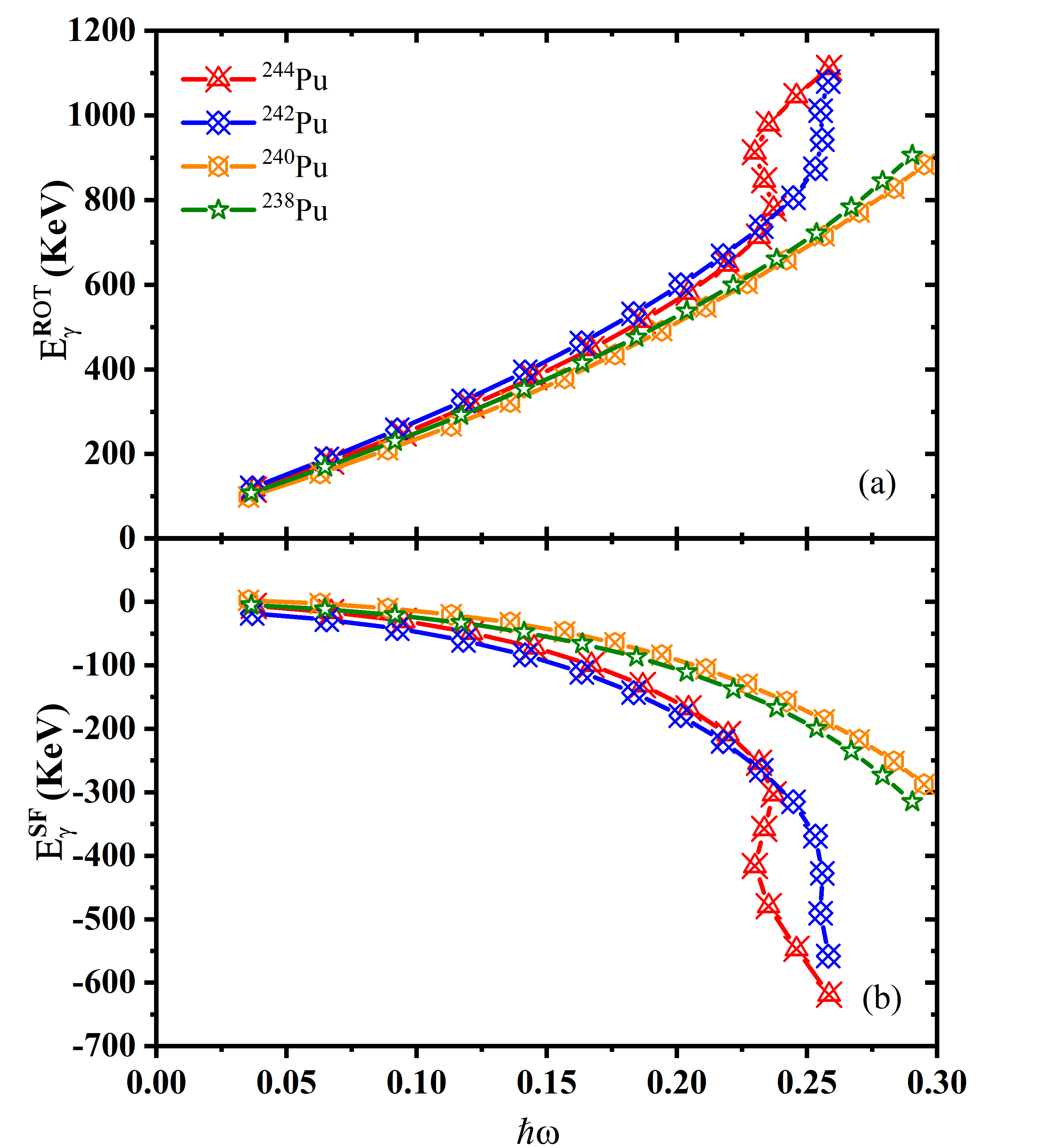}

\captionsetup{justification=raggedright, singlelinecheck=false}
\caption{The variation of (a) rotational energy (ROTE/$E_{\gamma}^{ROT}$) (b)
shape fluctuation energy (SFE/$E_{\gamma}^{SF}$) with rotational frequency for gsb/band-1 in even-A isotopes of Pu. }
\label{figPu-Isotopes}
\end{figure}

\begin{figure*}

\begin{subfigure}{.45\linewidth}
  \includegraphics[width=\linewidth]{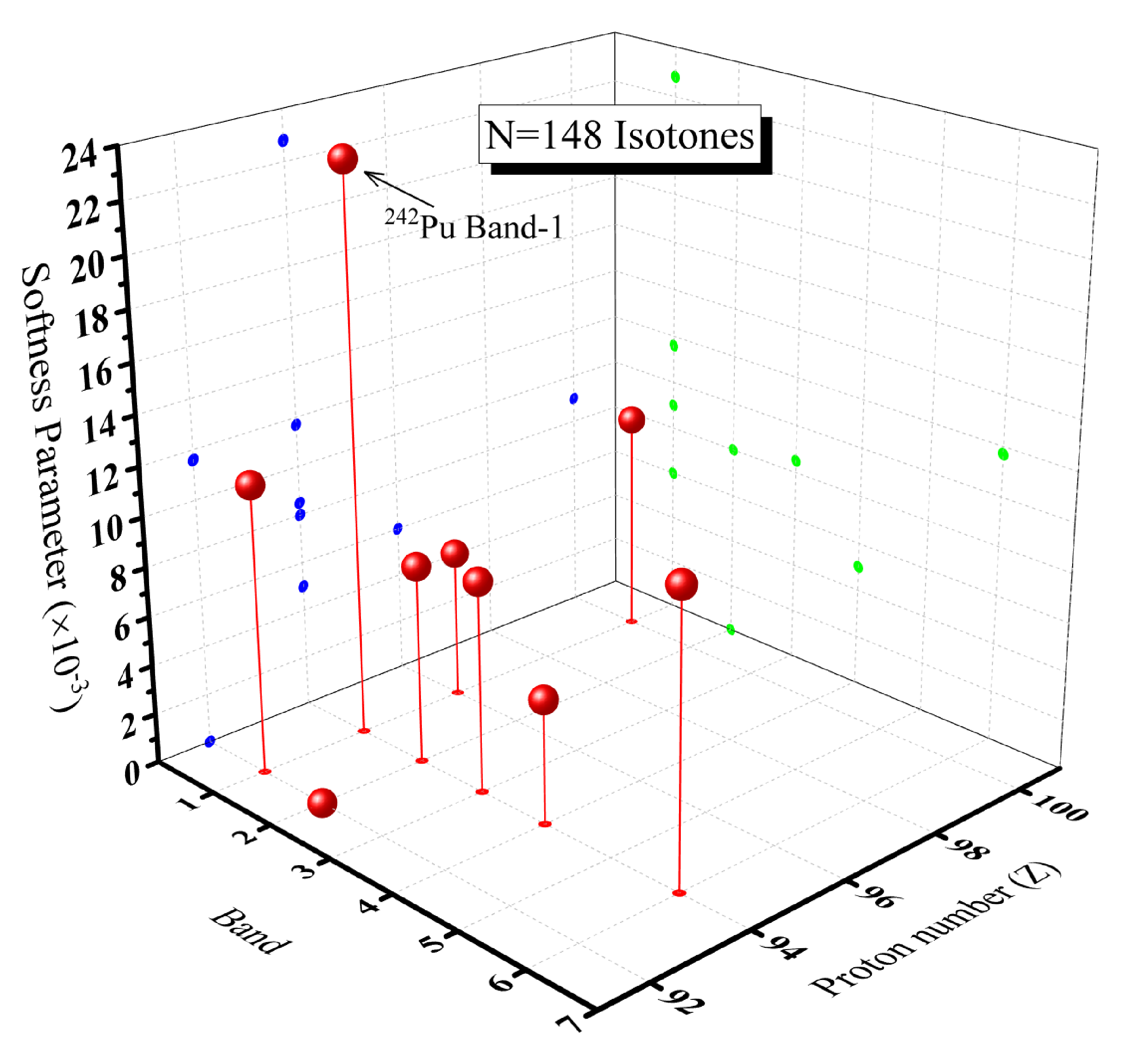}
  \caption{}
\end{subfigure}\hfill 
\begin{subfigure}{.45\linewidth}
  \includegraphics[width=\linewidth]{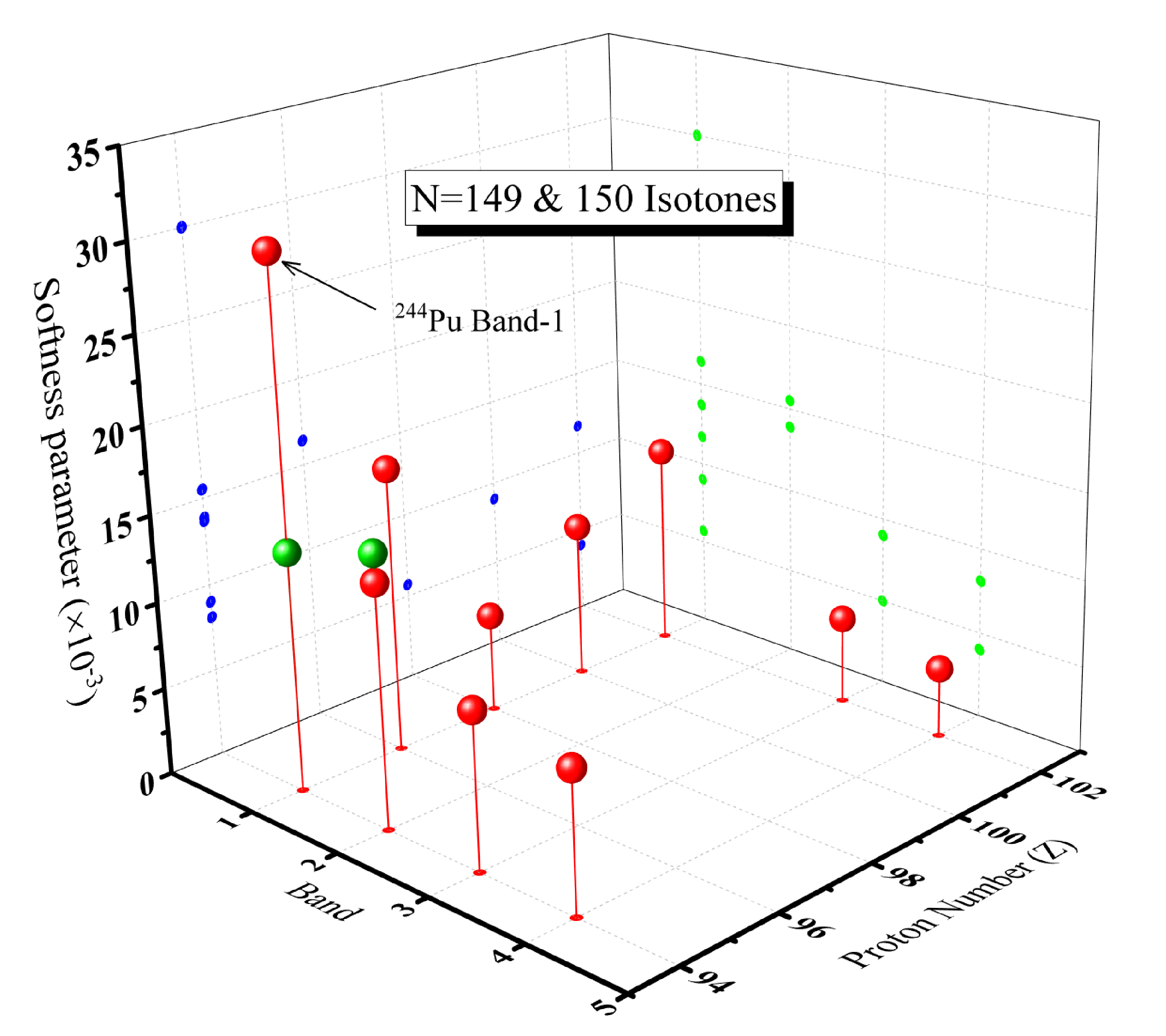}
  \caption{}
\end{subfigure}

\medskip 
\begin{subfigure}{.45\linewidth}
  \includegraphics[width=\linewidth]{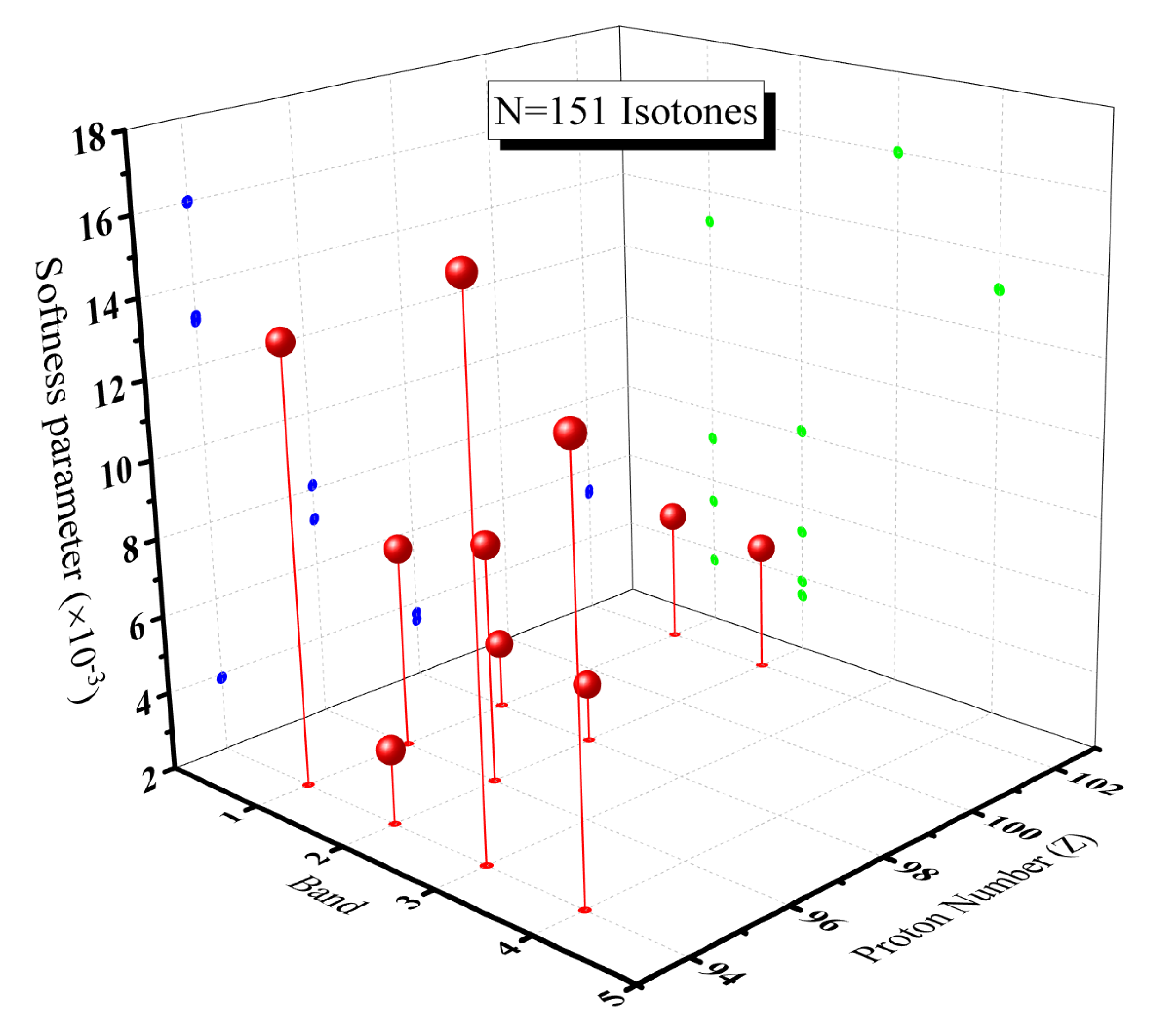}
  \caption{}
\end{subfigure}\hfill 
\begin{subfigure}{.45\linewidth}
  \includegraphics[width=\linewidth]{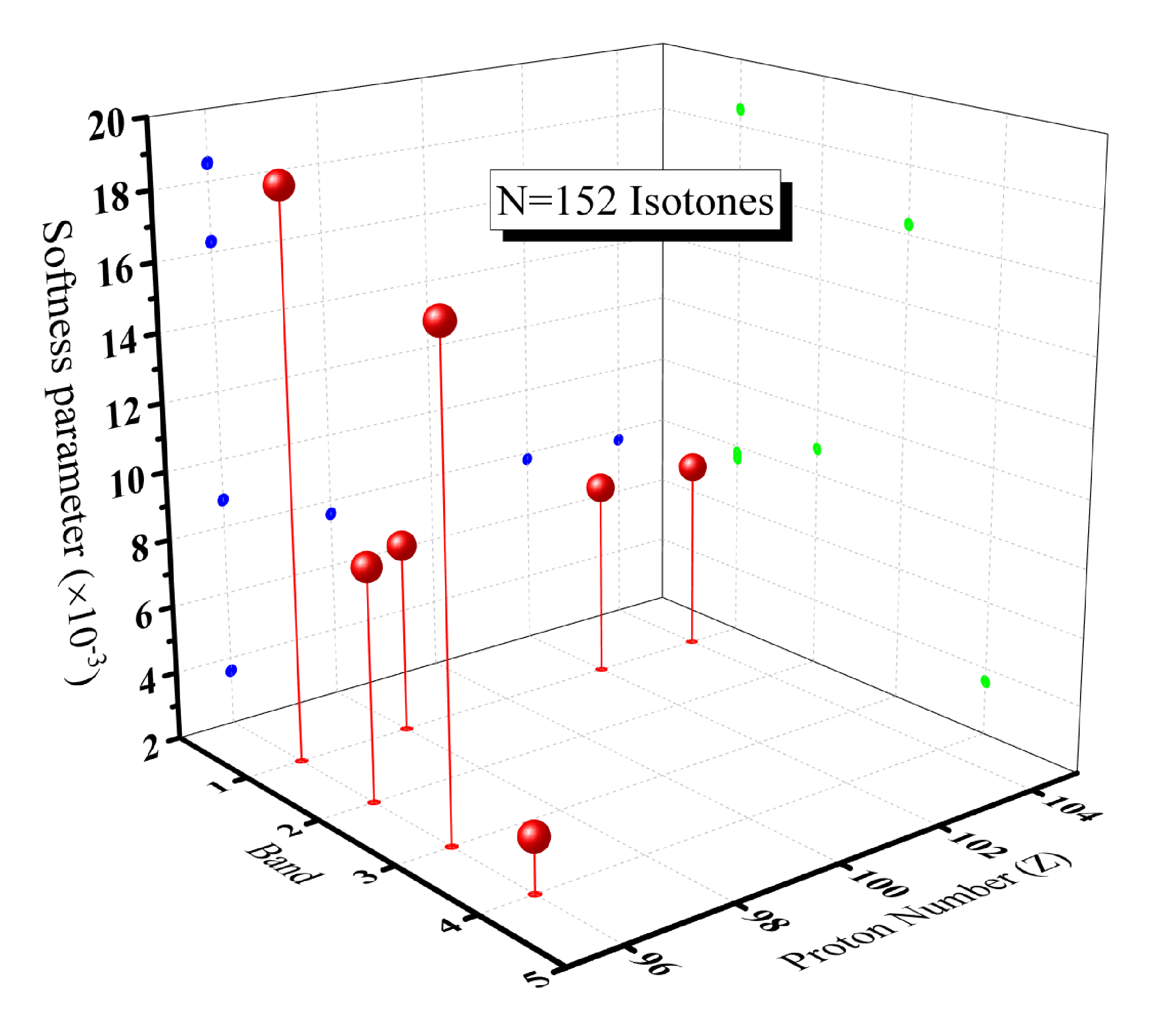}
  \caption{}
\end{subfigure}
\captionsetup{justification=raggedright, singlelinecheck=false}
\caption{ (a)-(d) Softness parameter vs the proton number (Z) and the band number
(a) for N=148 isotones, (b) for N=149 (green sphere), 150 isotones,
(c) for N=151 isotones, (d) for N=152 isotones.
The projection on XY, YZ and ZX is shown by red, blue and green dots, respectively.}
\label{fig16}
\end{figure*}

\begin{figure}
\centering

   \includegraphics[width=9cm]{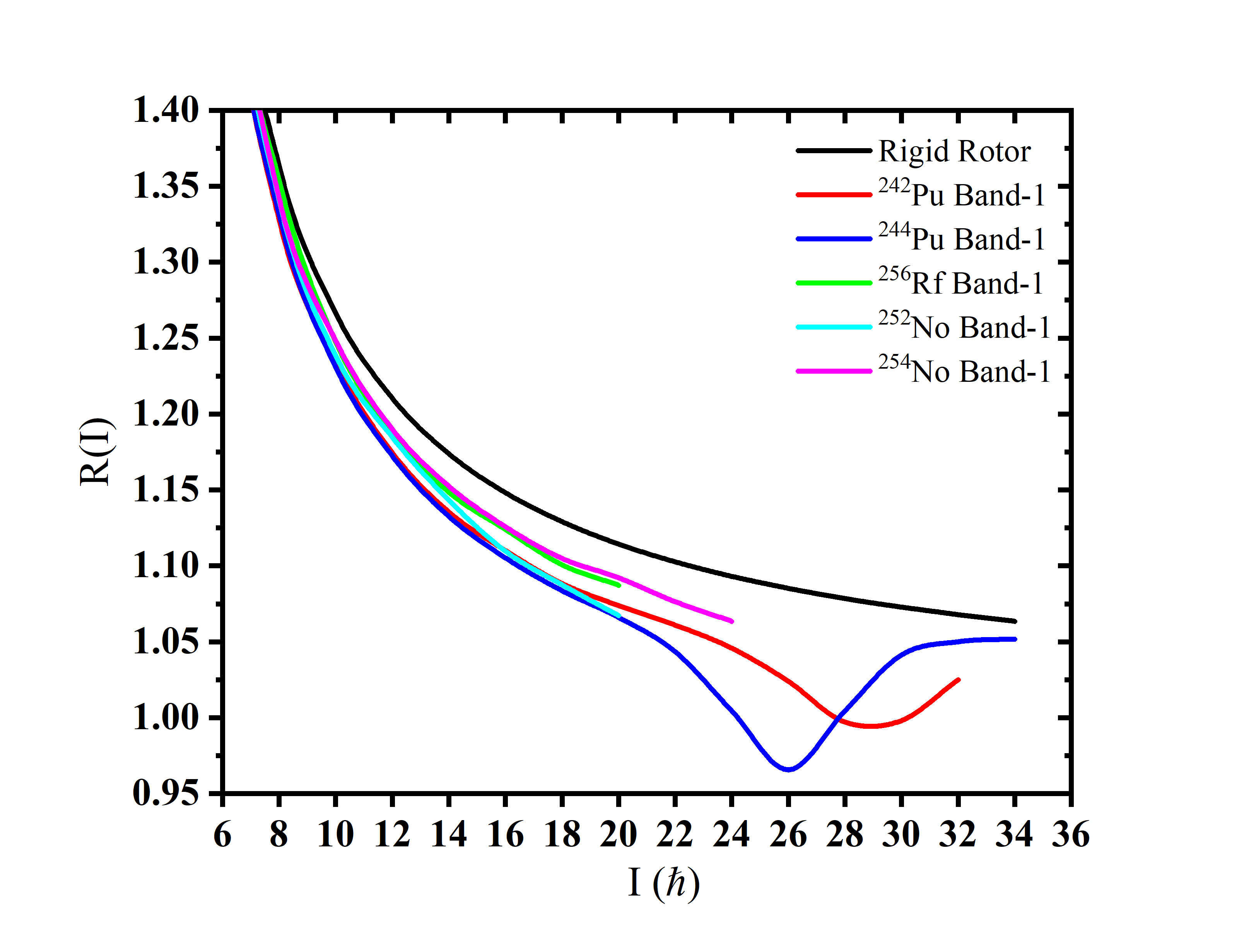}
\captionsetup{justification=raggedright, singlelinecheck=false}
\caption{The plot of $R(I)$ vs the spin ($I=6\hbar-36 \hbar$. The rigid rotor line is shown in solid black.}
\label{fig17}
\end{figure}

\begin{figure*}

\begin{subfigure}{.45\linewidth}
  \includegraphics[width=\linewidth]{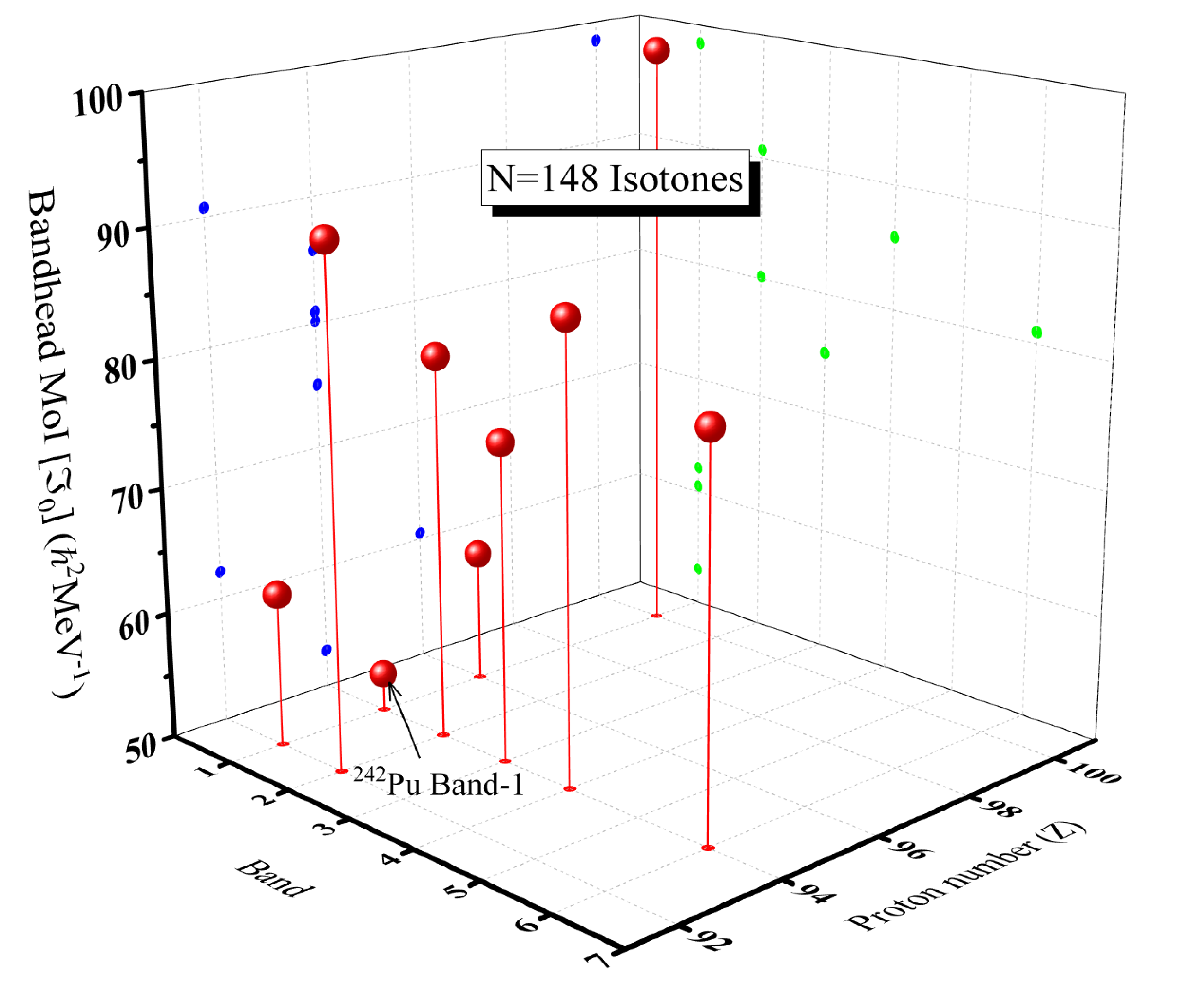}
  \caption{}
\end{subfigure}\hfill 
\begin{subfigure}{.45\linewidth}
  \includegraphics[width=\linewidth]{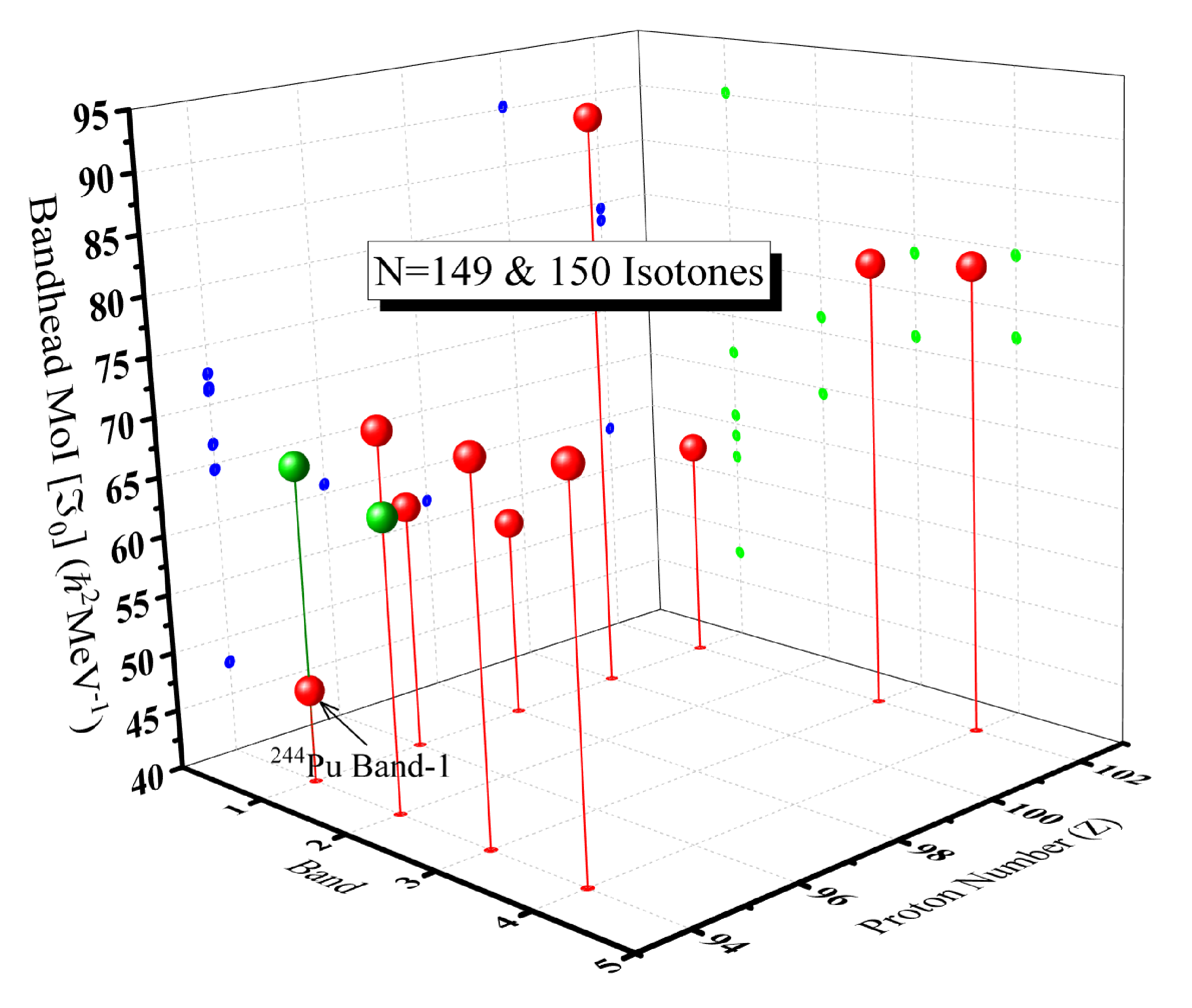}
  \caption{}
\end{subfigure}

\medskip 
\begin{subfigure}{.45\linewidth}
  \includegraphics[width=\linewidth]{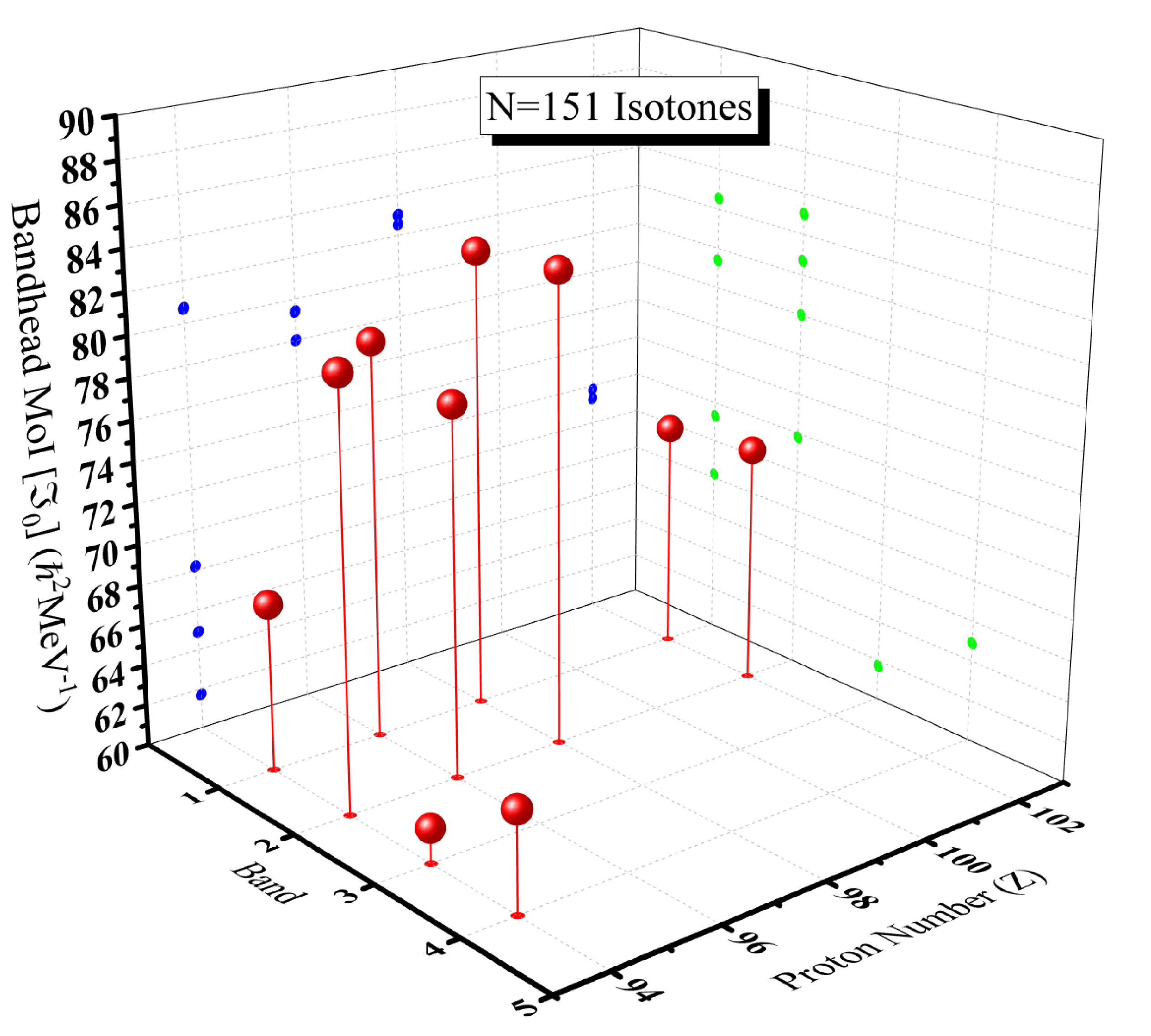}
  \caption{}
\end{subfigure}\hfill 
\begin{subfigure}{.45\linewidth}
  \includegraphics[width=\linewidth]{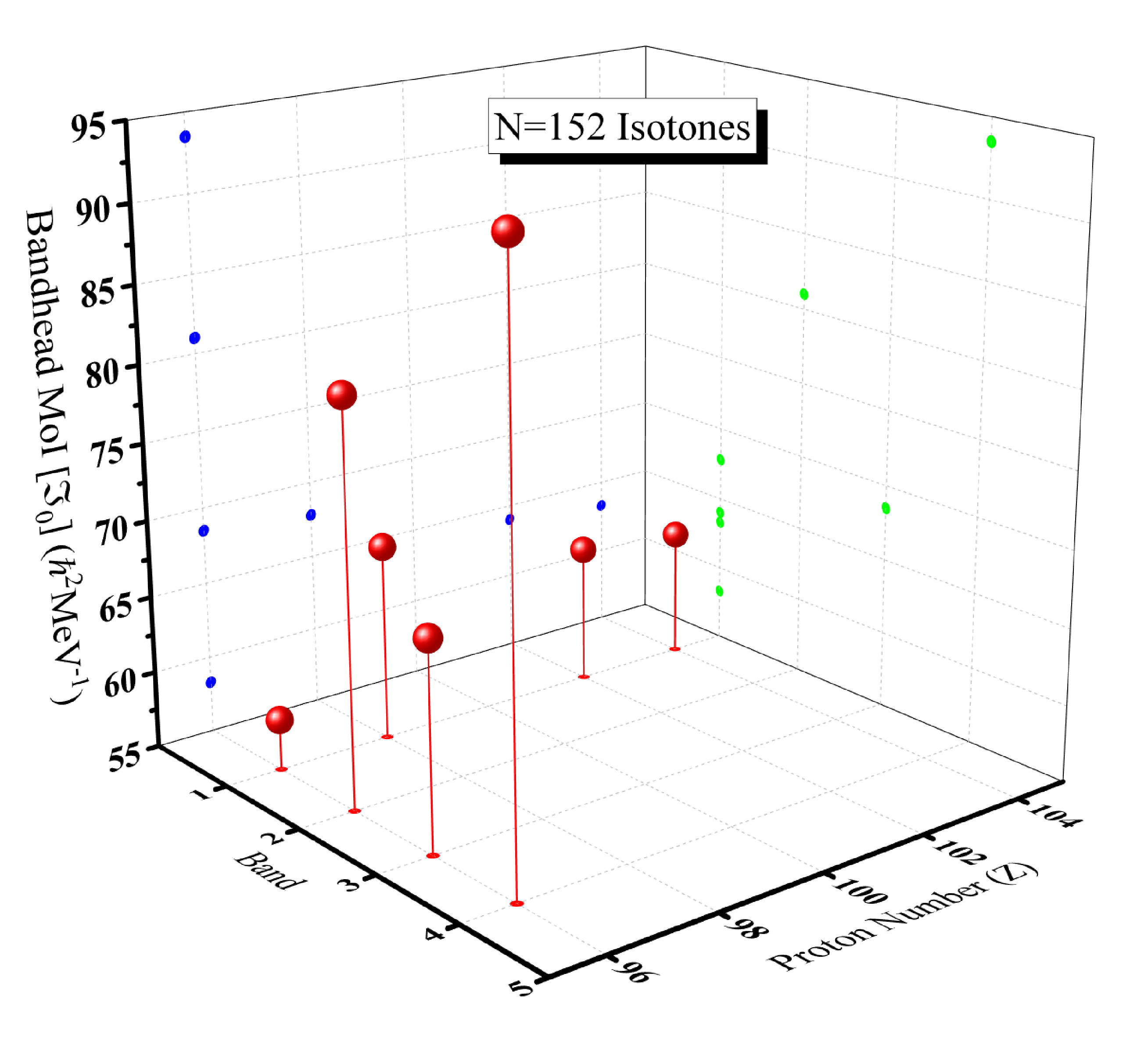}
  \caption{}
\end{subfigure}
\captionsetup{justification=raggedright, singlelinecheck=false}
\caption{ (a)-(d) Band-head moment of inertia ($\Im_{0}$) vs the proton number
(Z) and the band number (a) for N=148 isotones, (b) for N=149 (green sphere),
150 isotones, (c) for N=151 isotones, (d) for N=152 isotones. The projection
on XY, YZ and ZX is shown by red, blue and green dots, respectively.}
\label{fig18}
\end{figure*}

\begin{figure*}

\begin{subfigure}{.45\linewidth}
  \includegraphics[width=\linewidth]{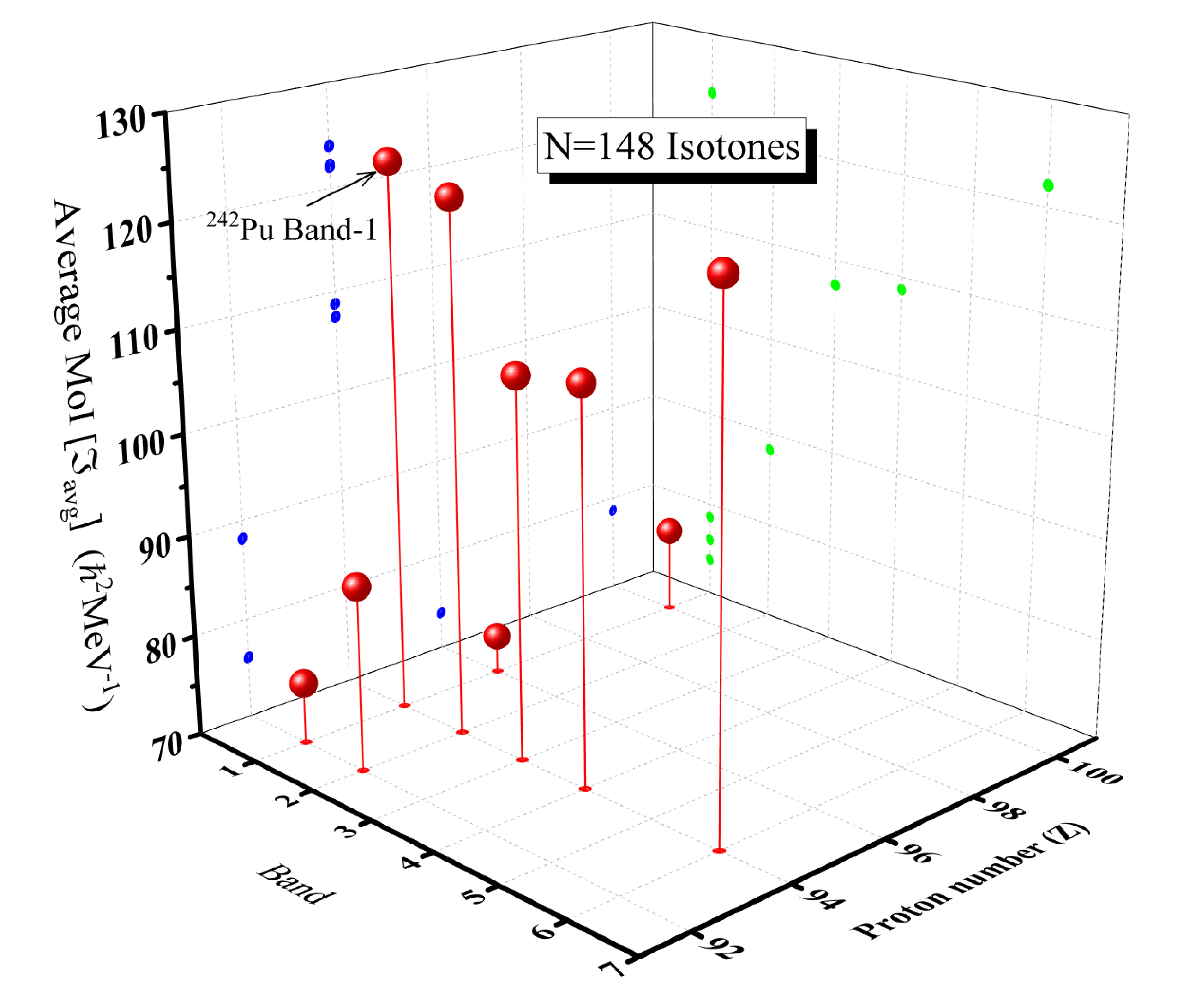}
  \caption{}
\end{subfigure}\hfill 
\begin{subfigure}{.45\linewidth}
  \includegraphics[width=\linewidth]{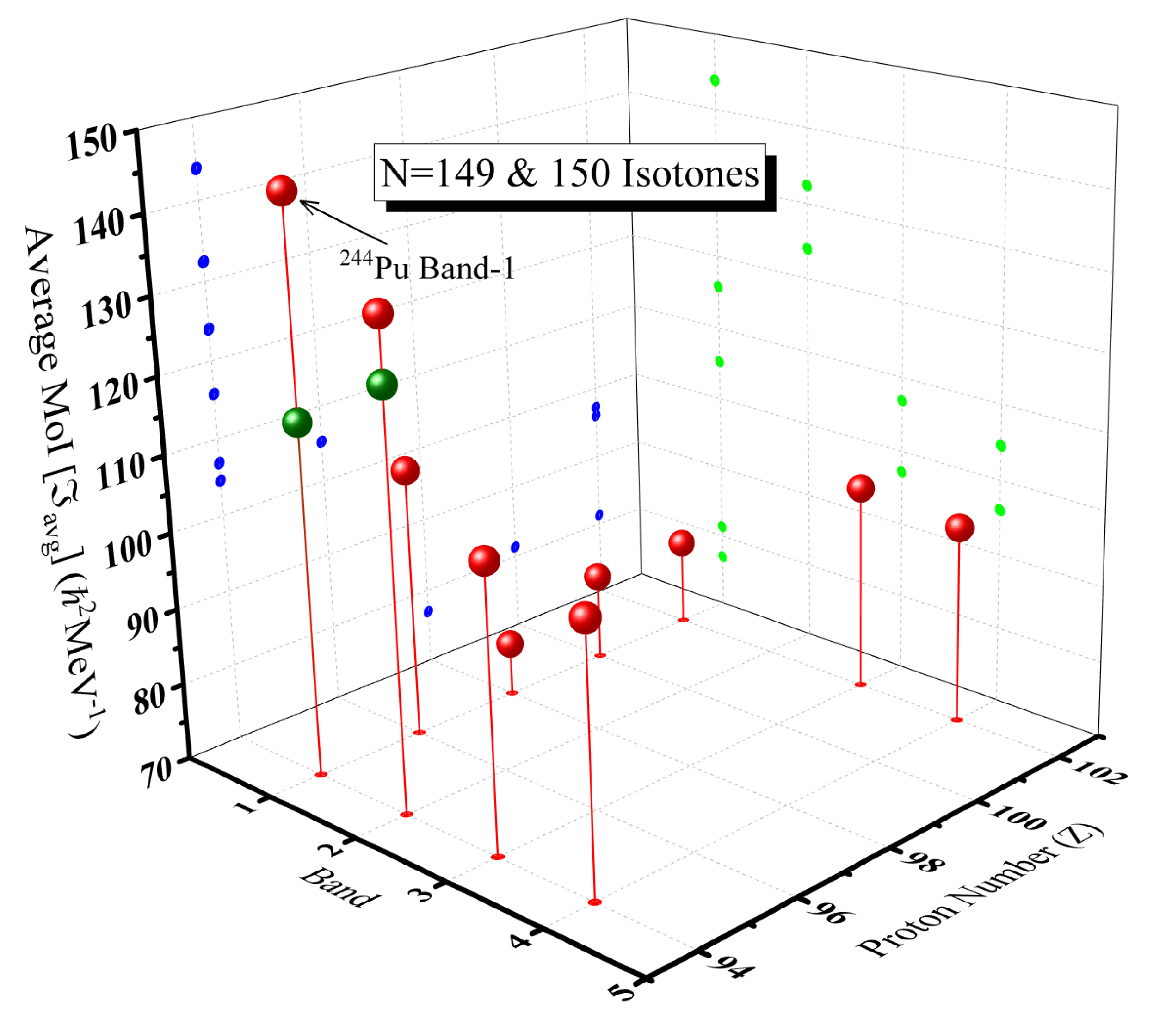}
  \caption{}
\end{subfigure}

\medskip 
\begin{subfigure}{.45\linewidth}
  \includegraphics[width=\linewidth]{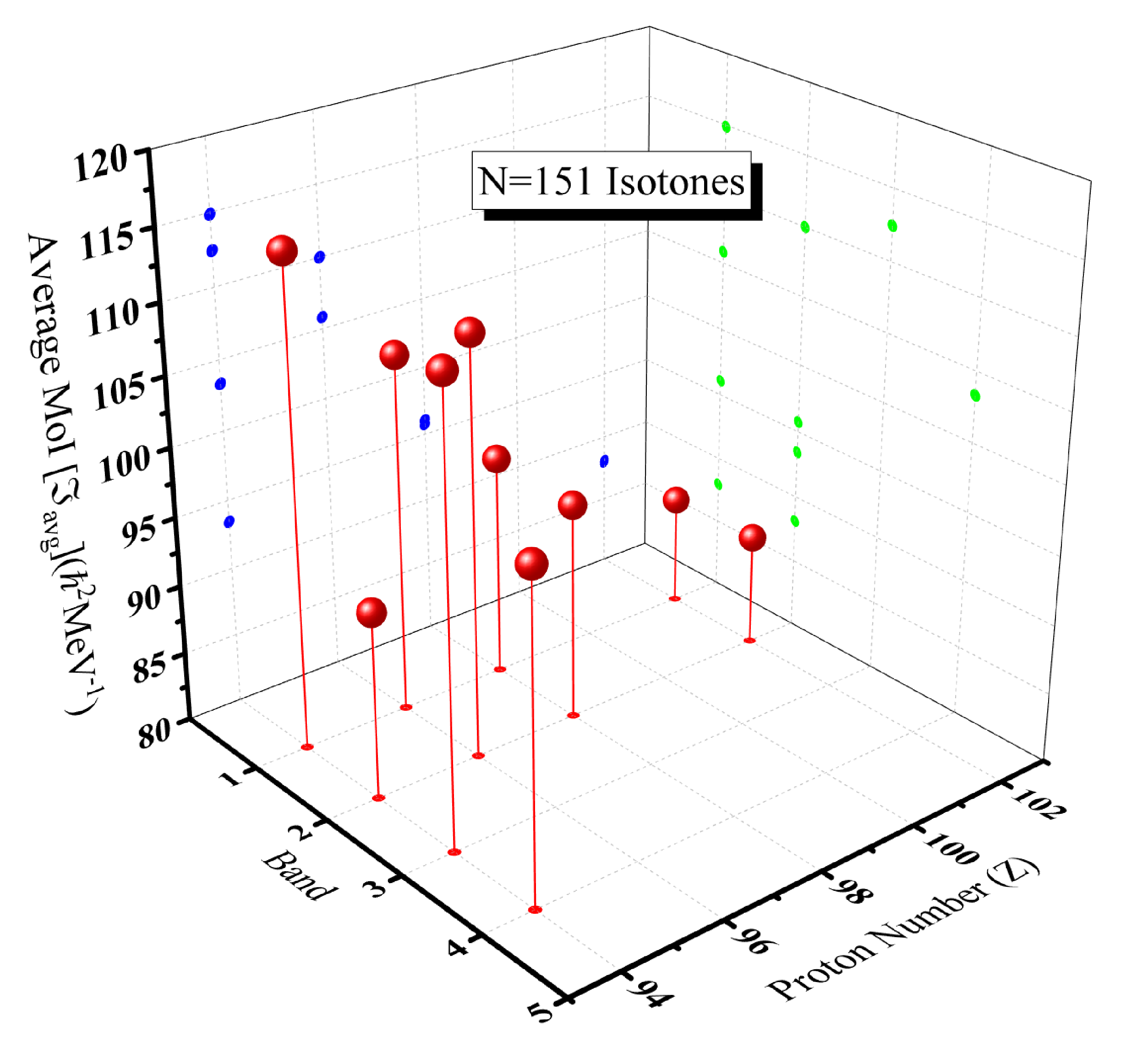}
  \caption{}
\end{subfigure}\hfill 
\begin{subfigure}{.45\linewidth}
  \includegraphics[width=\linewidth]{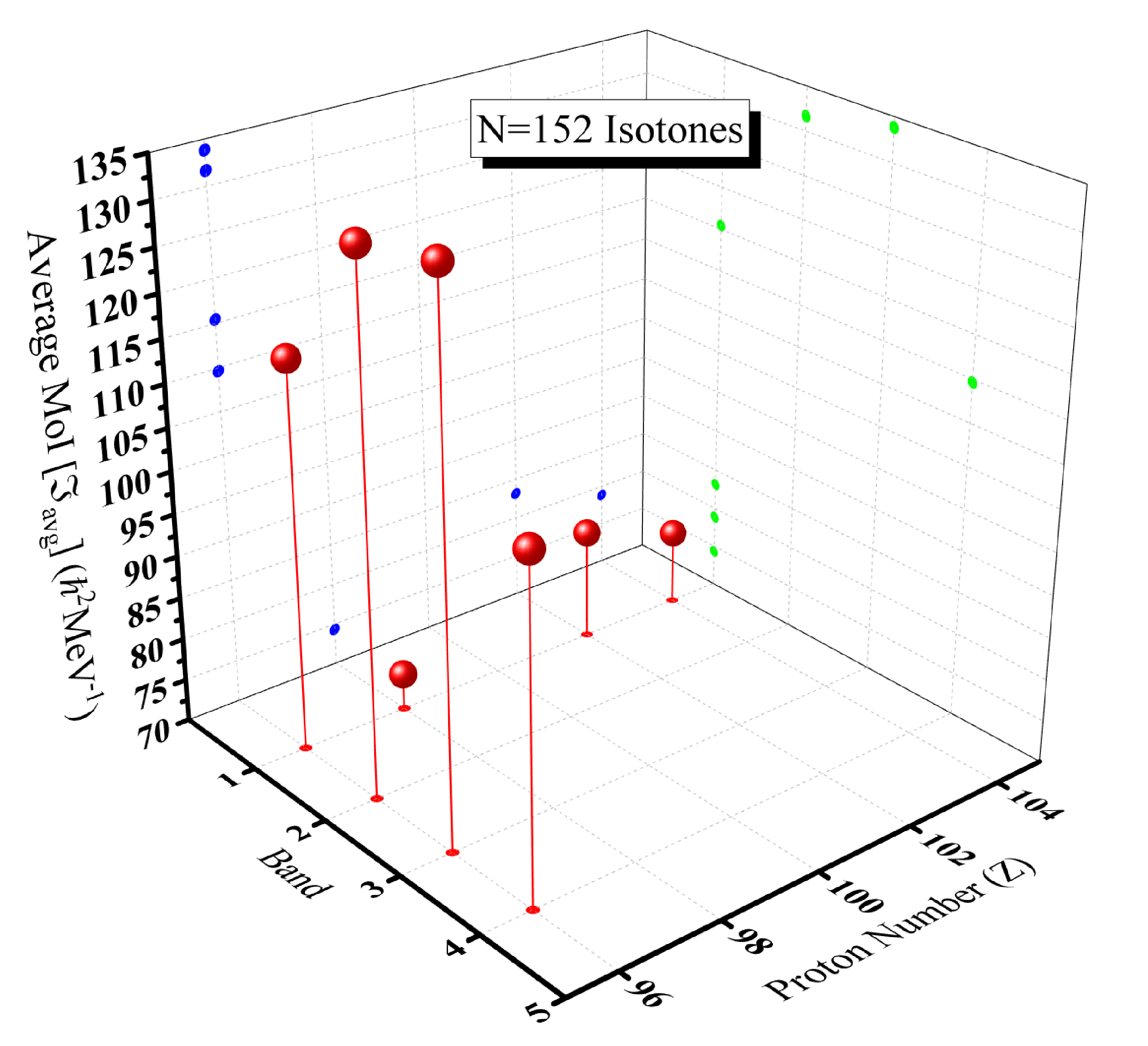}
  \caption{}
\end{subfigure}
\captionsetup{justification=raggedright, singlelinecheck=false}
\caption{ (a)-(d) Average moment of inertia ($\Im_{avg}$) vs the proton number (Z) and the band number (a) for N=148 isotones,
(b) for N=149 (green sphere), 150 isotones, (c) for N=151 isotones, (d) for N=152 isotones. The projection on XY, YZ and ZX is shown by red, blue and green dots, respectively.}
\label{fig19}
\end{figure*}

\begin{figure*}

\begin{subfigure}{.45\linewidth}
  \includegraphics[width=\linewidth]{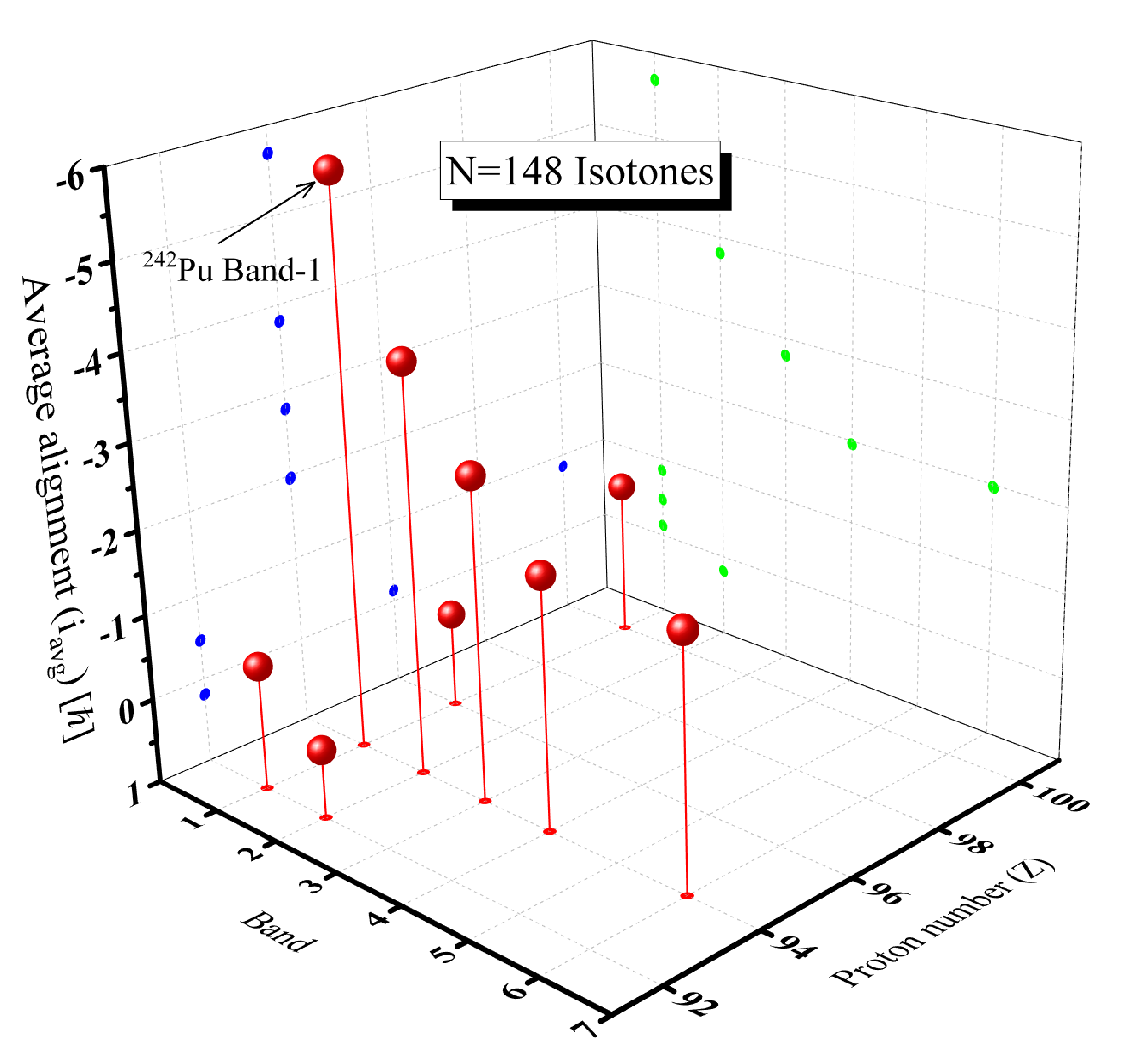}
  \caption{}
\end{subfigure}\hfill 
\begin{subfigure}{.45\linewidth}
  \includegraphics[width=\linewidth]{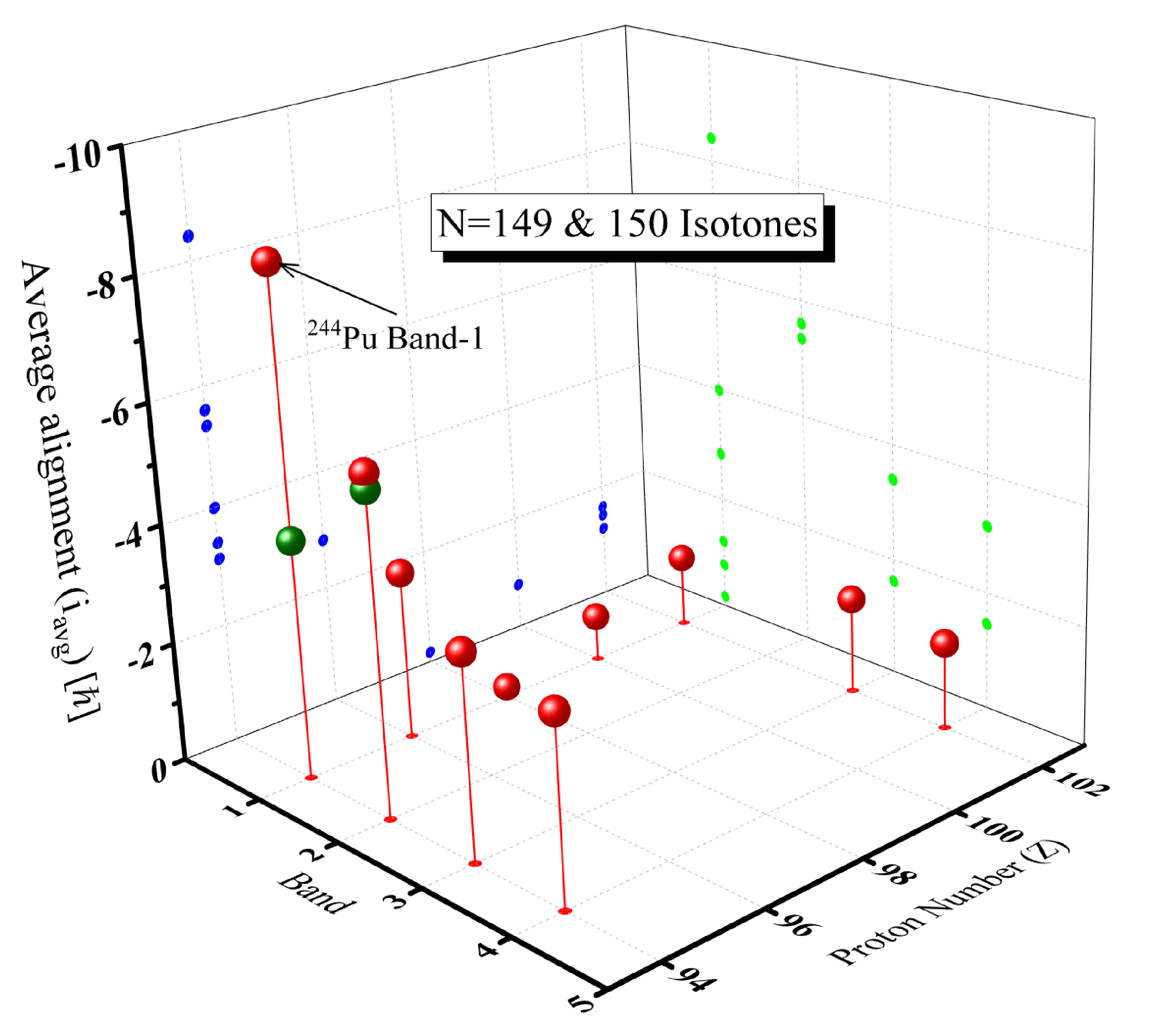}
  \caption{}
\end{subfigure}

\medskip 
\begin{subfigure}{.45\linewidth}
  \includegraphics[width=\linewidth]{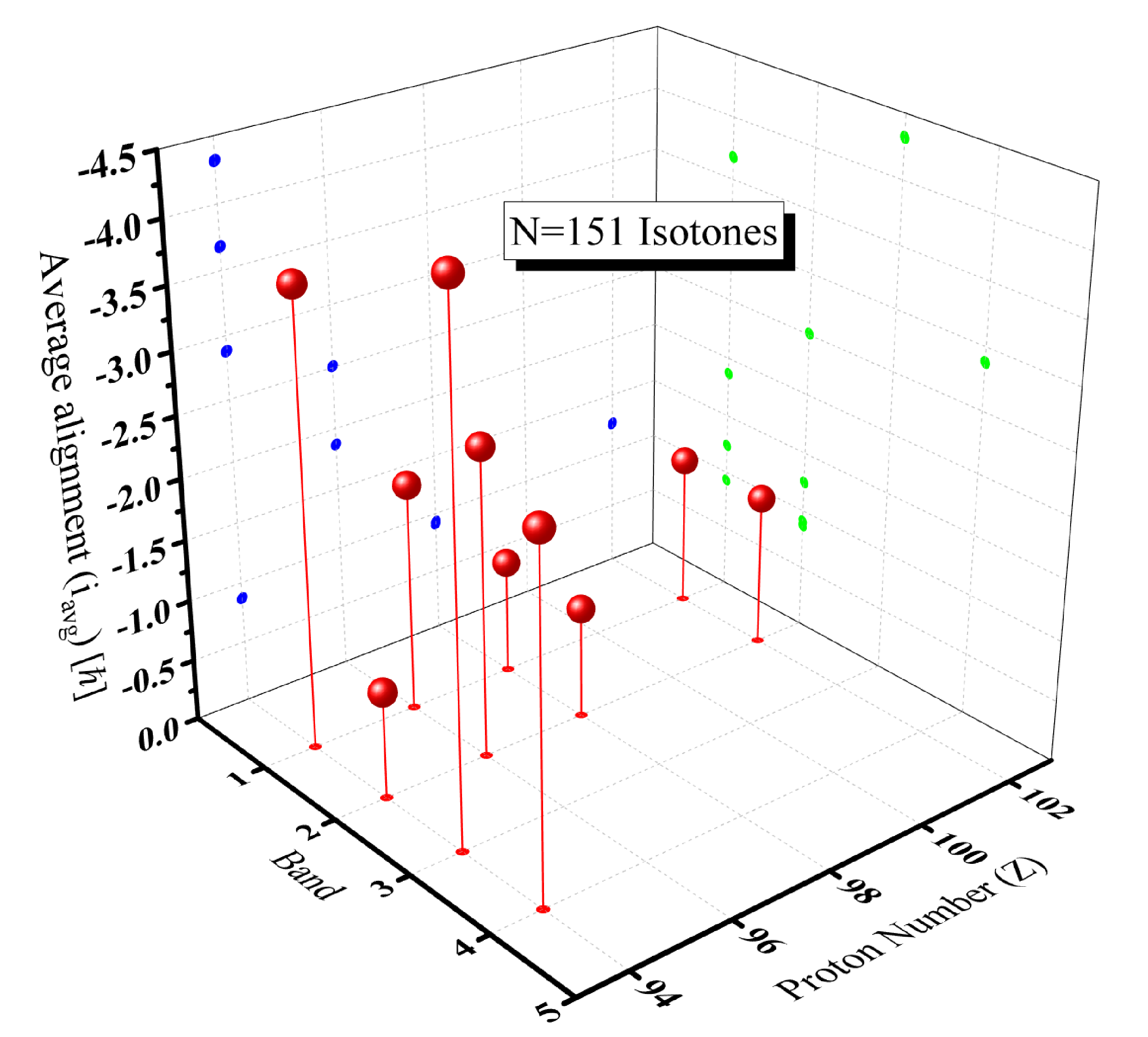}
  \caption{}
\end{subfigure}\hfill 
\begin{subfigure}{.45\linewidth}
  \includegraphics[width=\linewidth]{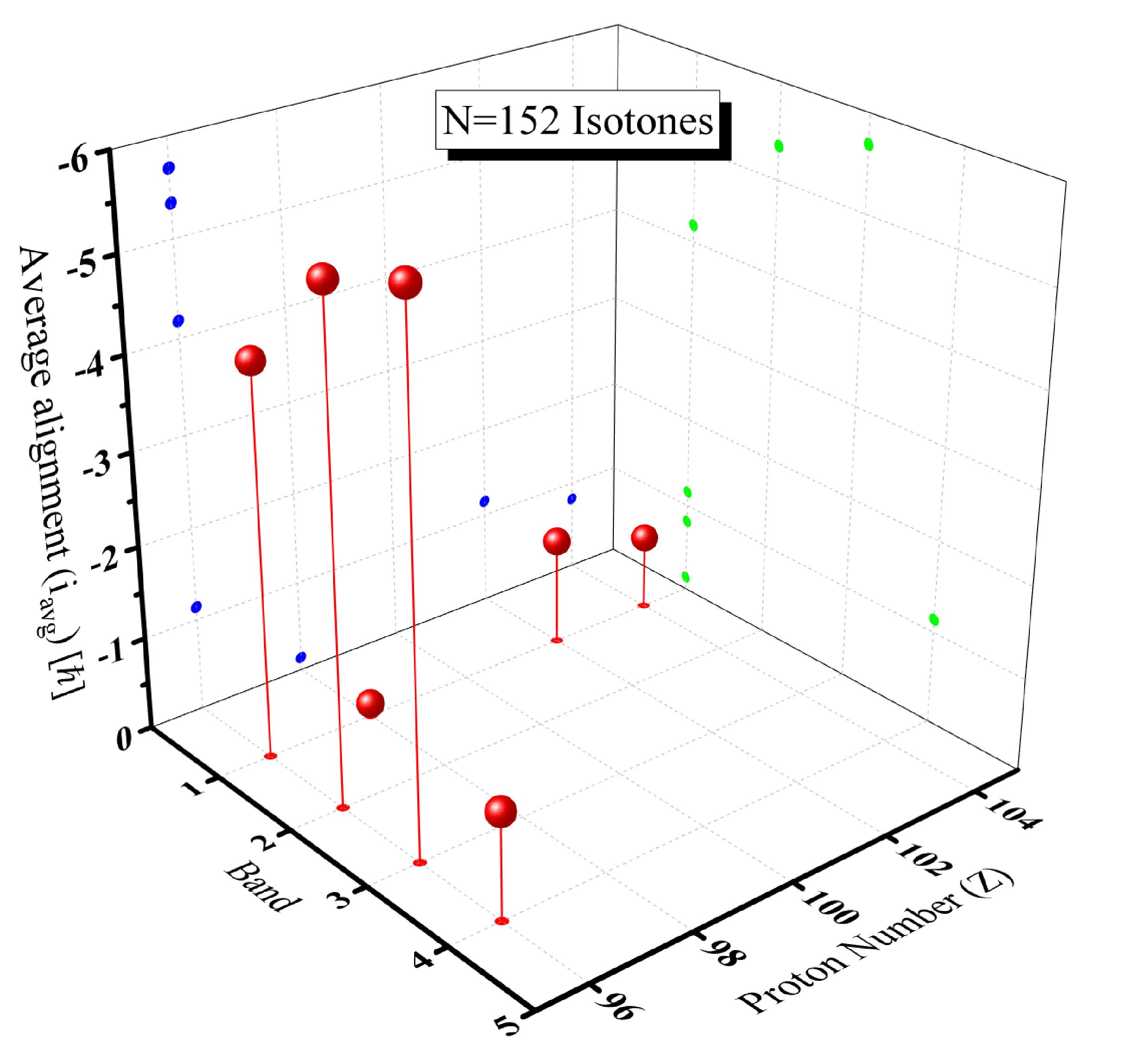}
  \caption{}
\end{subfigure}
\captionsetup{justification=raggedright, singlelinecheck=false}
\caption{ (a)-(d) Average alignment ($i_{avg}$) vs the proton number (Z) and the band number (a) for N=148 isotones, (b) for N=149 (green sphere), 150 isotones, (c) for N=151 isotones, (d) for N=152 isotones. The projection on XY, YZ and ZX is shown by red, blue and green dots, respectively.}
\label{fig20}
\end{figure*}

\section{Results and Discussion}

In the present paper, the rotational bands of the \(A \sim 250\) mass region have been systematically studied for the first time. The rotational bands from \(N=148\) to \(N=152\) isotones have been explored using various semi-classical approaches. The experimentally observed intraband \(\gamma\)-transition energies have been utilized, and a least-squares fitting method is employed. In this method, a comparison between the experimental and calculated data is conducted, a process known as the best-fit method (BFM). The degree of agreement between the calculated and experimental values is quantified using the root-mean-square (RMS) deviation.
\be
\chi = \left[ \frac{1}{n} \sum_{i=1}^{n} \left( \frac{E_{\gamma} ^{cal}(I_{i})-E_{\gamma} ^{exp}(I_{i})}
{E_{\gamma} ^{exp}(I_{i})}\right) ^{2}\right] ^{1/2},
\label{eq15}
\ee
(where $ n $ is the number of transitions involved in the fitting.) The paper has been distributed in the following ways: In Part A of
the paper, the dynamic MoI of the rotational bands has been examined employing an empirical semi-classical vibrational distortion model. The role of the vibrational distortion factor and its variation with rotational frequency is investigated. In Part B, we have explored the rotational bands with the shape fluctuation model. In Part C, the softness parameter, band-head moment of inertia, average moment of inertia, and average alignment are systematically studied. In Part D, the role of pairing and the anti-paring effect are examined with the variable moment of inertia inspired by the Interacting Boson model(VMI-IBM).
\subsection{Role of vibrational distortion term on dynamic moment of inertia}
Vibration and rotation are two modes of excitation commonly observed in atomic nuclei and molecules. In the nucleus, vibrational modes manifest as multipolar resonances, which are collective phenomena and demonstrate a mass number dependence of \(A^{-1/3}\). At low energies, these vibrational modes contribute to the rotational and vibrational effects on energy levels. Within this vibrational energy mode, the interaction between valence nucleons and the core generates distinct energy levels. In the region of low-lying nuclear structure, the interplay between rotation and vibration is frequently observed. Notably, increased vibration at a constant frequency often results in surface distortion or deformation of the nucleus, similar to a tidal wave rippling across its surface.\par

A vibration distortion model was proposed to study the superdeformed bands (SD) of the \(A \sim 150\) mass region \cite{roy}. This model is applicable to SD bands in other mass regions as well, where a smooth increase in the dynamic moment of inertia (\(\Im^{(2)}\)) is noted. Specifically, for the \(A \sim 150\) mass region, the vibrational distortion factor (\(\Im^{(2)}_{\text{vib}}\)) changes as a function of the rotational frequency (\(\hbar\omega\)). The expression \(((\omega_{\text{max}} - \omega)/\omega_{\text{max}})^{2}\) governs how \(\Im^{(2)}_{\text{vib}}\) varies with increasing rotational frequency. In the context of SD bands for the \(A \sim 150\) mass region, this expression leads to a decrease in \(\Im^{(2)}_{\text{vib}}\) as the rotational frequency increases. In contrast, for the \(A \sim 250\) mass region, where the deformation is about half that of the SD \(A \sim 150\) regions, the dynamic moment of inertia exhibits a marked increase after \(\hbar\omega \approx 0.2\) MeV. For the rotational bands in the \(A \sim 250\) mass region, we suggest adopting a \(((\omega_{\text{max}}-\omega)/\omega_{\text{max}})^{n}\) dependence on \(\Im^{(2)}_{\text{vib}}\), where \(n\) is a free parameter. For this particular mass region, we have selected \(n = -2\).\par

The dynamic MoI for rotational bands within \( N = 148 \) to \( 152 \) isotones has been thoroughly analyzed using the modified vibrational distortion model. This analysis allowed for the extraction of parameters \(\Im^{(2)}_{c}\) and \(\Im^{(2)}_{vib}\), which are detailed in Table \ref{tb1}. To validate the modified model incorporating a frequency-dependent vibrational distortion term, the dynamic MoI for 36 rotational bands in the \( A \sim 250 \) mass region, ranging from \( N = 148 \) to \( 152 \), was calculated. Figure \ref{fig1}(a) demonstrates the calculated dynamic MoI across varying values of \( n \) from \( +2 \) to \( -2 \), and includes a comparative analysis with the experimental dynamic MoI for \( ^{254}No \) Band-1. The results distinctly indicate that negative values of \( n \) more accurately reproduce the experimental curvature observed in the dynamic MoI. Specifically, for \( n = -2 \), there is a remarkable congruence between the calculated and experimental dynamic MoI. Moreover, the RMS deviation between the calculated and experimental dynamic MoI significantly diminishes, dropping from approximately \( 0.088 \) to \( 0.017 \) as \( n \) is adjusted from \( +2 \) to \( -2 \). Notably, the RMS deviation for \( n = +2 \) is approximately \( 5.4 \) times higher than that for \( n = -2 \), emphasizing the enhanced accuracy of the model with the selected negative parameter value.\par

The 36 rotational bands from $N = 148$ to $152$ isotones have been fitted with $n = -2$, and an excellent agreement with the experimental dynamic MoI is obtained. It is important to mention that in SD bands of the $A\sim 150$ mass region, only those bands are chosen for the fitting where there is no sudden increase in the dynamic MoI is observed \cite{roy}. It is encouraging to point out that by taking $n=-2$, the rotational bands, which show a sudden increase in dynamic MoI (such as $^{248}$Cm band-4, $^{249}$Cf band-1 and 2, $^{245}$Pu band-2, and so on), have been reproduced very well. By taking $n=-2$ in the $((\omega_{max}-\omega)/\omega_{max})^{n}$ term, we have obtained a lesser RMS deviation and, in most cases, an excellent agreement with the experimental trend of dynamic MoI, which validates our proposition of using a modified vibrational distortion model. The only exception where the vibrational distortion model (with $n=+2$ and $n=-2$) fails to reproduce the experimental trend is where the dynamic MoI is a very sharp downturn. The figure \ref{fig1}(b) shows the variation of calculated and experimental dynamic MoI for $^{248}$Cm band-1. It is evident from the figure \ref{fig1}(b) that up to the point where dynamic Moi increases smoothly, the calculated dynamic MoI with $n=-2$ is in excellent agreement with experimental dynamic MoI, however, the sharp downturn of dynamic MoI cannot be reproduced. For comparison, we have also shown the calculated dynamic MoI with $n=+2$. It is clear from the figure that with $n=+2$, neither the smoothly increasing trend nor the downturn is reproduced. In addition, the parameters obtained also seem unrealistic. Therefore, despite its overall effectiveness, especially with \( n = -2 \), the model has limitations in certain scenarios, particularly in accurately reproducing the sharp downturns in dynamic MoI for the \( A \sim 250 \) mass region. \par

In general, the constant part of the dynamic MoI varies from $50.24$ to $95.58$ $\hbar^{2} MeV^{-1}$, where the least value of $\Im^{(2)}_{c}$ is obtained for $^{240}$U band-1, and the highest is obtained for $^{242}$Pu band-1. The vibration part $\Im^{(2)}_{vib}$ inertia varies from $0.82$ $\hbar^{2} MeV^{-1}$ (for $^{243}Pu$ band-1) to $26.47$ $\hbar^{2} MeV^{-1}$(for $^{253}$No band-1) (see table \ref{tb1}). Also, a positive coupling between the constant and vibration part is considered for all the bands of the $A\sim 250$ mass region. For rotational bands of $N=148$ isotones and $N=149,150$ isotones, we observed two distinct patterns of dynamic MoI. The category-I belongs to the bands that start at $\approx 60 $ $\hbar^{2} MeV^{-1}$, and category-II of bands starts from relatively higher dynamic MoI $\approx 90$ $\hbar^{2} MeV^{-1}$. For both of these categories, the dynamic MoI starts with the different values, and after $\hbar\omega \approx0.2MeV$ starts to merge and increase (see figure \ref{fig2} and figure \ref{fig3}).The average values of the parameters obtained for N=148 isotones are $\Im_{c}^{(2)}= 57.0\hbar^{2}MeV^{-1}$ and $\Im_{vib}^{(2)}=9.0\hbar^{2}MeV^{-1}$ for category-I and $\Im_{c}^{(2)}= 88.6\hbar^{2}MeV^{-1}$ and $\Im_{vib}^{(2)}= 2.46\hbar^{2}MeV^{-1}$.The average values of parameters obtained for N = 149 and 150 isotones are $\Im_{c}^{(2)}= 59.1\hbar^{2}MeV^{-1}$ and $\Im_{vib}^{(2)}= 9.23\hbar^{2}MeV^{-1}$ for category-I and $\Im_{c}^{(2)}= 87.44\hbar^{2}MeV^{-1}$ and $\Im_{vib}^{(2)}= 1.25\hbar^{2}MeV^{-1}$.For illustrative purposes, figures \ref{fig2} and \ref{fig3} show the variation of the dynamic moment of inertia only up to the $0.20$ MeV rotational frequency. \par
For the \( N = 151 \) isotones, the analysis identifies three distinct patterns in the dynamic MoI. Category I encompasses bands where the dynamic MoI exhibits a gradual and smooth increase across the frequency range, exemplified by the rotational bands of \( ^{253}\text{No} \) (refer to Figure \ref{fig4}(a)). In this category, the average value for the constant component of dynamic MoI, denoted as \( \Im^{(2)}_{c} \), is calculated to be \( 47.53 \, \hbar^{2}\text{MeV}^{-1} \). This value is notably the lowest among all bands spanning from \( N = 148 \) to \( 152 \) isotones. Conversely, the vibrational component, \( \Im^{(2)}_{vib} \), averages at \( 22.6 \, \hbar^{2}\text{MeV}^{-1} \), which is the highest observed across the same range of isotones. This pattern highlights the unique behavior of the dynamic MoI in the \( N = 151 \) isotones, particularly in contrast to other categories within this nuclear mass region. In the N=151 isotones, Category II is characterized by bands where the dynamic MoI steadily rises, even beyond the rotational frequency of \( \hbar\omega \approx 0.2 \) MeV. Notably, the rotational bands of \( ^{249}\text{Cf} \) and \( ^{247}\text{Cm} \)are representatives of this category. The average constant component of the dynamic MoI for these bands, \( \Im^{(2)}_{c} \), is calculated to be \( 78.47 \, \hbar^{2}\text{MeV}^{-1} \), and the vibrational component, \( \Im^{(2)}_{vib} \), averages at \( 4.42 \, \hbar^{2}\text{MeV}^{-1} \) (refer to Figure \ref{fig4}(b)). Category III, on the other hand, includes bands like those of \( ^{245}\text{Pu} \), which exhibit a more abrupt increase in the dynamic MoI. This sharp rise is typically observed within the frequency range of \( \hbar \omega \approx 0.15 \) to \( 0.20 \) MeV. For these bands, the average values for the components of dynamic MoI are \( \Im^{(2)}_{c} = 83.64 \, \hbar^{2}\text{MeV}^{-1} \) and \( \Im^{(2)}_{vib} = 0.76 \, \hbar^{2}\text{MeV}^{-1} \) (as shown in figure \ref{fig4}(c)). These findings highlight the diversity in the behavior of the dynamic MoI across different bands within the N=151 isotones.

In the study of \( ^{245}\text{Pu} \) band-1 and band-3, it is noteworthy that the last two data points for band-1 and the last data point for band-3 were excluded from the calculations of dynamic MoI. This exclusion was necessary due to the sharp downturn observed in these points. The figure \ref{fig4}(d) shows the merged plot of categories I and III. The analysis of the rotational bands in categories I and III reveals an interesting observation. Initially, these bands begin with nearly identical values of dynamic MoI, indicating a close similarity in their initial rotational characteristics. Furthermore, the constant component of their dynamic MoI is very similar, suggesting comparable underlying nuclear structures in the initial phase of rotation. However, a notable divergence occurs after the rotational frequency (\(\hbar\omega\)) reaches 0.2 MeV. Beyond this point, the two categories start exhibiting distinct behaviors in terms of their dynamic MoI. This divergence implies that while the initial states of these categories are similar, the factors influencing their rotation evolve differently as the rotational frequency increases.

In the N=152 isotones, the behavior of the dynamic MoI has been categorized into two groups. The first category I shows a smooth and continuous increase in the dynamic MoI with the increase in rotational frequency (\(\hbar\omega\)) (figure \ref{fig5}(a)). For these bands, the average values of the constant and vibrational parts of the MoI are 73.34 \(\hbar^{2} MeV^{-1}\) and 7.49 \(\hbar^{2} MeV^{-1}\), respectively. Conversely, category II is characterized by a much slower and less pronounced increase in the dynamic MoI (figure \ref{fig5}(b)). The average values for this category are 54.44 \(\hbar^{2} MeV^{-1}\) for the constant part and 11.97 \(\hbar^{2} MeV^{-1}\) for the vibrational part. Interestingly, when the data for both categories are combined, a common starting point in dynamic MoI is observed (figure \ref{fig5}(c)). However, as the rotational frequency increases beyond about 0.15 MeV, the behaviors start to diverge. This divergence, similar to that observed in the N=151 isotones, suggests a shared structural or interactional feature in these isotones that responds uniquely to rotational forces. \par

The analysis of the dynamic MoI across various categories of N = 148 to 152 isotones, as shown in figure \ref{fig6}, reveals three distinct patterns. These categories reflect different nuclear behaviors in response to increasing rotational frequency (up to 0.20 MeV).\newline\textbf{1.} Category II of N = 148, 149, 150, and 151 Isotones (Red Curves): This category is characterized by a higher starting value of dynamic MoI ($\approx$ 90 \(\hbar^{2} \text{MeV}^{-1}\)), with a marked and abrupt rise after reaching a rotational frequency of 0.2 MeV. This steep increase suggests a significant structural or interactional change within the nucleus at higher rotational frequencies. This could be due to a phase transition or a change in the shape or configuration of the nucleus.\newline\textbf{2.} Category I of N = 148, 149, and 150 Isotones (Green Curves): In this pattern, the dynamic MoI begins at a lower value ($\approx$ 65 \(\hbar^{2} \text{MeV}^{-1}\)) and shows a more gradual increase with the rotational frequency. Interestingly, these bands converge with the red curves at $\sim$ 0.2 MeV frequency. Despite starting from different dynamic MoIs, this convergence suggests a similar nuclear response to rotational stress at higher frequencies.\newline\textbf{3.} Category I of N = 151 and Category II of N = 152 Isotones (Blue Curves): These bands initially follow a similar trend to Category I of N = 148, 149, and 150 isotones, exhibiting a smooth and quenched-type increase in dynamic MoI up to 0.2 MeV. They merge with the green lines at lower frequencies but eventually diverge due to variations in the vibrational distortion term.\par
To highlight the role of the constant and vibrational parts in producing different behaviors in dynamic MoI, we have calculated the variation of the constant part and the vibrational part of the dynamic MoI with increasing rotational frequency. The figure \ref{fig7} shows the variation of $\Im^{(2)}_{c}$ and $\Im^{(2)}_{vib}$ for $^{256}$Rf band-1, $^{253}$No band-1, and $^{245}$Pu band-2. These three rotational bands show a smoothly increasing, almost flat, and abrupt increase in the dynamic MoI. It is clear from the figure that for an almost flat dynamic MoI of the $^{253}$No band-1, the vibrational distortion term varies very smoothly with increasing rotational frequency. This smooth change reflects a gentle and predictable response of the nucleus to increasing rotational stress, maintaining a near-constant dynamic MoI. For $^{256}$Rf band-1, the increase in vibrational distortion term is more pronounced than $^{253}$No. The increasing vibrational distortion term indicates a more flexible response of the nucleus to rotational forces, deviating more from a rigid rotor model. The most intriguing behavior is observed in the \( ^{245} \)Pu band. Initially, up to a rotational frequency of approximately \( 0.15 \)MeV, the nucleus behaves like a rigid rotor, with the dynamic MoI being primarily contributed by the constant part associated with the rigid core of the nucleus. The vibrational term remains almost negligible, indicating minimal internal structural change or shape fluctuation. However, beyond \( 0.15 \) MeV, there is a sudden and rapid increase in the vibrational distortional term. This abrupt rise at higher frequencies suggests a significant shift in nuclear behavior. The fact that the vibrational contribution becomes almost half of the rigid MoI term at the maximum experimental frequency of \( \sim 0.22 \) MeV is particularly striking. It implies a strong rotational-vibrational interaction at higher rotational frequencies.This phenomenon, especially pronounced in Pu isotopes, where the dynamic MoI remains almost constant up to a certain rotational frequency and then sharply increases, points to unique nuclear behaviors in these isotopes. It highlights the critical role of the vibrational distortion term in the context of rotational-vibrational interactions and the transition from rigid rotor behavior to more complex dynamic responses.
\subsection{The systematics of shape fluctuation energy}
To corroborate our observations regarding an increase in the vibrational distortion term in the mass region around \( A \sim 250 \), we have calculated the rotational energy (ROTE or \( E_{\gamma}^{ROT} \)) and shape fluctuation energy (SFE or \( E_{\gamma}^{SF} \)) for the rotational bands in isotones from \( N = 148 \) to \( 152 \), using the shape-fluctuation model. Our primary focus centers on scrutinizing the behavior of ROTE and SFE in those specific bands where the vibrational distortion model anticipates a marked increase in the vibrational distortion component of the dynamic MoI. To affirm the applicability of the shape fluctuation (SF) model in the mass region around \( A \approx 250 \), we have calculated the RMS deviation of 36 rotational bands considered in this paper. It is noteworthy that for the majority of these bands, the RMS deviation between the calculated and experimental transition energies falls within the range of \( 10^{-2} \) to \( 10^{-3} \). This implies the validation of the SF model for the $A\sim250$ mass region. Since the SF model explicitly depends upon the band-head spins, we have adopted the band-head spins given in Ref. \cite{nndc}.\par
Using the parameters obtained from the BFM, we have calculated the intraband-$\gamma$ transition energy of the rotational bands in N = 148 to 152 isotones from the SF model. The SF model splits the $E_{\gamma}$ into two parts: rotational energy ($E_{\gamma}^{ROT}$) and shape fluctuation energy ($E_{\gamma}^{SF}$). The cumulative effect of these two energies yields the total intraband gamma transition energy of a nuclear state, expressed as \( E_{\gamma}^{ROT} + E_{\gamma}^{SF} = E_{\gamma} \). Notably, the shape fluctuation energy serves as a powerful indicator for identifying structural irregularities in nuclei across various mass regions, as evidenced in prior studies \cite{BINDRA201848,dadwal}.\par
The variation of SFE and ROTE with rotational frequency in the $A\sim250$ mass region displays several really interesting features.The variation of SFE and ROTE for N=148 isotones is shown in figure \ref{fig8}. For these isotones, ROTE generally exhibits a steady increase with rotational frequency. However, the bands of \( ^{242} \)Pu demonstrate notable deviations from this trend. It is evident from figure \ref{fig8}(a) that $^{242}$Pu bands 2 and 3 have a higher contribution from ROTE than other bands. For $^{242}$Pu band-1, the ROTE starts from a relatively lower value and merges with other rotational bands of $N=148$. isotones. Intriguingly, at around \( \hbar \sim 0.20 \) MeV, this band's ROTE diverges, marked by a pronounced upward bend at 0.25 MeV. The analysis of SFE shows that it decreases monotonically with rotational frequency for all the bands of N=148 isotones. It can be clearly seen in the figure that for $^{242}$Pu bands 2 and 3, the magnitude of the SFE is greater than the rest of the bands, which have a relatively low magnitude of SFE. Another astonishing observation is the rapid decrease in the SFE of the $^{242}$Pu band-1. For most of the bands in $N=148$ isotones, the magnitude of SFE is very less than the total energy however, for $^{242}$Pu bands 1, 2, and 3, the magnitude is very high. In particular, we noticed that the SFE of the $^{242}$Pu band-1 is almost half of ROTE at a higher rotational frequency ($\hbar\omega>0.2$MeV). The behavior of the SFE in \( ^{240} \)U band-2 presents an interesting case. Unlike typical trends observed in rotational bands, the SFE for this particular band does not diminish with an increase in spin values. Throughout its spin range, the SFE of \( ^{240} \)U band-2 remains remarkably stable, hovering close to zero. This is quite strange since the constant or negligible SFE variation was found in ``flat" superdeformed bands where the dynamic MoI remains mostly constant with increasing rotational frequency \cite{dadwal}. However, the lack of extensive experimental data for \( ^{240} \)U band-2, with only four energy levels identified so far \cite{nndc}, limits the ability to conclusively determine if this band exhibits a flat dynamic MoI.
\par
For the neutron number N = 149, only experimental data sufficient to study the systematics is available for $^{243}$Pu band-1 and 2 \cite{nndc}. Observations from figure \ref{fig9} reveal that both \( ^{243} \)Pu band-1 and band-2 exhibit a pronounced upward bend in ROTE around the rotational frequency of \( \hbar\omega \approx 0.22 \)MeV. Coinciding with this upbend in ROTE, the SFE demonstrates a sharp decline at approximately the same frequency. This pattern mirrors the behavior noted in \( ^{242} \)Pu band-1 of N = 148 isotones. Moving to N = 150 isotones, the majority of rotational bands display a consistent grouping in the ROTE versus rotational frequency plot, with notable exceptions being \( ^{244} \)Pu bands 1 and 2 (refer to figure \ref{fig10}). Throughout thefrequency spectrum, \( ^{244} \)Pu band-2 maintains a higher ROTE compared to other bands of N = 150 isotones. Notably, the ROTE of \( ^{244} \)Pu band-2 exhibits a slight upward trend beyond \( \hbar\omega = 0.2 \)MeV, as shown in figure \ref{fig10}(a). In contrast, \( ^{244} \)Pu band-1 displays a unique pattern, reminiscent of a backbending phenomenon. Up to \( \hbar\omega \approx 0.22 \)MeV, its behavior aligns with other N = 150 isotone bands. However, beyond this frequency, the ROTE of \( ^{244} \)Pu band-1 increases dramatically, almost doubling the value seen at \( \hbar\omega \approx 0.20 \)MeV. The behavior of SFE with rotational frequency for these isotones also follows a similar trend (figure \ref{fig10}(b)). The backbending-like shape is evident in the SFE versus \( \hbar\omega \) plot for \( ^{244} \)Pu band-1, where the SFE approximately doubles at \( \hbar\omega \approx 0.22 \)MeV. These findings indicate significant changes in nuclear structure and dynamics at these specific rotational frequencies, particularly in the \( ^{244} \)Pu isotopes.
Since the SF model depends explicitly on the band-head spins, we have deduced the band-head spins of the $^{253}$No bands 1 and 2. The methodology employed involves a comparison between calculated and experimental transition energies. Whenever an accurate band-head spin is assigned to the rotational band, the calculated and experimental transition energies coincide excellently \cite{dadwal3}. Conversely, any deviation in the band-head spin, such as \( \pm1\hbar \), results in a significant increase in the RMS deviation. The figure \ref{fig11} distinctly indicates that the band-head spins deduced for the \( ^{253} \)No bands 1 and 2 are lower by \( 1\hbar \) compared to the spins reported in Ref. \cite{nndc}. To corroborate these revised spin assignments, an analysis involving the plot of \( \Delta E_{\gamma}(I) \equiv E_{\gamma}(I+2)-E_{\gamma}(I) \) versus \( I-I_{0} \) was conducted for \( ^{253} \)No band-1, as illustrated in figure \ref{fig12}. In this figure, the calculated \( \Delta E_\gamma(I) \) values for various proposed band-head spin states are depicted by dashed lines. This analysis provides a clear and compelling visualization: the curve corresponding to a band-head spin \( I_{0} \) of \( 7/2\hbar \) aligns closely with the experimentally observed data. In contrast, the curves representing \( I_{0} \) values deviating by \( \pm 1\hbar \) (i.e., \( 7/2 \pm 1\hbar \)) exhibit significant divergence from the experimental trajectory.
A supplementary approach to validate band-head spin assignments involves the utilization of the ratio \( R \equiv \sqrt{[\Im^{(1)}]^{3}/\Im^{(2)}} \), as depicted in figure \ref{fig13}. This method, characterized by its simplicity and robustness, enables the determination of band-head spins without the need for parameter fitting. The distinctive advantage of this technique lies in its ability to confirm spin assignments through a clear empirical criterion that whenever the correct band-head spin is assigned to a rotational band, the Ratio-R exhibits a constant, spin-independent behavior. As evident in figure \ref{fig13}, the application of this method yields results that are in agreement with the spin assignments derived from different rotational energy formulae in figure \ref{fig11} . It was also proposed that $^{253}$No band-1 and 2 are signature partner bands \cite{Reiter3}. Previous systematic studies of band-head MoI reveal that the signature partner bands have a similar intrinsic structure \cite{dadwal3}. This is reflected in the very similar band-head MoI of signature partner bands. Since band-head MoI explicitly depends upon the spin assignment whenever the correct spin assignment is made, the band-head MoI of signature partners will be almost similar. In the case of \( ^{253}No \) bands 1 and 2, our proposed spin assignments have led to the determination of their band-head MoIs as approximately \( 75.36 \hbar^{2}MeV^{-1} \) and \( 75.45 \hbar^{2}MeV^{-1} \), respectively, calculated using the 'ab formula'. This close similarity in the MoI values of the two bands not only supports the notion that they are signature partners but also reinforces the validity of our spin propositions of \( I_{0}=7/2\hbar \) for band 1 and \( I_{0}=9/2\hbar \) for band 2. Previous level spin assignments for $^{253}$No are based on the spin dependence of $\Im^{(2)}$, which gives the correct spin when the quasiparticle alignment is almost zero, a condition that is fulfilled in this case \cite{Reiter3}. These spin assignments were further validated by comparing experimental $\Im^{(1)}$, which in low frequency is within $1\%$ of $7/2^+[624]$ band in the $^{249}$Cf \cite{Reiter3}. It is also encouraging to mention that our spin assignments for $^{253}$No band-1 and 2 also agree with the previous spin assignments using the Harris formula.
For neutron number N = 151 isotones, the trends observed in SFE and ROTE mirror those identified in Pu isotopes for N = 148, 149, and 150 isotones. In particular, for \( ^{245} \)Pu bands 1, 3, and 4, there is a notable pattern wherein the ROTE undergoes a sharp increase while the SFE exhibits a marked decrease (increase in magnitude) at a rotational frequency \( \hbar\omega \sim 0.22 \) MeV. This behavior is further illustrated in figure \ref{fig14}, where the SFE is observed to reach almost half the value of ROTE. A contrasting scenario is presented by \( ^{253}No \) bands 1 and 2, which display the smallest magnitudes of ROTE and SFE in comparison to other nuclei within the \( A\sim250 \) mass region. This is particularly intriguing considering that for these bands, the constant term \( \Im_{c}^{(2)} \) and \( \Im_{vib}^{(2)} \) calculated using the vibrational distortional model are also the lowest and highest, respectively, across the \( A\sim250 \) mass region. Moreover, the dynamic MoI of \( ^{253}No \) bands 1 and 2 exhibit the least increase in MoI among their counterparts. For N = 152 isotones, a similar pattern to that observed in Pu isotopes (however, more pronounced in Pu isotopes) is evident in \( ^{248}Cm \) bands 1-4, as evident in figure \ref{fig15}.

The systematics of ROTE and SFE of all the rotational bands in $N = 148$ to $152$ isotones reveal that for bands that have asharp upbend in the dynamic MoI, their SFE increases abruptly after $\hbar\omega 0.20$ MeV and attains $\approx 50\%$ of the ROTE. This result is also in accordance with our previous results, where the vibrational distortion term became half of the rigid core MoI in $\hbar\omega > 0.20$ MeV. This implies that SFE is decreasing, but its magnitude is increasing. The negative sign of the SFE represents that the core is de-exciting or losing its energy during the fluctuations \cite{sathpathy}. The absolute value of SFE (which ignores the negative sign) suggests that the shape of the nuclei is changing due to the fluctuations. This result complements our results obtained from the vibrational distortion model, where it was calculated that the abrupt increase in the dynamic MoI is due to the increase in the vibration distortion term. Also, we have observed that the constant part in the vibrational distortion model and rotational energy are particularly high in the bands where MoI shows an abrupt increase.
In contrast, the analysis of the $^{253}$No band-1,2 demonstrates a divergent trend where the dynamic MoI exhibits a consistent increase across the frequency spectrum, and both the constant term of MoI ($\Im_{c}$) and rotational energy ($E_{\gamma}^{ROT}$) maintain comparatively lower values. It is crucial to acknowledge that shape fluctuation energy is influenced not only by the rotational spectrum but also by contributions from the phonon spectrum. According to the insights gained from the vibrational distortion model and the shape fluctuation model, an increase in the rotation-vibration interaction is posited as a likely cause for the heightened MoI observed at $\hbar\omega > 0.20$ MeV. Furthermore, the observed anomalous behavior, which is more pronounced in the series of Plutonium (Pu) isotopes, suggests that these nuclei exhibit ``soft" characteristics. This softness implies a greater susceptibility to shape changes and deformations under rotational and vibrational forces, setting them apart from the trends observed in other isotopic sequences. \par
To gain further insight into the distinctive traits of the ground state bands (gsb or band-1) in $^{242}$Pu and $^{244}$Pu, we have extended our calculations to include the ROTE and SFE analyses for lighter Plutonium (Pu) isotopes with mass numbers $A \leq 240$. Figure \ref{figPu-Isotopes} in our study illustrates the variations in ROT and SFE for the gsb of even-A Pu isotopes. A notable observation from this figure is that the lighter Pu isotopes, specifically $^{238,240}$Pu, do not exhibit any abrupt increases in ROT or SFE in the higher frequency region ($\hbar\omega > 0.20$ MeV). Furthermore, the ROT and SFE values for $^{242,244}$Pu are consistently higher across the frequency spectrum in comparison to $^{238,240}$Pu. This difference is particularly pronounced in the SFE, where the SFE for $^{238,240}$Pu is approximately half that of $^{242,244}$Pu. The gradual nature of ROT and SFE changes in the $^{238,240}$Pu isotopes is intriguing, especially considering that the systematics of ROT and SFE in heavier Pu isotopes ($A \geq 242$) indicate an increase in these values in the higher frequency region. It has been hypothesized, as proposed by Wiedenhöver et al. \cite{wiedenhover}, that in lighter Pu isotopes there might be a transition from octupole vibration to stable octupole deformation. Our findings regarding ROT and SFE in $^{238,240}$Pu appear to support and complement the results presented in Ref. \cite{wiedenhover}, indicating a complementary understanding of the nuclear structural dynamics within these specific isotopes.

\subsection{The systematics of softness parameter and moment of inertia}
The softness parameter ($\sigma$) was obtained in the expansion of energy in terms of $I(I+1)$ \cite{bohr}. The softness parametermeasures the rigidity of bands \cite{neha}. For the bands that show small (or large) values of the softness parameter,the rigidity is high (or low). We have calculated the softness parameter of 36 rotational bands in $N = 148$ to $152$ isotones of$A\sim250$ mass region using the soft-rotor formula \cite{dadwal3}. It has been observed that the softness parameter is relativelyelevated for the Pu isotopes compared to other nuclei within the same isotonic series, as illustrated in figure \ref{fig16}. Notably, the softness parameter reaches its apex for the $^{244}$Pu band-1, representing the highest value among all the 36 bands considered in this study. Following closely, the $^{242}$Pu band-1 exhibits the second-highest value of the softness parameter. A high softness parameter value indicates that Pu isotopes exhibit increased sensitivity to shape alterations as angular momentum rises, signifying their inherent ``softness". This result aligns with our earlier findings derived from vibrational distortional and shape fluctuation models, revealing a swift transition in both vibrational distortion term and shape fluctuation energy within the $^{242,244}$Pu band-1.To further strengthen our proposition, we have used another approach that is simple yet robust \cite{neha}. In this approach, one of the most common signatures of rigidity, i.e., the ratio of $E_{4}/E_{2}$ is explored. We have calculated the ratio like $R(I)=E_{\gamma}(I\rightarrow I-2)/E_{\gamma}(I-2\rightarrow I-4)$ for the whole range of spin in rotational bands $^{244}$Pu band-1 and $^{242}$Pu band-1. For comparison, the ratio R (I) of band-1 of $^{252}$No, $^{254}$No, and $^{256}$Rf are also plotted. These R(I) vs. I plots are then compared with the ideal rigid-rotor over a similar spin range (figure \ref{fig17}). For both $^{244}$Pu and $^{242}$Pu within the band-1, there is an observable, smooth divergence from the characteristics of a rigid rotor, extending up to 20$\hbar$. As the spin increases to the 22-32$\hbar$ region, the deviation in the $^{242}$Pu band-1 becomes significantly more evident,indicating a departure from rigid rotor behavior. This deviation is further amplified in the case of $^{244}$Pu band-1, where it exceeds the deviation noted in the $^{242}$Pu band-1, suggesting a more pronounced departure from the rigid rotor model in $^{244}$Pu. As evident from figure \ref{fig17}, the peak deviation from the rigid-rotor curve occurs at 26$\hbar$. In comparison, $^{254}$No and $^{256}$Rf exhibit behaviors closely aligned with rigid rotor curves, maintaining a very smooth pattern across the entire spin range. The observed decrease in the R(I) values for $^{242,244}$Pu might be attributed to alterations in nuclear shape. This is consistent with the understanding that the emission of intraband $\gamma$-transitions are influenced by collective excitations, such as rotations or vibrations, which are known to affect nuclear shape. Also, at higher angular momentum ($\sim$ 28-34$\hbar$) in $^{244}$Pu band-1, the ratio R(I) starts to increase, and in the spin range of $30-34 \hbar$, it approaches very close to rigid rotor value. This increase in rigidity at higher angular momentum is also evident in the variation of $E_{\gamma}^{ROT}$ with rotational frequency at $\hbar\omega\sim 0.22 -0.275$ MeV (figure \ref{fig10}). For N=152 isotones, the softness parameter obtained is highest for the $^{248}$Cm band-1 and 3. This is again in agreement with the our previous calculations of the vibration distortion and shape fluctuation models. The systematic analysis of band-head moments of inertia (MoI) calculated using the soft-rotor formula unveils an intriguing characteristic.It is observed that the band-head MoI exhibits an inverse relationship with the softness parameter in Pu isotopes. The analysis of rotational bands reveals that those with a higher softness parameter possess the lowest values of band-head MoI, as illustrated in Figure \ref{fig18}. Notably, the softness parameter reaches its peak for the $^{244}$Pu band-1 ($\sigma \sim 30 \times 10^{-3}$), whichcorresponds to the smallest band-head MoI ($\Im_{0}\sim 48$ $\hbar^2 MeV^{-1}$) among the rotational bands of nuclei with $A\sim250$ considered in this study, as shown in figure \ref{fig18}. Similar results are also obtained for other Pu isotopes. From a classical perspective, a smaller value of the band-head MoI for nuclei suggests that these nuclei are more inclined to demonstrate faster rotational behavior within that specific band. To calculate the average MoI ($\Im_{avg}$) of rotational bands in the $A\sim250$ mass region, an empirical formula, which was supported by the dynamical features of the cranking model and particle rotor model, is used \cite{dudeja}. The empirical formula was based on the assumption that the total angular momentum consists of a rotational part $R$ and an aligned part $i$. Employing this model, the average moment of inertia ($\Im_{avg}$) calculated for Pu isotopes is found to be the highest when compared with other nuclei within the same isotonic series, as shown in figure \ref{fig19}. Moreover, it is noteworthy that the highest average MoI is observed for the $^{244}$Pu band-1, with $\Im_{avg}\sim 144$ $\hbar^2 MeV^{-1}$, making it the highest among all the rotational bands examined in this study. Analogous systematics are also evident in the calculated average aligned angularmomentum, showcasing its highest magnitude for Pu isotopes compared to nuclei with an equivalent neutron numbers. The $^{244}$Pu band-1 exhibits the most substantial value at $|i_{avg}| \sim 8.4 \hbar$, as illustrated in figure \ref{fig20}. It is important to mention that the aligned angular momentum extracted using this formula predicts a negative value. It was proposed that the calculated aligned angular momentum is only for a considerable range of angular momentum, and hence it is termed an average out value of alignment \cite{dudeja}. Consequently, the average value of the MoI is regarded as a more substantial quantity in comparison to the average aligned angular momentum. Moreover, this phenomenon of negative alignment is not unique to the current model but has been observed and discussed in earlier studies in context of s-bands of the $A\sim 150$ mass region \cite{ansari}. All the rotational bands of N = 148 to152 isotones revealed a negative value for average-aligned angular momentum $(i_{avg})$ except $^{240}$U band-2. For $^{240}$U band-2, the $i_{avg}=+0.19 \hbar$ is obtained. This finding is particularly noteworthy considering that calculations based on the shape fluctuation model also indicate that the $^{240}$U band-2 has an almost negligible contribution from shape fluctuation energy. This correlation suggests a unique structural characteristic of the $^{240}$U band-2, distinguishing it from its isotonic counterparts. Whencompared with $^{242}$Pu band-1 (isotone of $^{240}$U), $i_{avg}\sim -5.7$ is obtained and shows a pronounced increase in dynamic MoI. Since only three intraband $\gamma$-transitions are available experimentally, further investigations are required to determine whether the $^{240}$U band-2 exhibits flatness in its MoI.
\subsection{The systematics of pairing and anti-pairing favouring effect}
\begin{figure}
\centering
   \includegraphics[width=8cm]{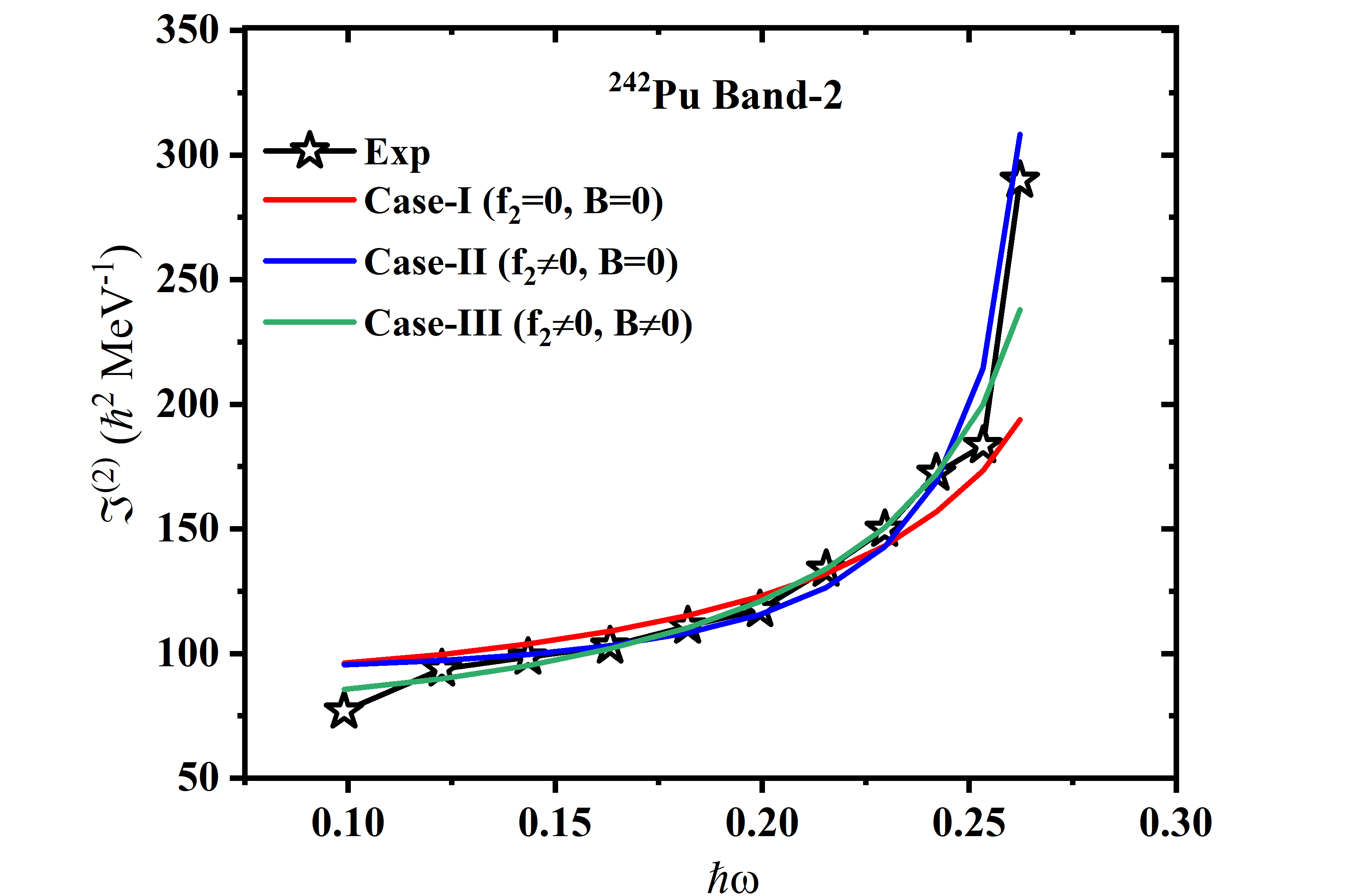}
   \includegraphics[width=8cm]{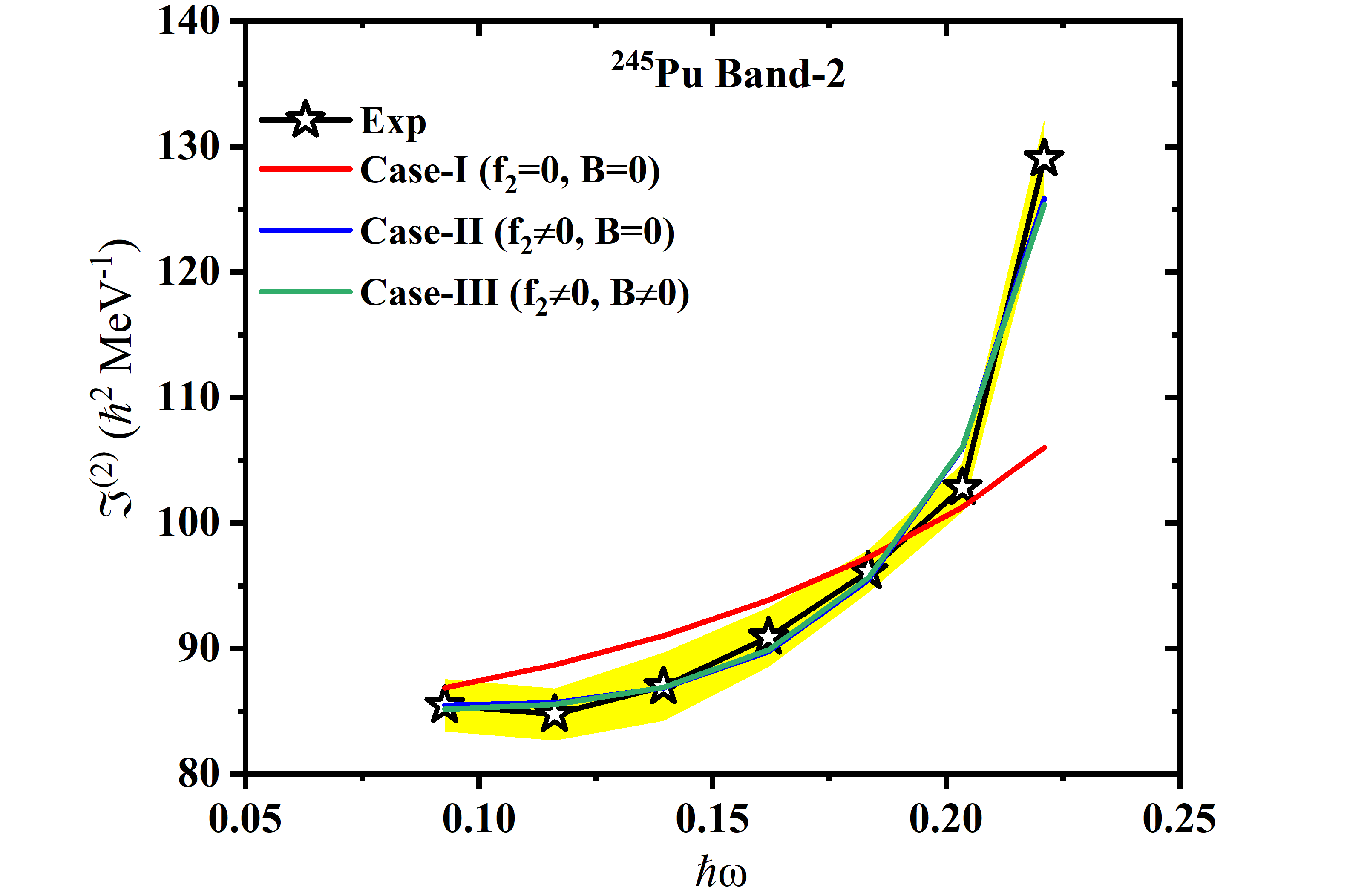}

\captionsetup{justification=raggedright, singlelinecheck=false}
\caption{(a) The variation of dynamic MoI with rotational frequency for $^{242}$Pu band-2, (b) $^{245}$Pu band-2. The experimental error in dynamic MoI is shown with yellow region.}
\label{fig21}
\end{figure}

In this section, we have studied the rotational bands of the $A\sim250$ mass region of N = 148 to 152 isotones with a four-parameter 
formula that includes a perturbation holding $SO_{sgd}(5)$ ($SU_{sdg}(5)$) symmetry in the supersymmetry scheme with many body 
interactions \cite{yuxin_prc_1}. This model is especially significant in determining the effect of pairing and anti-pairing effects in 
the rotational bands and its effect on the evolution of dynamic MoI. The model incorporates parameters $f_{1}$, and $f_{2}$, which are 
known as Arima coefficients, and determines the overall behavior of dynamic MoI \cite{Yuxin_JPG}. The systematics of Arima coefficients
in rotational bands of N = 148 to 152 reveal very interesting results. For N=148 isotones, $^{242}$Pu band-1 and 2 have both positive 
Arima coefficients ($f_{1}$, $f_{2}>0$) and $^{242}$Pu band-4 has $f_{1}<0$ and $f_{1}>0$. The systematics of other isotopes of Pu are 
also very similar in nature. For $^{243}$Pu bands 1 and 2, $^{244}$Pu bands 2 to 4, all have positive Arima coefficients $f_{1}$, 
$f_{2}>0$. The bands 1 to 3 of $^{245}$Pu have $f_{1}<0$ and $f_{1}>0$ and band-4 has $f_{1}$, $f_{2}>0$. The only bands with $f_{1}>0$ and $f_{2}<0$ in Pu isotopes are $^{242}$Pu band-3 and $^{244}$Pu band-1. The calculated dynamic MoI of $^{242}$Pu band 2 and 
$^{245}$Pu band 2 are shown in figure \ref{fig21}. We have taken three cases while calculating dynamic MoI.Case-I: The perturbation 
interaction on $SU_{sdg}(5)$ symmetry and parameter $f_{2}$ is switched off ($B=0, f_2=0$). Case-II: When $B=0, f_2\neq0$. Case-III, 
when $B\neq0, f_2\neq0$. From the figure \ref{fig21}, it is clear that the increasing dynamic MoI is reproduced very well with case-II, i.e., $B=0, f_2\neq0$. For case-I, the calculated dynamic MoI is relatively low in comparison to the experimental values. For case-III,
the calculated dynamic MoI is relatively low in comparison to the experimental values, or the same as in Case II. To discuss the 
significance of parameter $f_{2}$, a comparison with rigid rotor is made. The energy of the rigid rotor is $E_{rot}=(\hbar^{2}/2 
\Im)I(I+1)$ and its dynamical moment of inertia is constant $\Im$. When energy is written in the form of Eq. \ref{eq16}, the dynamic 
MoI of the state $I$ is
\begin{align}
\nonumber \Im^{(2)} \approx  \frac{\hbar^{2}[1+6f_{1}I(I+1)+15f_{2}I^{2}(I+1)^{2}]}{2C_{0}} \\
 \approx \frac{\hbar^2}{2C_0} \left\{1+\frac{3f_1\hbar^2\omega^2}{2C_{0}^{2}}[1+4f_1I(I+1)]\right.\nonumber \\
 + \left.\frac{15f_2\hbar^4\omega^4}{16C_{0}^{4}}[1+8f_{1} I(I+1)]\right\}
   \label{eq16}
   \end{align}
It is clear that if the parameter $f_{2}=0$, the angular momentum has a driving (restraining) effect on the dynamic MoI when $f_{1}>0$ ($f_{1} <0$). When the parameters are both positive $f_{1}$, $f_{2}>0$, or negative $f_{1}$, $f_{2}<0$, the angular momentum driving or restraining effects get enhanced \cite{Yuxin_JPG,yuxin_prc_1}. For $^{242}$Pu band-2, it is clear that when $f_{2}=0$, the calculated dynamic MoI obtained is much less than the experimental values. Only when both parameters contribute to the angular momentum driving
effect, the experimental data is reproduced well. This implies an enhanced anti-pairing effect is in play in this band. If the parameters take opposite values, i.e., $f_{1}>0$, $f_{2}<0$ and $f_{1}<0$ and $f_{2}>0$, both driving and restraining effects are included. For $|f_{1}|\gg|f_{2}|$, the $|f_{2}|[I(I+1)]^2$ term increases more rapidly than $|f_{1}|[I(I+1)]$ with increasing $I$, the two kinds of effects change from restraining dominant to driving dominant in $^{245}$Pu band-2. \par
\begin{figure}
\centering

   \includegraphics[width=8cm]{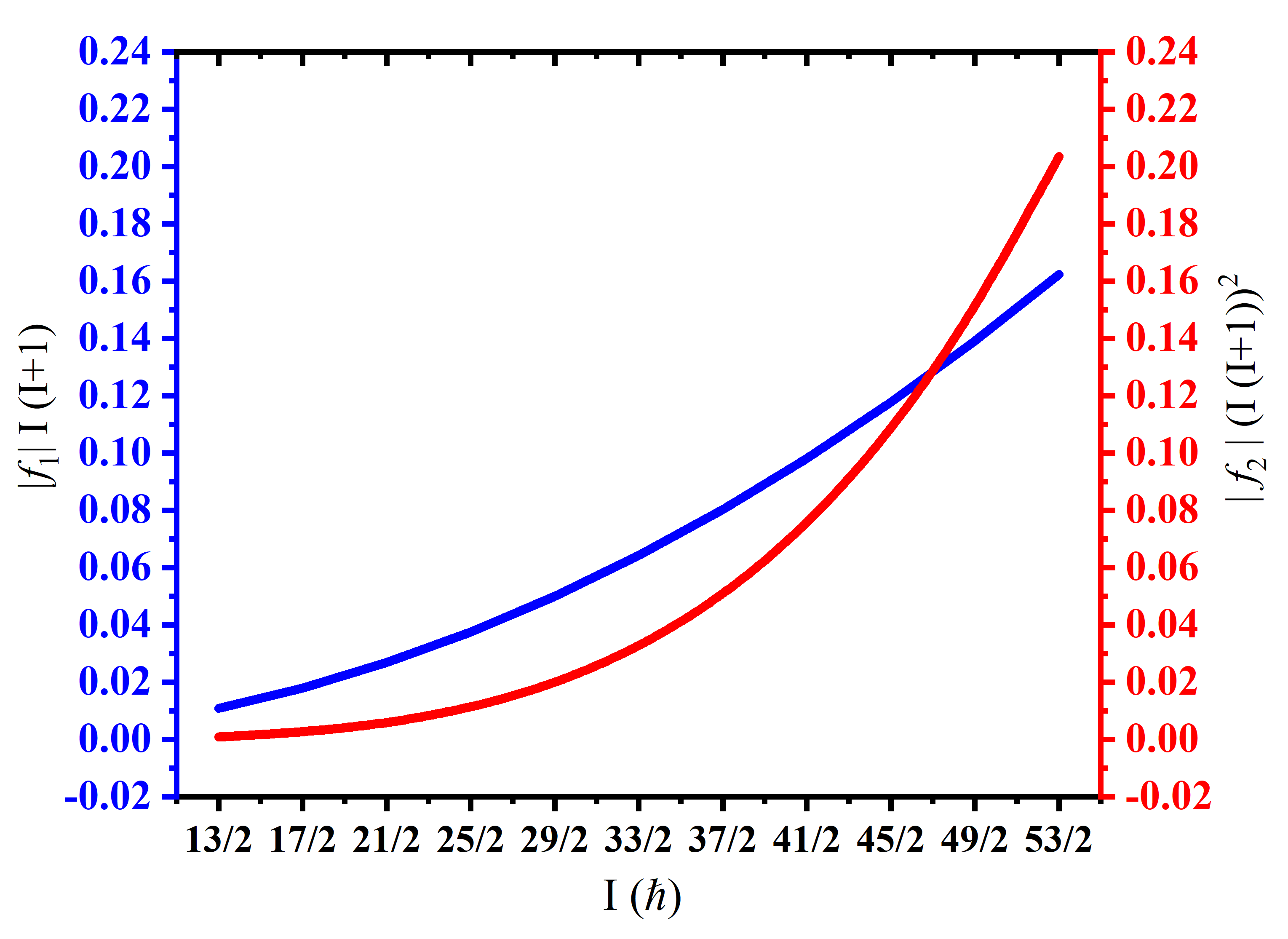}
\captionsetup{justification=raggedright, singlelinecheck=false}
\caption{The comparison of variation of parameters $|f_{1}|[I(I+1)]$ (blue Y-axis) and $|f_{2}|[I(I+1)]^2$ (Red Y-axis) with increasing spin for $^{245}$Pu band-1.}
\label{fig22}
\end{figure}
From the microscopic point of view, if the interaction is written as $f>0$, the $fI(I+1)$ term in the denominator plays a role in anti-pairing. On the same argument, if $f<0$, the $fI(I+1)$ term will play pairing favouring role. When both parameters take $f_{1}<0$, $f_{2}>0$ ($f_{1}>0$, $f_{2}<0$), both pairing and anti-pairing effects are taken into consideration \cite{Yuxin_JPG}. The switching of
pairing to anti-pairing (or anti-pairing-to-pairing) effect depends explicitly on the values of parameters $f_1$ and $f_2$. To illustrate further the change in effect, we have plotted the variation of terms $|f_{1}|[I(I+1)]$ and $|f_{2}|[I(I+1)]^2$ with increasing spin for $^{245}$Pu band-1 with $C_{0}=5.582$ KeV, $f_{1}=-2.227 \times 10^{-4}$ and $f_{2}=3.832 \times 10^{-7}$. It is clear from the figure \ref{fig22} that from spin $13/2 $ to $\sim 45/2 \hbar$ there is competition between the anti-pairing and pairing effects, and the magnitude of the pairing effect is dominating. However, in the higher spin region $\sim 45/2 $ to $ 53/2 \hbar$, the anti-pairing effect dominates over the pairing effect. The prevalence of the anti-pairing effect is evident in the energies associated with intraband gamma transitions. For instance, the energy for the transition \( E_{\gamma}(49/2 \rightarrow 45/2) \) is 464 keV,
which is lower than the 472 keV observed for \( E_{\gamma}(45/2 \rightarrow 41/2) \) \cite{nndc}. This decline in gamma-transition energy beyond the \( 45/2 \hbar \) state, coupled with the dominant influence of anti-pairing, is apparent in the plot representing the dynamic MoI. Notably, there is a marked increase in the MoI at higher rotational frequencies, as illustrated in figure \ref{fig4}(c), specifically for the category-III of N=151 isotones. The detailed analysis of the $A\sim250$ mass region where the downturn in the dynamic moment of inertia using this four parameter formula, which includes a perturbation holding $SO_{sgd}(5)$ ($SU_{sdg}(5)$)
symmetry in the supersymmetry scheme with many body interactions is underway and will be presented elsewhere \cite{AD_2024}.
\section{Summary}

A comprehensive analysis of rotational bands within isotones having neutron numbers \( N = 148 \) to \( 152 \) has been conducted using various semi-classical approaches and models. This research marks the first instance where the dynamic moment of inertia in the mass region around \( A \sim 250 \) is explored using a semi-classical, empirical model with two parameters, specifically tailored for vibrational distortion. In this context, we suggest a novel modification to the term representing vibrational distortion. For the mass region \( A \sim 250 \), it is proposed that this term be defined as \( ((\omega_{\text{max}} - \omega) / \omega_{\text{max}})^n \), where \( n \) represents an adjustable free parameter. For rotational bands of \( A \sim 250 \) mass region, \( n = -2 \) reproduces the experimental dynamic MoI extremely well. Employing the vibrational distortional model, we observed two distinct categories in the nuclei of \( N = 148 \), \( 149 \), and \( 150 \) isotones. Our findings demonstrate that in these isotones, category-I nuclei begin with a lower value of dynamic MoI, whereas category-II nuclei start with a relatively higher value of MoI. After \( \hbar\omega > 0.20 \) MeV, both categories merge. Similarly, for nuclei in \( N = 151 \) and \( 152 \) isotones, two categories are observed. However, in this case, the nuclei of categories I and II start with almost similar values, and then begin to diverge at \( \hbar\omega \approx 0.20 \) MeV. The investigation elucidates the roles of the constant term \( \Im_{c}^{(2)} \) and the vibrational distortion term \( \Im_{vib}^{(2)} \), particularly their dynamic evolution in response to increasing rotational frequencies. Our study sheds light on the behavior of Pu isotopes in this context. We recognize that in the lower frequency regions (\( \hbar\omega < 0.20 \) MeV), the contribution of the constant term \( \Im_{c}^{(2)} \) is pronounced, whereas the influence of the vibrational distortion term \( \Im_{vib}^{(2)} \) remains subdued. Conversely, in the higher frequency domain (\( \hbar\omega > 0.20 \) MeV), the vibrational distortion term assumes a dominant role, effectively driving the marked upbending observed in the dynamic MoI. This observation aligns with theoretical expectations, especially considering the higher mass region's significance on the vibrational distortion term \cite{roy_dae}. Furthermore, the interaction between the constant term \( \Im_{c}^{(2)} \) and the vibrational distortion term \( \Im_{vib}^{(2)} \) exhibits a positive coupling across all examined nuclei. This positive coupling indicates that the amplitude of the vibration aligns with the plane of rotation, thereby enhancing the magnitude of the moment of inertia.The efficacy of the vibrational distortional model likely stems from the cluster-like or binary structure of the nucleus and the coupling of octupole vibrational modes to rotational states \cite{roy_dae}. The credibility of this model is further reinforced by the observed $\alpha$-clustering behavior in actinides \cite{SHNEIDMAN1}. Given the strong $\alpha$ emission characteristics of transactinides, it is hypothesized that these nuclei also incorporate $\alpha$-cluster components in their wavefunctions \cite{ShneidmanPhysRevC.74.034316}.\par
Investigating the $A\sim250$ mass region through the shape fluctuation model uncovers comparable systematics, providing additional insights that complement the findings of the vibrational distortion model. The shape fluctuation model indicates a substantial rise in the shape fluctuation energy (SFE) for Pu isotopes at approximately $\hbar\omega\sim 0.20$ MeV. Furthermore, in the case of Pu isotopes, the SFE is roughly $\frac{1}{2}$ of the rotational energy (ROTE) at higher rotational frequencies. For other nuclei in the $A \sim 250$ mass region, the SFE constitutes approximately $\frac{1}{4}$ of the ROTE. The SFE is influenced by elements from the phonon spectrum as well as spin-dependent terms. An escalation in the magnitude of the SFE suggests a variation in the shape of Pu isotopes when subjected to higher rotational frequencies.\par
The systematic analysis of the softness parameter, computed using the nuclear-softness formula, also indicates anomalous characteristics for Pu isotopes. This parameter is notably elevated in Pu isotopes compared to other nuclei with an identical neutron count. The most pronounced values of the softness parameter are observed in the $^{242}\text{Pu}$ band-1 and $^{244}\text{Pu}$ band-1. The behavior of the softness parameter in Pu isotopes is indicative of their ``soft" nature. The literature also supports the notion of nuclei in this mass region displaying a soft behavior. Specifically, the observation of low-lying alternative parity states in both light and heavy actinides is linked to their soft nature, particularly in terms of reflection-asymmetry vibration around the equilibrium shape, as documented in studies \cite{ahmad,RevModPhys.68.349}. The observed band-head and average MoI are aligned with the distinctive characteristics of Pu isotopes, particularly in the case of \( ^{244}\text{Pu} \) band-1, where the band-head MoI is at its lowest and the average MoI is at its highest. From a classical perspective, the lower value of the band-head MoI implies that the nuclei are predisposed to exhibit higher rotational frequencies within that specific band. The unusual behavior of Pu isotopes has been explored through the use of the variable moment of inertia-inspired interacting boson model (VMI-IBM). The systematic analysis of parameters derived from VMI-IBM suggests that Pu isotopes either exhibit an enhanced anti-pairing effect or demonstrate a transition
from pairing to anti-pairing effects. Notably, the anti-pairing effect appears to be particularly significant at higher rotational frequencies.
\begin{acknowledgements}
AD would like to thank Prof. Ramon Wyss and  Prof. Timur Shneidman for many fruitful discussions. XTH would like thank the National Natural Science Foundation of China (Grant Nos. U2032138 and 11775112).
\end{acknowledgements}


\bibliography{apssamp}
\end{document}